\documentclass[11pt,a4paper]{article}
\pdfoutput=1
\usepackage{hyperref}
\usepackage{bm}
\usepackage{amsmath,amssymb}
\usepackage{cite}

\usepackage{graphicx}

\numberwithin{figure}{section}

\numberwithin{equation}{section}

\newcommand\ie{\textit{i.e.}\ }
\newcommand\eg{\textit{e.g.}\ }
\newcommand\cf{\textit{cf.}\ }

\newcommand{\viz}{{\it viz.}\ }
\newcommand{\half}{\tfrac{1}{2}}

\newcommand{\be}{\begin{equation}}
\newcommand{\ee}{\end{equation}}
\newcommand{\bea}{\begin{eqnarray}}
\newcommand{\eea}{\end{eqnarray}}

\newcommand{\ph}{\varphi}
\newcommand{\R}{{\cal R}}
\newcommand{\cT}{{\cal T}}
\newcommand{\eps}{\varepsilon}
\newcommand{\df}{\,\delta\! f}
\newcommand{\ddf}{\,\delta\!{\dot f}}
\newcommand{\Real}{{\rm Re}}

\begin{document}
\begin{titlepage}
\begin{flushright}
{\tt SHEP 12-26}
\end{flushright}

\begin{center}
{\huge \bf  Asymptotic safety in the $f(R)$ approximation.}
\end{center}
\vskip1cm


\begin{center}
{\bf Juergen A. Dietz and Tim R. Morris}
\end{center}

\begin{center}
{\it School of Physics and Astronomy,  University of Southampton\\
Highfield, Southampton, SO17 1BJ, U.K.}\\
\vspace*{0.3cm}
{\tt  J.A.Dietz@soton.ac.uk, T.R.Morris@soton.ac.uk}
\end{center}


\abstract{
In the asymptotic safety programme for quantum gravity, it is important to go beyond polynomial truncations.
Three such approximations have been derived where the restriction is only to a general function $f(R)$ of the curvature $R>0$. We confront these with the requirement that a fixed point solution be smooth and exist for all $R\ge0$. Singularities induced by cutoff choices force the earlier versions to have no such solutions. However, we show that the most recent version has a number of lines of fixed points, each supporting a continuous spectrum of eigen-perturbations. We uncover and analyse the first five 
such lines. Sensible 
fixed point behaviour may be achieved if one consistently incorporates geometry/topology change. As an exploratory example, we analyse the equations analytically continued to $R<0$,  however we now find only partial solutions. 
We show how these results are always consistent with, and to some extent can be predicted from, a straightforward analysis of the constraints inherent in the equations.}



\end{titlepage}

\tableofcontents
\vfill
\newpage

\section{Introduction}
\label{introduction}

One attempted route to a quantum theory of gravity is through the asymptotic safety programme \cite{Weinberg:1980gg,Reuter1,Niedermaier:2006wt,Percacci:2007sz,Litim:2008tt,Reuter:2012id}. Although quantum gravity based on the Einstein-Hilbert action is plagued by ultraviolet infinities that are perturbatively non-renormalisable (implying the need for an infinite number of coupling constants), a sensible theory of quantum gravity might be recovered if there exists a suitable ultraviolet fixed point \cite{Weinberg:1980gg}. 

The task is not just that of searching for an ultraviolet fixed point. `Conformal gravity', based on the square of the Weyl tensor, is perturbatively renormalisable, and corresponds to a Gaussian ultraviolet fixed point \cite{conformal}, for which the Einstein-Hilbert term is a relevant direction. It is apparently not suitable however, because the theory is not unitary.

Suitable fixed points, if they exist, have to be non-perturbative, and unitarity is
surely one of the more subtle properties that would need to be verified \cite{unitarity,Codello}. There are also phenomenological constraints, for example it has to allow a renormalised trajectory with classical-like behaviour in the infrared, as emphasised in \cite{Reuter:2004nx,Reuter:2001ag}. Of particular relevance to the current paper is that there should be a fixed point with a finite number of relevant directions (otherwise it would be no more predictive than the perturbatively defined theory). Preferably the theory should have only one fixed point, at  least only a finite number (otherwise again we lose predictivity). 

Of course the prize of finally combining quantum mechanics and general relativity into a complete predictive and unified theory cannot be overemphasised, and justify the considerable research already made into the possibility of asymptotic safety \cite{Niedermaier:2006wt,Percacci:2007sz,Litim:2008tt,Reuter:2012id}. It is also the case that functional renormalisation group\footnote{From here on we will use Wilson's own
terminology and call this the ``exact renormalisation group'' \cite{WilsonKogut}.} methods provide the ideal framework to investigate this possibility and by now provide powerful tools to attack in a different way  other longstanding problems of quantum gravity that have proved so mysterious and intractable in the past (for progress see \cite{Niedermaier:2006wt,Percacci:2007sz,Litim:2008tt,Reuter:2012id}). 

It is not possible to solve the full renormalisation group equations exactly however. In a situation such as this, where there are no useful small parameters, one can only proceed by considering model approximations which truncate drastically the infinite dimensional theory space. At least so far in such approximations, scheme independence (\ie independence on choice of cutoff) and the modified BRS Ward identities which encode diffeomorphism invariance for the quantum field \cite{Reuter1},
cannot then be recovered. Furthermore, aside from the three exceptions which are the subject of this paper, and a very recent conformal truncation study in three dimensions \cite{Demmel:2012ub}, only a finite number of operators are kept. Typically these are polynomial truncations, \ie where everything is discarded except powers of some suitable \emph{local} operators up to some maximum degree. Nevertheless,  confidence in the asymptotic safety scenario has grown strongly: one finds from these approximate models that qualitative features persist, and numerical values for universal quantities (renormalisation group eigenvalues) are reasonably stable, across many different choices of sets of operators, cutoffs and gauge fixings \cite{Reuter1,Niedermaier:2006wt,Percacci:2007sz,Litim:2008tt,Reuter:2012id}. 

This sounds very encouraging, however past studies in much simpler field theories have shown that it is easy to be misled, by even very high order polynomial truncations,  into thinking that fixed points exist \cite{Margaritis:1987hv}
 which can then be shown to be absent in a fuller description of the theory \cite{Morris:1994ki}. 

Another example may help to underline this point: in ref. \cite{Morris:1995he}, one of us analysed the exact renormalisation group for three-dimensional $U(1)$ theory in the general $f(F^2_{\mu\nu})$ approximation. Including such higher powers of $F^2_{\mu\nu}$ makes the theory perturbatively non-renormalisable. However, if any such fixed points existed, one would be able to construct new non-perturbative field theories about them. Similarities with the present case should not be lost on the reader. Despite the fact that such non-trivial fixed points \emph{do exist} in polynomial truncations of these equations \cite{Morris:1995he}, it was proven that no sensible non-trivial fixed point solution for $f(F^2_{\mu\nu})$ exists.\footnote{There is only the trivial Gaussian solution where $f$ is linear. As we review below, any sensible solution must be smooth and defined over the full domain of its argument, here $F^2_{\mu\nu}\ge0$.}

It is therefore necessary to go beyond polynomial truncations of the exact renormalisation group, in order to have confidence that the asymptotic safety scenario is realised. 
A natural starting point in the present case is to keep all powers of the scalar curvature and thus truncate the Wilsonian effective action to \cite{Machado,Codello,Benedetti}:
\be
\label{fRapprox}
\Gamma=\int\!d^4x\,\sqrt{g}\,f(R)\,.
\ee
We will refer to this as the ``$f(R)$ approximation''. As Benedetti and Caravelli
 \cite{Benedetti} have emphasised,
  it is as close as one can get to the Local Potential Approximation (LPA)  \cite{Morris:1994ki,Hasenfratz:1985dm}, a successful approximation for scalar field theory in which only a general potential $V(\ph)$ is kept for a scalar field $\ph$ (see \eg \cite{Morris:1994ki,Hasenfratz:1985dm,Morris:1994ie,Morris:1994jc,Morris:1996xq,Morris:1998da}). 

The  LPA can be viewed as the start of a systematic derivative expansion \cite{Morris:1994ie}, in which case this lowest order corresponds to regarding the field $\ph$ as constant. In rough analogy, an approximation of form \eqref{fRapprox} may be derived by working on a four-sphere where the curvature $R>0$ is constant, although this also has the effect of collapsing more general dependence on $R_{\mu\nu\lambda\sigma}$ and $R_{\mu\nu}$ to this form (since on a maximally symmetric space these are equal to $R$ times products of the metric). Using somewhat different methods, three versions of this type of $f(R)$ approximation have been derived \cite{Machado,Codello,Benedetti}. In all these cases the result is a non-linear partial differential equation which governs the evolution of $f(R,t)$ with respect to renormalisation group `time' $t=\ln (k/\mu)$. (Here $k$ is the cutoff value and $\mu$ is the arbitrary physical renormalisation scale.) For fixed points $f(R,t)\equiv f(R)$, they each reduce to third order non-linear ordinary differential equations. 

This paper is devoted to analysing these
equations and perturbations around their solutions, adapting techniques that have proved successful in scalar field theory \cite{Morris:1994ki,Morris:1994ie,Morris:1994jc,Morris:1996xq,Morris:1998da}. Despite the very different context, we will gain substantial insight by using these techniques. Indeed, our main conclusion is that
our simple parameter counting argument 
 remains a powerful way to determine what to expect from these equations in terms of fixed points and perturbations, and thus provide a first characterisation of the field theories that can be developed around them. This is particularly important given that in ref. \cite{Benedetti}, and very recently also in the three dimensional conformal truncation study of ref. \cite{Demmel:2012ub}, partial solutions are apparently found that disobey these simple rules, in particular resulting apparently in partial solutions where none are expected and a continuum of partial solutions where only a discrete set are expected. Concentrating on the equations of ref. \cite{Benedetti} we will show that after sufficient care, often requiring very high numerical precision, the parameter counting arguments actually do always work as expected.
 
Making the appropriate adaptations, the basic argument is as follows (the details are laid out in the following sections).
In the neighbourhood of a generic point $R$, the fixed point equation for $f(R)$, being a third order ordinary differential equation, has a continuous three-dimensional parameter-space of solutions. However as a consequence of the particular non-linearity of these equations, almost all of these solutions end at singularities. True fixed points can only be uncovered by insisting on the physical requirement that these solutions exist and are smooth\footnote{In fact to accommodate Litim's cutoff \cite{Litim:2001up}, we will need to relax this at certain critical points. This does not alter the analysis of whether fixed points exist or not, so we skip over this for now and incorporate it in the full analysis later.} (\ie infinitely differentiable) for all $R\ge0$. In this case, any so-called fixed singular point of the ordinary differential equation imposes a constraint that reduces the parameter space dimension.  
Furthermore the asymptotic behaviour of the solution $f(R)$ for large $R$ can be found analytically and this provides further constraints.  

Unless a miracle occurs, which would imply some hidden regularity or symmetry in the equations, all these constraints act independently. It is then a simple matter to count them up and figure out whether we should expect any smooth solutions to exist for all $R$, and if so, whether they are discrete or continuous in number.
Once we know what to expect we can also use these insights to underpin a careful numerical analysis of these equations and as we will see, confirm these expectations. 

What if quantum gravity exactly treated, does indeed have powerful hidden regularities or symmetries? We can be confident that if so, they would not survive the approximations we described above or indeed others specific to the derivations of the three $f(R)$ approximations \cite{Machado,Codello,Benedetti} (again see later). 

We will see that we encounter only the generic case for the singular points where a single pole reduces the parameter space dimension by one.
We will see that the two earliest versions of the $f(R)$ approximation \cite{Machado,Codello} have four singular points.  The equations are over constrained. Three singular points are already enough to eliminate all the parameters and leave us with at most a discrete set of fixed point solutions.  It is therefore not credible that these equations have global solutions. 

The origin of some of these singular points are single poles in $R$ due to unphysical divergences that occur for the cutoffs at these points (as has been noted in \cite{Benedetti}).   If ultraviolet fixed points do exist in this truncation, either they cannot be reached with these choices of cutoff, or such fixed points are formally described by certain singular solutions $f(R)$. Either way, we cannot make further progress mathematically.

A significant advance was made in ref. \cite{Benedetti} by choosing the cutoffs so as to avoid these singularities, however at the expense of not integrating out the lowest scalar mode on the sphere. Together with the expected singular point at the origin $R=0$, this results in one more singular point in the applicable range, at $R=R_+>0$.

In the LPA approximation for scalar field theory, an asymptotic expansion can be found for $V(\ph)$ at large $\ph$, and one finds that number of parameters is again less than would na\"ively be expected \cite{Morris:1994ki,Morris:1994ie}, although this can be understood by considering small perturbations around it. The situation here is more involved but we show how the full asymptotic expansion can again be understood in this way and we show how to derive it by breaking the equation down into smaller parts according to the order of asymptotic behaviour.
However  here we encounter a surprise. Contrary to the conclusions of ref \cite{Benedetti}, we find that the full asymptotic series solution for $f(R)$ at large $R$, has three parameters, thus providing no restriction on the dimension of the parameter space. Instead these parameters are restricted only to a three-dimensional domain (the inside of a cone). 

Since only the two singular points reduce the dimension of the parameter space, we expect one-parameter sets, \ie   lines, of smooth fixed point solutions defined for all $R\ge0$. After some extensive numerical searching, this is exactly what we find. In fact we find five
separate lines of fixed point solutions, and have evidence for at least one more.

So far this would be interesting if somewhat problematic for the asymptotic safety scenario. However, the parameter-counting is much the same for the eigen-perturbations and it shows that the result is unacceptable. The eigen-perturbations obey a \emph{linear} third-order differential equation for any given renormalisation group eigenvalue $\lambda$. The singular points at $R=0$ and $R=R_+$ each continue to reduce the parameter space by one, whilst the asymptotic series solution does not reduce the dimension. Linearity allows a further condition (normalisation) which thus reduces the dimension of the parameter space to zero. (If we include $\lambda$ in the parameter-count, we are left with one dimension.)  In the LPA approximation for scalar field theory, the extra property of linearity
over constrains the equations thus ensuring that solutions exist only for certain discrete values of $\lambda$. Here however we can expect to find a solution for a range, or ranges, of $\lambda$. Therefore if we have any eigen-operators, we have uncountably many, and operator dimensions  are not quantized. 

If this is the real physics of the situation, fixed points for quantum gravity are unlike any fixed points discovered before in quantum field theory. Furthermore the situation is hopeless for asymptotic safety, since we now have an uncountable infinity of fixed points each with either no renormalised coupling constants or an uncountable infinity of them.

Fortunately, sensible and familiar fixed point behaviour can be recovered if we find another constraint to reduce by one the dimension of the parameter space. Although the equations were derived by assuming a space of constant positive curvature (the four-sphere), evidently quantum gravity has to make sense also for negative curvature. Furthermore, one should be able to pass smoothly from one to the other, since a four-sphere with vanishing curvature, should be physically indistinguishable from flat space or a hyperbolic space with vanishing curvature. Therefore for both mathematical and physical reasons, we see that we need to extend the equations so as to incorporate consistently spaces of different topology, or at least to incorporate spaces where the local curvature can have either sign. 

Although the $f(R)$ approximation so far derived has assumed the background space is a four-sphere, we explore the above proposal by considering its analytic extension to $R<0$. Indeed in this case, we get one further singular point: at $R=R_-=-R_+$ \cite{Benedetti}. We should now thus expect at most a discrete set of fixed point solutions, and around these we should expect to find a discrete set of eigen-values and their associated eigen-perturbations. 

In sec. \ref{numerics}, we explain in detail our numerical techniques. The existence of some large numbers (in particular twice the volume of the four-sphere $2V=768\pi^2/R^2\approx 7580/R^2$), together with the singular points, which control an intricate pattern of partial solutions  as we uncover, makes the non-linear fixed point equation challenging to analyse numerically. We show how to test that sufficient accuracy has been achieved, and how this relates to the various numerical tolerances. As we will see, expanding into
some regions of parameter space requires increasing accuracy; we have had to utilise
25 significant figures.
After extensive such numerical searches however, we uncovered four lines
 of solutions in the domain $R_-\le R\le 0$ but none of these extend to $R=-\infty$. Although at first sight, a discrete set of four solutions match onto those found for $R\ge0$, we will see that, after a careful analysis of these four solutions, only three of them actually do.

We should not be too quick to draw conclusions from the null result however.
  Expecting the equations to make sense for $R<0$ is probably too optimistic, especially given the nature of the approximations used to derive them. What is required is to formulate equations that correctly take into account throughout that $R$ can be negative, however this task is beyond the scope of the present paper.

Parameter-counting is only the beginning of a mostly-asymptotic analysis of the LPA in scalar field theory, which allows one to \emph{prove} many of the expected features \cite{Morris:1996xq,Morris:1998da}.  Perturbations around the fixed point either have \emph{real} quantised eigen-values and power-law behaviour at large $\ph$, with a power that follows from dimensional reasoning  (these are the eigen-perturbations), or the perturbations grow faster than a power at large $\ph$. One can prove that the latter perturbations play no r\^ole since one can show that any solution $V(\ph,t)$ can be written as a convergent
expansion over the eigen-perturbations at any scale below the overall ultraviolet cutoff. One can prove  that only the (marginally) relevant perturbations that grow as a power of the field at large $\ph$, are associated with renormalised couplings. One can also prove that if the physical scale introduced by these renormalised couplings as they leave the fixed point, is finite, the evolution in $t$ to the far infrared, tends to a finite limit (at least in the regime of large $\ph$). In sec. \ref{perturbationanalysis}, we consider the present case. Although  all perturbations seem to behave in a reasonable way in the asymptotic regime,  the fact that quantum corrections do not decouple here limits the power of these arguments. We will show that small finite irrelevant perturbations shrink into the fixed point, although not quite in the way that would na\"\i vely be expected, and that ultimately the evolution under $t$ of finite relevant perturbations can be expected even in the asymptotic regime to be controlled by the equations at finite $R$. 

Finally in sec. \ref{conclusions}, we discuss some of the open questions and our findings in more detail, and highlight our main conclusions.

\section{A very brief review}
\label{review}
Here we introduce the $f(R,t)$ renormalisation group equations in order to set the stage for their analysis in subsequent sections. The purpose of the review is to expose as briefly as possible the key assumptions and methods that go in to deriving them. It is not our intention in the present paper to explore these in detail although we do make some comments here and in the conclusions.

The starting point for the derivation of the $f(R)$ approximation is the exact renormalisation group flow under the cutoff $k$ \cite{WilsonKogut,wegho}, written in terms of  the Legendre effective action \cite{Nicoll,Wetterich,Morris1,Morris:1998da} where it is commonly known as the effective average action \cite{Wetterich}. For a single scalar field $\ph$, it takes the form:
\be
\label{erg}
{\partial\over\partial k} \Gamma[\ph] ={1\over2}\, {\rm tr}\ \left[\R +{\delta^2 \Gamma\over\delta \ph\delta \ph}\right]^{-1}\!\!\! {\partial\over\partial k} \R\,.
\ee
It is the continuum expression of Kadanoff blocking, the first step in constructing a Wilsonian renormalisation group \cite{WilsonKogut}. The computation is performed in Euclidean signature, tr standing for a functional trace over the space-time coordinates. 
 $\R$ is some infrared cutoff function written as a momentum dependent effective mass term $\half\ph\cdot \R\cdot \ph$. Its purpose is to suppress momentum modes $p<k$. Evidently dependence of $\R$ and thus $\Gamma$ on $k$, should be understood, even though we do not indicate it explicitly.

This equation needs to be significantly adapted to cope with quantum gravity, but the basic features, the inverted two-point function and differentiated cutoff function, remain. Therefore \eqref{erg} can be interpreted schematically as the flow equation for quantum gravity, if $\ph$ is taken to stand for all the required fields. As we will see shortly, not only will we need the metric $g_{\mu\nu}$ on some Euclidean-signature manifold, but also associated ghost fields and auxiliary fields.

In the approximations we will be discussing \cite{Machado, Codello, Benedetti}, the first step is to use the background field method, where a background field ${\bar g}_{\mu\nu}$ is split off  by writing $g_{\mu\nu}={ \bar g}_{\mu\nu}+h_{\mu\nu}$, the second order functional differentiation in \eqref{erg} now being performed with respect to $h_{\mu\nu}$ \cite{Reuter1}.
Gauge fixing  and cutoff terms for $h_{\mu\nu}$ are introduced in a way that leaves diffeomorphism invariance for the background field undisturbed.  

Although some computations are carried through in general dimensions, using various choices of gauge fixing and cutoff functions, we will work exclusively in four dimensions, since the final equations for $f(R,t)$ are only reported for this case, and similarly we will quote only the final choices of gauge fixing and cutoff functions used in refs. \cite{Machado,Codello,Benedetti} to derive these equations. 

The gauge choice implemented is a limit such that Landau gauge with 
\be
{\bar\nabla}^\mu h_{\mu\nu} ={1\over4}{\bar\nabla}_\nu h\,,
\ee
is satisfied,
where ${\bar\nabla}_\mu$ is the covariant derivative with respect to the background field. Here $h=h^\mu_{\phantom{\nu}\mu}$, and indices are raised and lowered using the background field. With the gauge fixing come ghosts. They appear bilinearly in the action and these kinetic terms in turn need a cutoff to suppress their modes $p<k$. In standard fashion,
diffeomorphism invariance of the total field $g_{\mu\nu}$ becomes BRS invariance, which however is then broken by the cutoff. In principle it can be recovered once the flow $k\to0$ is complete, providing modified Ward identities are satisfied \cite{Reuter1}.

In order to facilitate the computation of the inverse of the two-point function, a transverse-traceless decomposition of the metric fluctuations is made \cite{York:1973ia,Dou:1997fg,Lauscher:2001ya,Codello:2007bd}:
\be \label{TT}
h_{\mu\nu} = h_{\mu\nu}^{T} + {\bar\nabla}_\mu \xi_\nu + {\bar\nabla}_\nu \xi_\mu + {\bar\nabla}_\mu {\bar\nabla}_\nu \sigma + \frac{1}{4} { \bar g}_{\mu\nu} {\bar h}\, ,
\ee
where $\xi_\mu$ and $\sigma$ are the gauge degrees of freedom to be distinguished from the physical traceless-transverse mode $h^T_{\mu\nu}$ and physical scalar mode, and
\be
\label{TTdefs}
h_{\mu}^{T\,\mu} = 0 \, , \quad {\bar\nabla}^\mu h_{\mu\nu}^{T} = 0
\, , \quad {\bar\nabla}^\mu \xi_\mu = 0 \, , \quad {\bar h} = h -{\bar\nabla}^2 \sigma \, .
\ee
The vector ghosts are also decomposed into their longitudinal and transverse parts. These changes of variables have associated Jacobians which then become bilinear kinetic terms for some (unphysical) auxiliary vector and scalar fields. These in turn need cutoffs to suppress modes $p<k$.

The authors of refs. \cite{Machado, Codello, Benedetti} then make three types of approximation. Firstly, $k$ dependence in the ghost \cite{Reuter1} and auxiliary terms on the left hand side of \eqref{erg} is neglected. This means that the ghost and auxiliary field terms only contribute in \eqref{erg} through inversion of their two-point functions.
Secondly, mixed terms depending on both $h_{\mu\nu}$ and $g_{\mu\nu}$ in the left hand side are neglected. This means that the $h_{\mu\nu}$ functional derivatives in \eqref{erg} can be regarded as evaluated at $h_{\mu\nu}=0$, and afterwards the background field ${\bar g}_{\mu\nu}$ and full metric $g_{\mu\nu}$ may be identified. Finally, the dependence on the full metric is truncated to that of \eqref{fRapprox}, which means that on evaluating the right hand side of \eqref{erg}, only terms that take this general form are kept.

Since it greatly helps in moulding and extracting such terms to specialise to a maximally symmetric space, all the authors choose to work on a sphere. In addition all authors choose to work with Litim's optimised cutoff profile \cite{Litim:2001up} since this also leads to technical simplifications:
\be
\label{Daniel}
r(z) = (1-z)\theta(1-z)\,.
\ee
In particular in the numerator in \eqref{erg}, one appearance of this can
appear differentiated, thus as $dr/dz=-\theta(1-z)$.

From here on there are differences in implementation. We discuss the earliest case, ref. \cite{Machado}, first. It uses a cutoff of ``type Ib''. The ``b'' refers to the use of \eqref{TT}, while ``Type I'' \cite{Codello} refers to the fact that in \eqref{erg}, $\R$ is chosen so that it modifies all appearances of the Laplacian $\Delta:=-{\bar \nabla}^2$ in the two-point functions by the replacement:
\be
\label{typeI}
\Delta \mapsto \Delta + k^2 r(\Delta/k^2)\,.
\ee
Combined with \eqref{Daniel}, it can be seen
that the numerator restricts the eigenvalues of $\Delta$ to be less than $k^2$, and in the two-point function itself this means that $\Delta$ is just replaced by $k^2$.

The space-time trace in \eqref{erg} is taken from a heat-kernel asymptotic expansion, however certain unphysical modes for the component fields in \eqref{TT} are then explicitly subtracted, namely those that give zero in \eqref{TT} and thus do not contribute to $h_{\mu\nu}$. These correspond to the constant mode of $\sigma$, vectors $\xi_\mu$ satisfying Killing's equation, and since $h$ is taken as the physical scalar component mode, also the modes ${\bar\nabla}_\mu\sigma$ that satisfy the conformal Killing equation 
\be
\label{cfk}
{\bar\nabla}_\mu{\bar\nabla}_\nu\sigma = \tfrac{1}{4}{\bar g}_{\mu\nu}{\bar\nabla}^2\sigma\,.
\ee
These excluded modes correspond to zero modes for the auxiliary field two-point functions which are also subtracted. A similar procedure is carried out for the vector ghost fields \cite{Machado}.

Finally the rescaling step of Wilson's renormalisation group is achieved by writing everything in dimensionless variables using appropriate powers of the cutoff $k$. From here on that is what we will mean by $f$ and $R$. If we rename the dimensionful curvature and Lagrangian by using a tilde, these dimensionless variables are given by
\be
\label{rescale}
{R} := {\tilde R}/k^2\qquad{\rm and}\qquad {f}({R}) := {\tilde f}({\tilde R})/k^4\,.
\ee
The result is:
\bea\label{flow1} 
&& 384 \pi^2   \left( \partial_t f + 4 f - 2 R f^{\prime} \right) = 
\\ \nonumber
&& \quad \Big[ 5 R^2 \theta\left(1-\tfrac{R}{3}\right) 
-  \left( 12 + 4 \, R - \tfrac{61}{90} \, R^2 \right)\Big]
\Big[1 - \tfrac{R}{3} \Big]^{-1}  
+ \Sigma \\ \nonumber
&& + \Big[ 10 \, R^2 \, \theta(1-\tfrac{R}{4}) - R^2 \, \theta(1+\tfrac{R}{4}) 
-   \left( 36 + 6 \, R - \tfrac{67}{60} \, R^2 \right) \Big] 
\Big[ 1 - \tfrac{R}{4}\Big]^{-1} \\ \nonumber
&& +  \Big[ (\partial_t f' +2f'-2Rf'') \, \left( 10 - 5  R - \tfrac{271}{36}  R^2 + \tfrac{7249}{4536}  R^3 \right) 
+ f'\left( 60 - 20  R - \tfrac{271}{18}  R^2 \right)
\Big] \left[ f+f'(1  - \tfrac{R}{3}) \right]^{-1} \\ \nonumber
&& + \tfrac{5R^2}{2} \, \Big[ 
 (\partial_t f' +2f'-2Rf'') \left\{r(-\tfrac{R}{3})+2r(-\tfrac{R}{6}) \right\}
+ 2 f'\theta(1+\tfrac{R}{3}) + 4 f'\theta(1+\tfrac{R}{6}) 
\Big]  
 \left[ f+f'(1  - \tfrac{R}{3}) \right]^{-1} \\ \nonumber
&& + 
\Big[
(\partial_t f' +2f'-2Rf'') f^{\prime}
\left(6 + 3 R + \tfrac{29}{60} R^2 + \tfrac{37}{1512} \, R^3 \right)\\ \nonumber
&& \qquad+ \left( \partial_t f^{\prime \prime}  - 2 R f^{\prime \prime \prime} \right) 
\left( 27 - \tfrac{91}{20} R^2  - \tfrac{29}{30} \, R^3 - \tfrac{181}{3360} \, R^4 \right) 
 \\ \nonumber
&&  \qquad + f^{\prime \prime} \left( 216 - \tfrac{91}{5}  R^2 - \tfrac{29}{15}  R^3 \right)   
+  f^\prime \left( 36 + 12  R + \tfrac{29}{30}  R^2 \right)
\Big] \Big[ 2 f + 3 f^\prime (1-\tfrac{2}{3}R) + 9 f^{\prime \prime} (1-\tfrac{R}{3})^2 \Big]^{-1} \, ,
\eea
where
\be
\label{Sigma1}
\Sigma = 10 \, R^2 \, \theta\left(1 -\tfrac{R}{3}\right)\,.
\ee
Here $f$ means $f(R,t)$, and primes indicate differentiation with respect to $R$. The explicit appearance of $r$ from \eqref{Daniel} and
the Heaviside $\theta$ functions, arise via isolated modes, either directly or as a consequence of re-expressing the heat kernels in terms of unconstrained fields. The first two lines of the right hand side are independent of $f(R)$ and encapsulate the contributions from the ghosts, auxiliaries, $\xi_\mu$ and $\sigma$. The third and fourth line arises from $h^T_{\mu\nu}$, whilst the final ratio is the contribution from ${ h}$.

In ref. \cite{Codello}, the same cutoff is used but the equation is adapted to polynomial truncations only, which means that the Heaviside $\theta$ functions are all set to one.
Unphysical modes are isolated  differently. The authors first consider a way of accounting for these excluded modes which amounts to setting $\Sigma=0$ \cite{Codello:2007bd}. Then for a different way of isolating modes, they supply results for two ways of implementing the Landau gauge limit (the second involving higher derivatives in the 't Hooft averaging procedure). The resultant effect is to change $\Sigma$ to one of three options:\footnote{According to refs. \cite{Machado,Codello} there are not meant to be any further differences, however the equations actually differ by a volume factor, and also differ in the coefficient of the $R^3 (\partial_t f' +2f'-2Rf'')/(3f+3f''-Rf')$ term. Presumably these differences are typographical errors. They do not alter our conclusions.}
\be
\label{Sigma2}
\Sigma = 0\,,\qquad - \frac{10 R^2 (R^2-20R+54)}{(R-3)(R-4)}\qquad{\rm or}\qquad   \frac{10(11R-36)}{(R-3)(R-4)}\,.
\ee

We should make some further comments on the use of the cutoff profile \eqref{Daniel} together with heat-kernel methods of evaluation as in refs. \cite{Machado,Codello}. As noted in footnote 15 of ref. \cite{Machado}, it results in appearances of $\delta^{(n)}(k)$, which were neglected. It is not clear how these should be treated or interpreted. It also results in one of the major simplifications in that the heat kernel expansion is then truncated at a finite order. However the heat kernel expansion is an asymptotic expansion for small $R_{\mu\nu\sigma\lambda}$, which means that we can expect it to have zero radius of convergence. The fact that the cutoff then truncates it to finite order does not therefore mean that the result is exact. In fact the result is smoothed implicitly by this procedure as remarked in ref. \cite{Benedetti}. The exact result has a stair-case behaviour as we will shortly review. Although in principle divergences render space-time traces ambiguous,  traces that are properly regulated by the cutoff, have an exact answer, and this is the one in principle that should be used. In practice using some smoothing approximation  is presumably as legitimate as the truncations imposed in order to make progress with the exact renormalisation group equations in the first place. 

Benedetti and Caravelli \cite{Benedetti} made a number of significant advances. They define $\Delta_0 :=\Delta-{\tilde R}/3$, $\Delta_1:=\Delta-{\tilde R}/4$ and $\Delta_2:=\Delta-{\tilde R}/6$, for scalar, vector and traceless-tensor component fields respectively. They observe that the two-point functions for the unphysical scalar and vector fields have factors of $\Delta_0$ and $\Delta_1$ respectively, causing them to vanish whenever the corresponding $\Delta_i$ has a vanishing eigenvalue. After replacement \eqref{typeI} and rescaling \eqref{rescale}, these two-point functions thus have a factor of $1-R/3$ (respectively $1-R/4$). These denominator factors  then appear through the isolated modes as poles in the first two lines of the right hand side of \eqref{flow1}.

To eliminate these singular points, they use a cutoff of type II \cite{Codello} in the following sense. For the scalar, vector and traceless tensor component fields, the Laplacian is eliminated in favour of the $\Delta_i$ respectively. Then the cutoff $\R$ is chosen differently for these component fields so that in their two-point functions the replacement \eqref{typeI} is made not for $\Delta$ but for $\Delta_i$. 

The ghost action is chosen following ref. \cite{Benedetti:2011ct}. It is formally equivalent to the standard implementation but has in particular the merit of realising exact cancellations between ghost and gauge degrees of freedom in the Landau gauge. This greatly simplifies the contribution from the unphysical scalar and vector sectors. In \eqref{TT}, ${\bar h}$ is chosen as the physical scalar field in preference to $h$. This means that the $\sigma$ modes satisfying \eqref{cfk} now do contribute to $h_{\mu\nu}$. This in particular has the merit of removing a further singular point from the renormalisation group equations. 

The space-time traces are computed exactly by recognising that use of \eqref{Daniel} on a four-sphere collapses this evaluation to a sum over certain powers of the eigenvalues $k^2\lambda_i$ of $\Delta_i$ up to the maximum $\lambda_i<1$ (where we have written the eigenvalues in scale-free form). However in order to avoid the resulting ``staircase'' behaviour, the results are smoothed by replacing them with the simplest functions matching the leading behaviour at small $R$ and also as $R\to\infty$. Multiplying through by $R^2$, the result is \cite{Benedetti}:
\be \label{flow2}
{384\pi^2} \left( \partial_t f + 4 f - 2 R f'    \right) = \cT_2 + \cT_1 +\cT^{\text{np}}_0+\cT^{{\bar h}}_0 \, ,
\ee
where the contribution from $h^T_{\mu\nu}$ is
\be
\label{T2}
\cT_2 = -\frac{20 \left(\partial_tf'-2 R f''+8
   f'\right)}{(R-2) f'-2
   f} \, ,
\ee
from the vector modes is
\be
\label{t1}
\cT_1 = -36 \, ,
\ee
from the non-physical scalar modes:
\be
\label{tnp}
\cT^{\text{np}}_0 = - 12-5 R^2 \, ,
\ee
and finally from ${\bar h}$:
\begin{multline}
\label{Tbh}
\cT^{{\bar h}}_0 =\\
 {(R^4-54 R^2-54)( \partial_tf'' -2R  f''') - (R^3+18 R^2+12)(  \partial_tf'     -2R f'' + 2    f') -36(R^2+2)(f'+6f'')
\over 2 (-9 f''+(R-3) f'-2f)}\,.
\end{multline}
Again, $f$ means $f(R,t)$, and primes indicate differentiation with respect to $R$. 

Although the singular behaviour at $R=3$ and $R=4$ has thus been avoided, singular behaviour still remains in the ${\bar h}$ sector although for different reasons. These singular points appear as algebraic zeroes in front $f'''$.

Firstly, introducing the cutoff $\R$ in order to modify the Laplacian as in \eqref{typeI}, using either $\Delta$ or the $\Delta_i$, means that $\R$ is forced to depend on the other terms in the two-point function. In the ${\bar h}$ sector this includes an $f''$ term. In \eqref{erg}, when the cutoff gets differentiated with respect to $k$, and after the change of variables \eqref{rescale}, this turns into the $\partial_t f'' -2R f'''$ term present in \eqref{Tbh} and the penultimate line of \eqref{flow1}. As Benedetti and Caravelli have noted \cite{Benedetti}, the modification \eqref{typeI} is thus responsible for turning a second-order differential equation into a third-order one (through introducing $f'''$), and at the same time removing the extra parameter by introducing a singular point at $R=0$ (through the $-2R$ coefficient).

Secondly, the lowest eigenvalue in the ${\bar h}$ sector corresponds to the constant mode. For the modified Laplacian $\Delta_0$, this eigenvalue is now negative: $\lambda_0=-R/3$. This mode is then never eliminated by the cutoff requirement $\lambda_0<1$. The $f'''(R)$ piece turns out to be multiplied by $\sum_{\lambda_0<1}(1-\lambda_0^2)$. Note that the negative eigenvalue contributes increasingly negatively to this sum once $R>3$, whilst the contribution of the positive eigenvalues falls to zero as $R$ increases. Thus there is always a positive value of $R$ for which this sum changes sign. After smoothing, this turns in to the factor $R^4-54R^2-54$ in \eqref{Tbh}, which changes sign by vanishing at the points $R=R_\pm$, where 
\be
\label{Rplus}
R_+=-R_-= \sqrt{27+3\sqrt{87}}\,.
\ee

\section{Basic requirements on the solutions to the fixed point equations}
\label{fpeqns}
 To obtain the corresponding fixed point equations we simply set $\partial_t f\equiv0$, and $f$ now stands for the $t$-independent fixed point solution $f(R)$. As already advertised, these are then third order non-linear ordinary differential equations, the first being
\bea\label{fp1} 
&& 768 \pi^2   \left( 2 f -  R f^{\prime} \right) =
\\ \nonumber
&& \quad \Big[ 5 R^2 \theta\left(1-\tfrac{R}{3}\right) 
-  \left( 12 + 4 \, R - \tfrac{61}{90} \, R^2 \right)\Big]
\Big[1 - \tfrac{R}{3} \Big]^{-1}  
+ \Sigma \\ \nonumber
&& + \Big[ 10 \, R^2 \, \theta(1-\tfrac{R}{4}) - R^2 \, \theta(1+\tfrac{R}{4}) 
-   \left( 36 + 6 \, R - \tfrac{67}{60} \, R^2 \right) \Big] 
\Big[ 1 - \tfrac{R}{4}\Big]^{-1} \\ \nonumber
&& +  \Big[ (2f'-2Rf'') \, \left( 10 - 5  R - \tfrac{271}{36}  R^2 + \tfrac{7249}{4536}  R^3 \right) 
+ f'\left( 60 - 20  R - \tfrac{271}{18}  R^2 \right)
\Big] \left[ f+f'(1  - \tfrac{R}{3}) \right]^{-1} \\ \nonumber
&& + \tfrac{5R^2}{2} \, \Big[ 
 (2f'-2Rf'') \left\{r(-\tfrac{R}{3})+2r(-\tfrac{R}{6}) \right\}
+ 2 f'\theta(1+\tfrac{R}{3}) + 4 f'\theta(1+\tfrac{R}{6}) 
\Big]  
 \left[ f+f'(1  - \tfrac{R}{3}) \right]^{-1} \\ \nonumber
&& + 
\Big[
(2f'-2Rf'') f^{\prime}
\left(6 + 3 R + \tfrac{29}{60} R^2 + \tfrac{37}{1512} \, R^3 \right)\\ \nonumber
&& \qquad - 2 R f^{\prime \prime \prime}
\left( 27 - \tfrac{91}{20} R^2  - \tfrac{29}{30} \, R^3 - \tfrac{181}{3360} \, R^4 \right) 
 \\ \nonumber
&&  \qquad + f^{\prime \prime} \left( 216 - \tfrac{91}{5}  R^2 - \tfrac{29}{15}  R^3 \right)   
+  f^\prime \left( 36 + 12  R + \tfrac{29}{30}  R^2 \right)
\Big] \Big[ 2 f + 3 f^\prime (1-\tfrac{2}{3}R) + 9 f^{\prime \prime} (1-\tfrac{R}{3})^2 \Big]^{-1} \, ,
\eea
where $\Sigma$ is given by \eqref{Sigma1}. The second is given by the above with all $\theta$ functions set to one, and $\Sigma$ given by \eqref{Sigma2}. And the third is given by
\be \label{fp2}
{768\pi^2} \left( 2 f - R f'    \right) = {\tilde \cT}_2 + \cT_1 +\cT^{\text{np}}_0+{\tilde \cT}^{{\bar h}}_0 \, ,
\ee
where
\be
\label{t2}
{\tilde \cT}_2 = \frac{40 \left(R f''-4
   f'\right)}{(R-2) f'-2
   f} \, ,
\ee
and
\be
\label{th}
{\tilde \cT}^{{\bar h}}_0 = {R\left(R^4-54 R^2-54\right)  f''' - \left(R^3+18 R^2+12\right) \left(  R f'' -     f' \right) +18\left(R^2+2\right)\left(f'+6f''\right)\over  9 f''-(R-3) f'+2f}\,.
\ee

Since the flow equations were derived on a four-sphere it is natural to assume that the domain over which solutions should exist is $0\le R<\infty$, where we include $R=0$ via the limit $R\to0$. As already mentioned in the introduction, we should insist that solutions to \eqref{fp1} or \eqref{fp2}, and all derivatives of the solutions, exist throughout this domain. Since these fixed point solutions already give us massless continuum limits, failure to satisfy these requirements would translate into unacceptable singularities in the physics implied by these fixed points. Also these singularities would be inherited by the more general field theories which would need to be built using the renormalised trajectories emanating from these fixed points.\footnote{See for example further discussion and references  in the review ref. \cite{Morris:1998da} about all these issues.} Note that the limit $R\to0$ is included in the domain, because a fundamental requirement of the Wilsonian renormalisation group is that the effective Lagrangian $f(R,t)$ must be quasi-local (for non-vanishing $k$). This means that for non-vanishing $k$, the effective Lagrangian must have a derivative expansion to all orders. Since $R$ contains two derivatives, this requirement implies that $f(R,t)$ must have a Taylor expansion in $R$ about $R=0$. 

Again, as touched on in the introduction (and see also the conclusions), these requirements can fail to be satisfied, not because of underlying difficulties in the physics, but because a poor choice of cutoff results in an ill-defined Wilsonian renormalisation group. Generally in this case it will not be clear how to match across the singularities caused by the cutoff; we then have no option but to abandon this choice of cutoff and search for a better one. 

The first fixed point equation, \eqref{fp1}, is a case in point. The appearance of explicit Heaviside $\theta$ functions means that any fixed point solutions cannot possibly have well defined derivatives to all orders. Clearly this is inherited from the fact that the cutoff profile \eqref{Daniel} itself incorporates the $\theta$ function, resulting in it having singular second and higher derivatives. However, we do not yet have to abandon this equation. We can continue if we assume that the correct matching across a $\theta(R-R_c)$ is provided by absorbing this change in a jump in $f'''(R)$ at $R=R_c$.  The equation can then be solved separately in the two regions $R<R_c$ and $R>R_c$, thus also leading to discrete changes in all the higher derivatives as we cross the boundary from one region to the other.

\section{Counting parameters in the fixed point solutions}
\label{parameters}

We are now ready to address a central theme of the paper: the counting of parameters in the solutions to \eqref{fp1} or \eqref{fp2}. We start by recalling some textbook mathematics adapted for the present purpose. We will often need to consider these equations cast in \emph{normal} form:
\be
\label{normal}
f'''(R)=rhs\,,
\ee
where $rhs$ contains no $f'''$ terms. A Taylor expansion about some generic point $R_p$ takes the form:
\be
\label{Taylor}
f(R)=f(R_p)+(R-R_p)f'(R_p)+\frac{1}{2}(R-R_p)^2f''(R_p)+\frac{1}{6}(R-R_p)^3 f'''_\pm(R_p)+\cdots\,,
\ee
where the differing values for $f'''(R_p)$ and higher derivatives are required if $R_p$ coincides with the jump in a $\theta$ function. Since \eqref{normal} determines the fourth coefficient in terms of the first three, we see that typically \eqref{Taylor} provides a series solution depending on three continuous real parameters, here 
\be
\label{generic}
f(R_p)\,,\quad f'(R_p)\quad{\rm and}\quad f''(R_p)\,,
\ee 
with some finite radius of convergence $\rho$ whose value also depends on these parameters. Therefore we recover the standard result that around a generic point $R_p$ there is some domain ${\cal D}=(R_p-\rho,R_p+\rho)$ in which we have a three-parameter set of well-defined solutions. From here we can try to extend the solution to a larger domain, \eg by matching to a Taylor expansion about another point within ${\cal D}$. A typical problem, seen also in the LPA and the derivative expansion \cite{Morris:1998da}, is that eventually, at some point $R=R_c$, dependent on the parameters, the denominator of $rhs$ develops a zero, so that as $R\to R_c$, \eqref{normal} implies 
\be
\label{movsing}
f'''(R)\sim 2c^\pm/(R-R_c)\,,
\ee
where $c^\pm$ is also some function of the three parameters (again two values are required if $R_c$ coincides with a jump in a $\theta$ function) and ``$\sim$'' means we are neglecting less singular parts.\footnote{Exceptionally,
under appropriate conditions higher powers of the linear factor $R-R_c$ can appear in \eqref{movsing}.}
Thus we see that the solution typically ends in a \emph{moveable singularity}, of form 
\be
\label{sing}
f(R)\sim c^\pm(R-R_c)^2\ln|R-R_c|\,.
\ee
The fixed point equations \eqref{fp1} and \eqref{fp2} present another challenge in that they also have \emph{fixed singular points}. These correspond in $rhs$ to explicit algebraic poles in $R$. 

\subsection{Counting parameters of the fixed point solutions of refs. \cite{Machado,Codello}}

Looking at the structure of \eqref{fp1}, we see that from the first two lines there are single poles at $R=3$ and $R=4$, as already discussed in the previous section. It is straightforward to check that they are not cancelled by factors in the numerator on either side of the relevant $\theta$ functions, \ie the first line does indeed diverge for $R\to 3$ from above, and also from below, and similarly the second line diverges for $R\to4^\pm$. 
$f'''$ appears once in the penultimate line. Rearranging the equation into normal form it is clear by inspection that the explicit single poles at $R=3,4$ still do not disappear, and indeed we also gain poles from inverting the polynomial 
\be 
R \left( 27 - \frac{91}{20} R^2  - \frac{29}{30} \, R^3 - \frac{181}{3360} \, R^4 \right)\,, 
\ee
since this appears as a factor multiplying $f'''$. The term in brackets has four single zeroes and one of them is real and positive: $R=2.0065$. We therefore have four fixed single poles in $rhs$ in the applicable domain, namely at $R=R_c=0,2.0065,3,4$. 

At these points, $f$ will end at a singularity of form \eqref{sing} unless the $f$--dependent parts in $rhs$ conspire to cancel the pole. Substituting the Taylor expansion \eqref{Taylor}, with $R_p=R_c$, we see that this requirement forces some generally non-linear combination of $f(R_c)$, $f'(R_c)$ and $f''(R_c)$ to vanish. Actually, at two of the singular points the constraints simplify to become linear:
\be
\label{34}
f'(3)=\frac{2}{3}f(3)\qquad{\rm and}\qquad f''(4)=5f'(4)-2f(4)\,,
\ee
but the constraint at the origin is:
\be
f''(0)=-\frac{2}{9}{\frac {192{\pi }^{2}{{ f(0)}}^{3}+6{{ f(0)}}^{2}+480{
\pi }^{2}{{ f(0)}}^{2}{ f'(0)}+2{ f(0)}{ f'(0)}+288{\pi }^{2}{{ f'(0)}}^
{2}{ f(0)}-9{{ f'(0)}}^{2}}{3{ f(0)}+192{\pi }^{2}{{
 f(0)}}^{2}+192{\pi }^{2}{ f(0)}{ f'(0)}-7{ f'(0)}}}\,,
\ee
and the constraint at $R=2.0065$ is also non-linear and even longer. As we argue in the next subsection, these four constraints provide four independent boundary conditions on the three parameters in our solution. Therefore the equation is over-constrained and it is not credible that any non-singular solutions exist over the range $0\le R\le4$ let alone over the full domain $0\le R<\infty$. 

Choosing $\Sigma$ to be one of the expressions in \eqref{Sigma2}, setting all $\theta$ functions to one, and repeating the analysis, we still have the same number and positions for the fixed singularities and moreover these yield exactly the same constraints. Therefore the second equation also is over constrained and thus has no smooth solutions in the range $0\le R\le4$.

\subsection{Why the counting arguments should be trusted}
\label{whycount}

If the equation \eqref{fp1} could nevertheless furnish a smooth solution compatible with these four constraints then it must be that somewhere in parameter space these constraints are no longer independent. Consider two fixed singularities bounding a domain ${\cal D}$ which is free of fixed singularities  (\eg $3\le R\le 4$). If the corresponding constraints are not always independent, instead of obtaining the expected one-parameter sets of solutions within ${\cal D}$, we will also find regions of parameter space where there are still two-parameter sets of solutions. We now explain why this is too much to expect from these equations. It is straightforward to extend the arguments to the full set of singular points and this is one reason why we are confident in these counting arguments, and the conclusion that \eqref{fp1} has no global solutions.

From the previous subsection we know that in the neighbourhood of each of the fixed singular points bounding ${\cal D}$, we have an analytic solution (\viz a convergent Taylor expansion) which, in order to avoid a singularity in $f'''(R)$, has only two parameters. Let us choose a generic point $R_p$ inside ${\cal D}$. If we are lucky, we can extend our two-parameter solution at the upper singular point down to $R_p$. In the space of parameters at $R_p$, namely \eqref{generic}, we see that the set of such solutions form a two-dimensional surface. If we are unlucky, the solution will fail at a moveable singularity of form \eqref{sing} before we reach this point. Therefore this two-dimensional surface has boundaries (which may thus also result in disconnected pieces). Now we develop the solution up to $R_p$ from the lower singular point. This will result in another two-dimensional surface with boundaries. Given that there is nothing special about the point $R_p$, unless there is some hidden simplicity or symmetry in the equations,\footnote{hidden because such regularity is clearly not evident in \eqref{fp1}} these two surfaces, if they meet at all, do so on surfaces of one less dimension, \ie on curves (with possibly disconnected pieces). Note that we can place the  point $R_p$ anywhere inside the domain, but the topology of the space of matching solutions must be independent of this choice. The only way a solution space of dimension two could then result, is if the equations already have this regularity built in across the whole domain. Although we will not directly need it, we note that these remarks apply equally well if one of the singular points is sent to $R=\infty$.

Let us explain further by considering a counter-example. Consider the third order differential equation:
\be
\label{counter}
f'''=\frac{R^2f''-2Rf'+2f}{R(R-1)D(f,f',f'',R)}\,,
\ee
where $D$ is any combination of its arguments assumed to be well behaved in the region $0\le R\le1$. This region is bounded by two fixed singularities which impose on a smooth solution the conditions:
\be
f(0)=0\qquad{\rm and}\qquad f''(1)-2f'(1)+2f(1)=0\,.
\ee
Na\"\i vely such a solution has thus only one parameter in this domain, however the solution actually has two parameters ($a,b$) and is given by the vanishing of the numerator in $rhs$ throughout the domain:
\be
f(R)= a R^2 +b R\,.
\ee
Of course the reason this is allowed as a solution to \eqref{counter} is that this is also separately a solution of the left hand side: $f'''(R)=0$. This is an example of the type of hidden simplicity that would be required to evade the counting argument. 

We can check that this sort of factorisation does not happen in \eqref{fp1}, for example functions given by the vanishing of the final denominator over some range of $R$:
\be
2 f + (3-2R) f^\prime  + (3-R)^2 f^{\prime \prime} =0\,,
\ee
do not also satisfy the final numerator. If this were the case, these functions would be solutions of the normal form \eqref{normal}. Of course, there is nothing in the derivation \cite{Machado,Codello} which would lead one to expect such a remarkable coincidence, and had it been the case it would point to a much deeper understanding of the renormalisation group in quantum gravity than currently gained from these fixed point equations.

In the next subsection, we will see what to expect from the latest fixed point equation \eqref{fp2} \cite{Benedetti}. We will then follow this up with an extensive numerical investigation in sec. \ref{numerics} from which we will confirm these expectations. This is another reason for our confidence in these counting arguments. We will also show how easy it is to be misled by the numerics if one is willing to abandon these counting arguments too quickly. 

\subsection{Counting parameters in the fixed point solutions of ref. \cite{Benedetti}}
\label{counting}

The fixed point equation is \eqref{fp2}, where the terms on the right hand side are given by \eqref{t2}, \eqref{t1}, \eqref{tnp} and \eqref{th}. $f'''$ appears once: in \eqref{th}. After casting in normal form \eqref{normal}, and using \eqref{Rplus}, we see that we have just two fixed singular points corresponding to the algebraic single poles at $R=0$ and $R=R_+$. Since we take the domain of the solution to be $0\le R<\infty$, the algebraic single pole at $R=R_-$ need not concern us. Requiring that a Taylor expansion at $R=0$ exists, we find that $f''(0)$ is determined in terms of $f(0)$ and $f'(0)$ as:
\be
\label{expra2}
a_2 = -{2\over9}\,{\frac {6{a_0}^{2}-9{a_1}^{2}+480{\pi }^{2}{{
 a_0}}^{2}a_1+2a_0a_1+192{\pi }^{2}{a_0}^{
3}+288{\pi }^{2}a_0{a_1}^{2}}{3a_0-7a_1+192{\pi }^{
2}{a_0}^{2}+192{\pi }^{2}a_0a_1}}\,.
\ee
where it will be convenient to define $a_n:=f^{(n)}(0)$. Similarly requiring a Taylor expansion at $R=R_+$, turns out to imply a quadratic constraint on $f''(R_+)$ in terms of $f'(R_+)$ and $f(R_+)$.
As before, we expect these singular points to provide independent boundary conditions, therefore we are left with at most a one-parameter set of solutions -- with possibly disconnected pieces.

As we discussed in the Introduction, another constraint is usually supplied by the asymptotic behaviour at large field. As we will see from sec. \ref{asymptotics} however, the asymptotic behaviour does not result in a reduction of parameters. Instead we find that as $R\to\infty$,
\begin{multline}
\label{asymp}
f(R) = A \,R^2 + R\left\{\frac{3}{2}A+B\cos\ln R^2 + C\sin\ln R^2\right\}\\ 
 -{\frac {3}{68}} (103B-149C) \cos \ln R^2 
-{\frac {1728}{17}}A (11C+7B ) {\pi }^{2}\cos
 \ln R^2 
 -{\frac {192}{37}} ({B}^{2}+12BC-{
C}^{2} ) {\pi }^{2}\cos \ln R^4 \\
 -{\frac {3}{68}}
 (149B+103C ) \sin \ln R^2
  +{\frac {1728}{17}
}A (-7C+11B ) {\pi }^{2}\sin \ln R^2
+{\frac {384}{37}} (3{B}^{2}-BC-3{C}^{2} ) {\pi }^{2
}\sin \ln R^4\\
+{\frac {63}{4}}A-96 (9{A}^{2}+2{B}
^{2}+2{C}^{2} ) {\pi }^{2}+
O(1/R)\,,
\end{multline}
where $A,B,C$ are three real parameters subject only to the constraint that they lie within a cone given by the inequality \eqref{safedisc} 
(otherwise $f(R)$ develops periodic singularities, as shown in sec. \ref{asymptotics}).
Therefore we should still expect one-parameter set(s) of solutions, \ie lines of fixed points. As we will see in sec. \ref{numerics}, that is precisely what we find.

If true, this would be a surprising outcome for the asymptotic safety scenario. We would need to use other arguments to decide which line of fixed points corresponds to the one realised by quantum gravity. On the chosen line of fixed points, we would have an exactly marginal coupling (\ie dimensionless non-running parameter) labelling the position of the experimentally realised fixed point on this line, and an exactly marginal operator corresponding to perturbations that move the theory along this line. We would need to use experimental measurements to determine the value of this coupling. The real problem with this conclusion however is that, as we will see in sec. \ref{perturbationanalysis}, it implies that the renormalisation group eigenspectrum is not quantised, \ie each fixed point supports a continuous infinity of operators with a continuum of scaling dimensions.

As already argued in the introduction, the resolution must lie in the fact that we have not yet considered the full domain over which quantum gravity must be defined. Although the flow equations in sec. \ref{review} were derived by working on a four-sphere, quantum gravity must also make sense for spaces where $R\le0$. In fact our insistence on quasi-locality (\cf sec. \ref{fpeqns}) means that our solution $f(R)$ already exists  in some small neighbourhood of $R=0$, and thus also for $R<0$ to some extent. As explained in the introduction, we will therefore explore this option in this paper by analytically continuing the flow equation \eqref{flow2} to $R<0$, \ie we will regard this equation and the corresponding fixed point equation \eqref{fp2}, as valid for all real values $-\infty<R<+\infty$, despite the explicit use of a four-sphere in their derivation.

It is easy to see what happens to the parameter-counting arguments now. The singular point at $R=R_-$, \cf \eqref{Rplus}, imposes a quadratic constraint on $f''(R_-)$, determining it to be one of the roots in terms of $f'(R_-)$ and $f(R_-)$. This constraint is independent of the one at $R_+$ because the fixed point equation \eqref{fp2} is not symmetric under $R\mapsto -R$. The three singular points $R=0,R_\pm$ together thus ensure that we have at most a discrete set of solutions. These solutions are however subject to the asymptotic constraint \eqref{safedisc} which now operates independently on both sides, and each branch can be expected to eliminate some of these solutions. We have checked that the asymptotic expansion \eqref{asymp} is valid also for $R<0$. However, because \eqref{fp2} is not symmetric under $R\mapsto -R$, there is no simple relation expected between the values of the asymptotic parameters $A_\pm,B_\pm,C_\pm$ for positive/negative $R$, \cf sec. \ref{asymptotics}.

Although the asymptotic behaviour is unusual, we have arrived at the expected conclusion that there exists at most a discrete set of global solutions. We can hope that, when properly derived and understood, the equations furnish just one solution: the fixed point about which asymptotically safe quantum gravity can be constructed. As we will see in sec. \ref{perturbationanalysis}, parameter-counting now shows that the eigen-spectrum behaves properly too: the renormalisation group eigenvalues are quantised. Therefore we can also hope that the relevant ones are only finite in number.

Unfortunately, as described in sec. \ref{numerics}, despite an extensive numerical investigation we found only partial solutions and no solution defined for all $R$. 

\section{Asymptotic expansion of the fixed point solutions}
\label{asymptotics}
We have seen in sec. \ref{fpeqns} that the solution must be Taylor expandable about $R=0$ as a consequence of the requirement of quasi-locality. It is for a very different reason that constraints arise as $R\to\infty$. We have already noted that partial solutions typically end at a moveable singularity, this being of form\footnote{the $\pm$ labels are not applicable now} \eqref{sing}. This is the typical situation one finds in the LPA and derivative expansions \cite{Morris:1998da}. The mere fact that the solution is well defined for all $R>R_f/\eps$, where $R_f$ is some fixed finite value and $\eps$ is small, would therefore be expected to impose some strong constraints on the solution. Furthermore, one can hope that now we are supplied with the small parameter $\eps$, we can develop the solution analytically. This is the asymptotic expansion for $f(R)$.

Note that there is no reason to expect that $f(R)$ has a Taylor or Laurent expansion in $1/R$ as $R\to\infty$. Indeed in the LPA for a single scalar field $\ph$ in two dimensions the result is a sinusoidal $V(\ph)$ corresponding to critical Sine-Gordon models \cite{Morris:1994jc}, whilst at higher orders in the derivative expansion,  $V(\ph)$ for large $\ph$ grows as some irrational power of $\ph$ \cite{Morris:1994ie,Morris:1994jc,Morris:1998da}. In both these cases, $V(\ph)$ has an essential singularity in $1/\ph$, at $\ph=\infty$.

Benedetti and Caravelli found the asymptotic expansion \cite{Benedetti}:
\be
\label{Aexp}
f(R) = A R^2 + \frac{3}{2} A R +{\frac {63}{4}}A-864{\pi }^{2}{A}^{2}
+O(1/R)\,,
\ee
(and computed to high order further coefficients as a Taylor series in $1/R$).\footnote{Actually their numerical coefficients are wrong as a result of the error in \eqref{Tbh}, but their series has the same qualitative behaviour as the one displayed here.} 
From this it would appear that requiring a fixed point solution to exist for large $R$ results in two conditions on the third order differential equation (and thus only one parameter $A$). 

So far this is very similar to the behaviour found in the LPA  in greater than two dimensions: the leading power being determined by dimensions as would be expected. However there is still an important test that should be carried out \cite{Morris:1994ki}. By linearising around the solution, one can understand what has happened to the `missing' parameters. In the LPA for scalar field theory, one finds that the asymptotic expansion indeed describes an isolated one-parameter set because perturbations corresponding to the missing parameter, grow much faster than the asymptotic expansion (in fact as the exponential of a power of $\ph$). No analytic or numerical solution is found that corresponds to such behaviour at the non-linear level. This can be understood heuristically because the linearised solution for such behaviour involves balancing the same terms that result in the moveable singularities, and thus can be expected to suffer this fate at the full non-linear level. 

Carrying out this test on the present case, we will arrive at very different conclusions. Writing $f(R)\mapsto f(R)+\df(R)$ in \eqref{fp2} and expanding to first order in $\df$ gives us just the right hand side of \eqref{perturbation} (with the left hand side set to zero). Substituting the expression \eqref{Aexp} for $f(R)$ (actually we need only the first two terms) and keeping only the leading terms in large $R$ in front of the $\df^{(n)}$, we obtain the linear ordinary differential equation:
\be
R^3\df''' -R^2\df''+6R\df'-10\df= 0\,.
\ee
This equation is invariant under scale changes $R\mapsto s R$ and thus has power law solutions. The complete solution is given in terms of a linear combination of powers $R^p$, where $p=2, 1+2i,1-2i$. Expressing the complex powers in terms of their real and imaginary parts: 
\be
\label{apart}
\df(R) = \delta A\, R^2 + \delta B\, R \cos\ln R^2 +\delta C\, R \sin\ln R^2\,,
\ee
where we have introduced two infinitesimal parameters $\delta B$ and $\delta C$, and recognised that the first term results from perturbing $A$ in \eqref{Aexp}. We have therefore discovered that \eqref{Aexp} is not an isolated one-parameter set: perturbations in the missing parameters correspond to sub-leading corrections in the large $R$ limit. 

We also see that the full asymptotic expansion including these terms cannot be made to correspond to a Taylor expansion in $1/R$. Given the complexity of \eqref{fp2} and the anticipated complexity of the series solution following \eqref{apart}, it will prove fruitful to separate it into the various orders of the asymptotic expansion by introducing a small parameter $\eps$. For this purpose define:
\be
\label{eps-scaling}
f_\eps(R)=\eps^2 f(R/\eps)\,.
\ee
where $R$ is kept finite, and we are taking the hint from \eqref{Aexp} and \eqref{apart}, that $\lim_{\eps\to0} f_\eps(R) = f_0(R)$ exists. 

(In principle there could also be  asymptotic solutions at large $R$ other than the assumed $f\propto R^2$ however no other singularity-free behaviour has been found numerically. It is straightforward to rule out analytically various ans\"atze: for example suppose that at large $R$, $f\propto R^p$ for some $p\ne2$. If $p<2$ then the $R^2$ asymptotic will dominate almost always (\ie unless $A=0$). If $p>2$ then since the right hand side of \eqref{fp2} is homogeneous in $f$, it is easy to see that to leading order the left hand side must be satisfied on its own, unless the denominator of ${\tilde \cT}_2$ or  ${\tilde \cT}^{{\bar h}}_0$ vanishes to leading order. However satisfying any of these conditions to leading order requires $p=2$, thus ruling out this ansatz.) 

Now we write:
\be
\label{defasymp}
f_\eps(R) = f_0(R) +\eps f_1(R)+\eps^2 f_2(R)+\cdots\,.
\ee
However, because a Taylor expansion in $\eps$ does not exist, the higher $f_n(R)$ actually still depend on $\eps$, although we will see that ultimately this implicit dependence can be absorbed in the parameters. The division into orders as above will still make sense because we will insist that the implicit dependence of $f_n(R)$ on $\eps$ is not sufficient to cause it to vanish as fast as $\eps$, or to cause it to diverge as fast as $1/\eps$. (This is in effect the Poincar\'e definition of an asymptotic series.) It will be helpful to write 
\be
\label{defgn}
f_n(R)=R^{2-n}g_n(R)\,. 
\ee
By \eqref{eps-scaling} and \eqref{defasymp}, the $g_n(R)$ are then invariant under the rescaling $R\mapsto s\, R$ and $\eps\mapsto s\,\eps$. Note that the higher $g_n$ will achieve this through their implicit dependence on $\eps$. It follows that a solution for $g_n$ is at the correct order in the asymptotic expansion if its behaviour at large $R$ is bracketed such that 
\be
\label{asympconds}
1/R \lesssim g_n(R) \lesssim R\,.
\ee
At lowest order we have $g_0(R)=A$, \ie 
\be
\label{f0}
f_0(R) = AR^2\,.
\ee
Rewriting the fixed point equation as a differential equation for $f_\eps(R)$ using \eqref{eps-scaling}, amounts to the substitutions $R\mapsto R/\eps$ and $f^{(n)}\mapsto \eps^{n-2} f^{(n)}_\eps$ in \eqref{fp2}, \eqref{t2}, \eqref{t1}, \eqref{tnp} and \eqref{th}. We substitute \eqref{defasymp} and expand the differential equation as a series in the explicit dependences on $\eps$. 

The lowest powers come from $O(\eps^{-2})$ terms in \eqref{tnp} and \eqref{th}. Requiring that the $O(\eps^{-2})$ part vanishes gives us a differential equation for $g_1(R)$:
\be
\label{g1eq}
R^3 g'''_1+2R^2g''_1+4Rg'_1-4g_1=-6A\,.
\ee
The equation has the expected invariance under scale changes $R\mapsto s R$ (explicit dependence on $\eps$ having dropped out). After shifting $g_1\mapsto3A/2+{ g}_1$,
this equation is also solved by power ansatz, this time with $p=1,\pm2i$. Recalling the extra power of $R$ from \eqref{defgn}, we see that the first power just perturbs the leading order solution and is rejected by the conditions \eqref{asympconds}. Expressing $R^{\pm2i}$ in terms of real and imaginary parts, we recognise that we have recovered in full form the linearised sub-leading corrections in \eqref{apart}, thus the next order term in the asymptotic series is:
\be
\label{f1}
f_1(R) = R\left\{ \frac{3}{2} A + B  \cos\ln R^2 + C\sin\ln R^2 \right\}\,.
\ee
We prefer to write the sinusoidal parts in this way but note that there are a
number of alternatives, for example we can exchange $B$ and $C$ for parameters $E$ and $\Lambda$:
\be 
B  \cos\ln R^2 + C\sin\ln R^2 = E\cos\ln(R^2/\Lambda)\,.
\ee
This form makes it clear that a dimensionful parameter ${\tilde \Lambda} =k^4\Lambda $ has crept in, which thus is multiplicative in the implicit $\eps$ dependence and scaling ($\Lambda\mapsto s^2\Lambda$).

Had we not assumed the form \eqref{f0} for the lowest order, we would have found that the dominant terms come from the same $O(\eps^{-2})$ terms as above, namely from the $-5R^2$ term in \eqref{tnp}, and the terms at this order from \eqref{th}:
\be
\label{theps}
R^2 {R^3  f''' - R \left(  R f'' -     f' \right) \over  2f-R f'}\,,
\ee
but also including the left hand side of \eqref{fp2}. This limit is actually singular:  substituting $f=f_0=AR^2$ results in  the left hand side vanishing, whilst \eqref{theps} gives $0/0$. Adding in the $\eps f_1$ dependence also, results in equations that are no longer singular, but now the left hand side of \eqref{fp2} is neglected because it is of higher order in $\eps$, whilst \eqref{th} continues to contribute at $O(\eps^{-2})$. Rearranged this is the equation \eqref{g1eq} when written in terms of $g_1$. Although the leading part \eqref{f0} is as expected by dimensions, we see that in fact the reason for its appearance is more subtle than in the LPA for scalar field theory. In the LPA, the reason is that the quantum corrections can be neglected for large field. In contrast here, the dependence at large $R$ is actually dominated by balancing the physical scalar quantum corrections against the non-physical scalar quantum corrections.

The $O(1)$ contribution comes from the next power in $\eps$: the $O(\eps^{-1})$ terms. These are found in the left hand side of \eqref{fp2}, and in \eqref{th}. Changing variables\footnote{the modulus is included so that the derivation will be applicable also when $R<0$} to $|R|=\exp u$, we obtain:
\begin{multline}
\left\{ \partial^3_u -4\partial^2_u+9\partial_u -10 \right\} g_2(u)= 
{-{315}A/{2}}+960( 9{A}^{2}+2{B}^{2}+2{C}^{2} ) {\pi }^{2} \\
-3 ( 2C+31B ) \cos2u
+6912A ( B-2C ) {\pi }^{2}\cos2u 
 -384 ( B+3C )  ( 3B-C ) {\pi }^{2}\cos 4u  \\
+3 ( 2B-31C ) \sin 2u 
+6912A ( 2B+C ) {\pi }^{2}\sin 2u 
+768 ( 2B+C )( B-2C ) {\pi }^{2}\sin 4u\,.
 \end{multline}
We see that the left hand side is translation invariant under $u\mapsto u +\ln s$, as required. The general solution is just given by \eqref{apart}, and does not satisfy \eqref{asympconds}. 

At higher orders $n\ge3$ also, the linear homogeneous part is always of this form when expressed in terms of $f_n(u)$, and thus the general solution is always given by the perturbation of parameters in \eqref{apart}. It differs for the $g_n(u)$ only through the change of variables \eqref{defgn}; these are all excluded by \eqref{asympconds}. It will therefore be the special solution we are interested in from now on.

The special solution of the above equation is lengthy but readily found. Expressed in terms of $R$:
\begin{multline}
\label{f2}
f_2(R) = g_2(R)= {\frac {63}{4}}A-96 (9{A}^{2}+2{B}
^{2}+2{C}^{2} ) {\pi }^{2}\\
 -{\frac {3}{68}} (103B-149C) \cos \ln R^2 
-{\frac {1728}{17}}A (11C+7B ) {\pi }^{2}\cos
 \ln R^2 
 -{\frac {192}{37}} ({B}^{2}+12BC-{
C}^{2} ) {\pi }^{2}\cos \ln R^4 \\
 -{\frac {3}{68}}
 (149B+103C ) \sin \ln R^2
  +{\frac {1728}{17}
}A (-7C+11B ) {\pi }^{2}\sin \ln R^2
+{\frac {384}{37}} (3{B}^{2}-BC-3{C}^{2} ) {\pi }^{2
}\sin \ln R^4
\end{multline}
This $O(1)$ part of the asymptotic expansion is the last piece that can be derived in closed form. 

The asymptotic expansion is retrieved by substituting back into \eqref{defasymp} and setting the formal parameter $\eps=1$. We see that, together with \eqref{f0} and \eqref{f1}, we have now derived the already advertised form \eqref{asymp}.

The $O(1/R)$ term comes from $O(\eps^0)$ in the fixed point equation. All parts of \eqref{fp2} now contribute. This gives us a differential equation of the form:
\be
\label{g3eq}
\left\{ \partial^3_u -7\partial^2_u+20\partial_u-24\right\} g_3(u) = {\cal N}(u)/d(u)\,,
\ee
where the numerator ${\cal N}$ is too long to display. It is quartic in the parameters, and linear in sines and cosines with arguments $2nu$ ($n=0,1,2,3,4$).  In terms of $R$, the denominator is given by:
\be
\label{singpoint}
d(R)=11A/2+(B-2C)\cos\ln R^2 + (2B+C)\sin\ln R^2 \,.
\ee
The general solution of \eqref{g3eq} just reproduces \eqref{apart} for $f_3(R)$ and is disallowed by \eqref{asympconds}, as already mentioned. The special solution  can be expressed in integral form. The important feature is that it manifests moveable singularities \eqref{sing} whenever $d(R)$ has a zero.\footnote{We have verified that ${\cal N}$ does not contain $d(u)$ as a factor. (In fact $d(u)$ first appears in the derivation of the $g_2$ equation where however it is a factor that appears also in the numerator.)} It is straightforward to see that $d(R)$ has real zeroes if and only if the following is violated:
\be
\label{safedisc}
\frac{121}{20}A^2 > B^2 + C^2\,,
\ee
\ie the inequality is violated if the parameters lie outside the cone with axis $A$, vertex at the origin, and opening angle $2\tan^{-1}\left(\frac{11}{2\sqrt{5}}\right)\approx 135.8^\circ$. 
We see that if the inequality is violated the asymptotic series predicts an infinite series of such singularities $R_c$ with multiplicative periodicity $R_c\mapsto R_c\, {\rm e}^\pi$, although only those with large enough $R_c$ can be trusted. Indeed we find that, integrating out numerically from the fixed singular point $R_+$, if we can reach large enough $R$, we can extract reliable values for $A,B,C$ by using the first three orders of the asymptotic expansion \eqref{asymp}.

At $O(1/R^n)$ with $n\ge2$, the equations take a similar form to \eqref{g3eq} and can also be solved in terms of integrals of ratios of sine and cosine series. The denominator has a factor which is an increasing power of $d(R)$, however a new factor also arises, coming from the denominator in \eqref{t2}:
\be
15A/2+(B-2C)\cos\ln R^2 + (2B+C)\sin\ln R^2\,.
\ee
Note that this is of the same form as \eqref{singpoint} but is a cone with larger opening angle. 

The asymptotic series solution as $R\to-\infty$ is the same \ie as given by \eqref{asymp}, developing moveable singularities \eqref{sing} at the zeroes of \eqref{singpoint}, if \eqref{safedisc} is violated. The derivation is exactly as above, since we have been careful to ensure that all the intermediate steps are valid also for $R<0$.  As mentioned at the end of sec. \ref{counting}, we expect for any given fixed point solution that the values of $A,B$ and $C$ are different in the two sectors $R\to\pm\infty$.  As a function of complex $R$, this can be expected to arise because the asymptotic series is multivalued and thus subject to the Stokes phenomenon.

\section{Analysis of eigen-operators}
\label{perturbationanalysis}
In this section we develop a similar analysis to secs. \ref{parameters} and \ref{asymptotics} for the perturbations around a fixed point solution $f(R)$ of \eqref{fp2}. We first take the perturbations to be strictly infinitesimal and then go on to investigate asymptotic properties of finite perturbations. To begin, we substitute $f(R,t)=f(R)+\df(R,t)$ into \eqref{flow2} and expand to first order in the infinitesimal $\df(R,t)$. Writing $ \delta\!{\dot f} \equiv \partial_t\df$, the result is:
\begin{multline}
\label{perturbation}
384\pi^2\ddf +\frac{20\ddf'}{(R-2)f'-2f} + 
 \frac{(R^4-54R^2-54)\ddf''-(R^3+18R^2+12)\ddf'}{2\left(9f''-[R-3]f'+2f\right)}
=\\
768\pi^2\left(R\df'-2\df\right)+\frac{40 \left(R \df''-4\df'\right) - {\tilde\cT}_2\left( [R-2]\df'-2\df \right)}{(R-2)f'-2f}+\\
\frac{(R^4-54R^2-54)R\df'''-(R^3+18R^2+12)\left(R\df''-\df'\right)+18(R^2+2)(\df'+6\df'')}{9f''-[R-3]f'+2f}\\
-{\tilde\cT}^{\bar h}_0\frac{9\df''-[R-3]\df'+2\df}{9f''-[R-3]f'+2f}\,,
\end{multline}
where we have utilised the expressions (\ref{t2},\ref{th}). Note that we have split the equation so that the $t$-derivative terms appear exclusively on the left hand side. The left hand side pieces of \eqref{flow2} give the first pieces appearing on the left and right of this equation, then in order we have the parts from perturbing \eqref{T2} and \eqref{Tbh}. 

We search for perturbations with a characteristic scaling dimension under the exact renormalisation group \ie for separable solutions of the form:
\be
\label{eigenpert}
\df(R,t) = v(R)\, \alpha\exp\left(-2\lambda t\right)\,,
\ee
where here $\alpha$ is some dimensionless normalisation constant (and the factor of $-2$ has been included for convenience). The expectation is that the $v(R)$ are the eigen-operators for this fixed point, \ie the operators with well defined scaling dimension.\footnote{Actually since these are perturbations of  \eqref{fRapprox} they only yield the integrated operators. Operators that can be written as a total covariant derivative are excluded.} It is also expected that the solutions are quantised so that we obtain a countable infinity of such operators, and that these span the space of all possible interactions at this level of truncation. The associated couplings follow from the $t$-dependent coefficients after untieing the transformation \eqref{rescale}. Recalling that $t=\ln(k/\mu)$, where $\mu$ is some finite physical scale, we see that they have definite mass-dimension at the fixed point and are given by  ${\tilde g} = \alpha\mu^{2\lambda}$. The couplings with $\lambda<0$, $\lambda=0$ and $\lambda>0$ are respectively irrelevant, marginal and relevant. It is expected that the last, together with any marginally relevant couplings, evolve into the running renormalised couplings which parametrise the general continuum limit associated to this fixed point, \ie the most general renormalised trajectory out of this fixed point, and that these couplings are finite if the corresponding $\alpha\mu^{2\lambda}$ are finite \cite{Morris:1996xq,Morris:1998da}. Although these are natural expectations for a Wilsonian renormalisation group \cite{WilsonKogut}, they are not necessarily true in the present context. In this section we will see how far we can go in verifying or refuting these expectations.

Substituting \eqref{eigenpert} into \eqref{perturbation} we obtain:
\begin{multline} 
\label{eigen}
-\lambda\left\{ 768\pi^2v +\frac{40v'}{(R-2)f'-2f} + 
 \frac{(R^4-54R^2-54)v''-(R^3+18R^2+12)v'}{9f''-[R-3]f'+2f}\right\}
=\\
768\pi^2\left(Rv'-2v\right)+\frac{40 \left(R v''-4v'\right) - {\tilde\cT}_2\left( [R-2]v'-2v \right)}{(R-2)f'-2f}+\\
\frac{(R^4-54R^2-54)Rv'''-(R^3+18R^2+12)\left(Rv''-v'\right)+18(R^2+2)(v'+6v'')}{9f''-[R-3]f'+2f}\\
-{\tilde\cT}^{\bar h}_0\frac{9v''-[R-3]v'+2v}{9f''-[R-3]f'+2f}\,,
\end{multline}
This is a linear third order ordinary differential equation for $v$. Thus in the neighbourhood of a generic point $R_p$, we can construct a Taylor expansion solution with three parameters. However, since the equation is linear, one of these parameters is trivial, being just the overall scale. Choosing some convenient normalisation, we are left with a two-dimensional parameter space. Since $f(R)$ is now a fixed global solution to the fixed point equation \eqref{fp2} and $v(R)$ does not appear in the denominators or in the coefficient of $v'''$, there is no opportunity for moveable singularities. Therefore the constraints on finding a global smooth solution for $v$ can only arise from the fixed singularities and its behaviour as $R\to\infty$.

\subsection{Asymptotic analysis of eigen-operators}
\label{asymptoticeigen}

We can analyse what constraints arise as $R\to\infty$, by developing the asymptotic solutions of \eqref{eigen}. To find the asymptotic expansion we follow a similar analysis to that of sec. \ref{asymptotics}. We can use the freedom to choose a normalisation to eliminate any power of $\eps$ in front of $v$. Therefore we define:
\be
\label{veps-scaling}
v_\eps(R) = v(R/\eps)\,,
\ee
where $R$ is kept finite, and expand as 
\be
\label{defvasymp}
v_\eps(R) = v_0(R)+\eps v_1(R) + \eps^2 v_2(R) + \cdots\,,
\ee 
where the $v_n$ also have implicit dependence on $\eps$. The combinations $v_n R^n$ will be invariant under $R\mapsto s R$ and $\eps\mapsto s\,\eps$, after multiplication by some power of $s$  to be determined. [This is the missing power in \eqref{veps-scaling}.] We rewrite the eigen-operator equation \eqref{eigen} as a differential equation for $v_\eps(R)$ with coefficients dependent on $f_\eps(R)$. This amounts to the substitutions $R\mapsto R/\eps$, $f^{(n)}\mapsto \eps^{n-2} f^{(n)}_\eps$ and $v^{(n)}\mapsto \eps^n v^{(n)}_\eps$. We substitute \eqref{defasymp} together with (\ref{f0},\ref{f1},\ref{f2}), and \eqref{defvasymp}, and then expand the differential equation as a series in the explicit dependences on $\eps$. As before the equations are most easily solved by changing variables as $|R|=\exp u$. Here we report the results exclusively in terms of $R$ however. 

The lowest power comes from $O(\eps^{-1})$ pieces in the third terms on the left and right of \eqref{eigen}, \ie originally from \eqref{Tbh}, yielding:
\be
\label{v0eq}
R^3 v_0'''-R^2v_0''+6Rv_0'-10v_0 +\lambda\left\{ R^2v_0''-Rv_0'\right\}=0\,.
\ee
As expected, this equation is invariant under scaling $R\mapsto s R$. It is therefore solved by a power ansatz. We find that for generic $\lambda$, the three linearly independent solutions are $v_0=R^2$, and $v_0=R^{p}$, where $p$ can take two values: 
\be
\label{pdef}
p = 1-\frac{\lambda}{2}\pm\frac{1}{2}\sqrt{\lambda^2-4\lambda-16}\,.
\ee
These give the power of $s$ required for invariance of the solutions, namely $s^{-2}$ and $s^{-p}$ respectively. 

From the perspective of the LPA and scalar field theory, these power law solutions are surprising because they are not the ones expected by dimensional reasoning. Given that the mass-dimension of the associated coupling ${\tilde g}$ is $2\lambda$, this reasoning would lead us to expect that $v_0\sim R^{2-\lambda}$. However this reasoning is justified when the quantum corrections can be neglected in the large field regime. Here in contrast we see that the asymptotic behaviour of eigen-operators in this regime is controlled by the physical scalar quantum corrections \eqref{Tbh}. 

The fact that one of the solutions, namely $v_0=R^2$, is independent of $\lambda$,  is also novel. The reason for this is that it solves separately both the term in curly brackets in \eqref{v0eq} and the rest of this equation, \ie to leading order separately both the left hand side and the right hand side of \eqref{eigen} vanish. If this were to persist at sub-leading orders it would mean that there is a solution to \eqref{perturbation} whose  $t$-dependence is undetermined since $\ddf = {\dot g}(t)\, v_0$ would solve the left hand side and separately $\df = g(t)\,  v_0$ would solve the right hand side, for any choice of $g(t)$. We will see however that at sub-leading orders the two sides are re-coupled and furthermore, all three linearly independent solutions are coupled together by the constraints stemming from the finite singular points.

Expanding to next order, we look for a $v_1(R)$ that is $\sim O(1/R)$ down (the precise definition  being phrased as in \eqref{asympconds} for $R\, v_1$). The next order comes from $O(\eps^0)$ pieces which involve again the third terms on each side of \eqref{eigen} and now also the first terms, originally from the left hand side of \eqref{fp2}.
 The first terms just receive the now known $v_0$. In the third terms, $d(R)$ appears as a factor which cancels on top and bottom just as in the derivation of $f_1$. The result is 
\be
\label{v1eq}
R^3 v_1'''-R^2v_1''+6Rv_1'-10v_1 +\lambda\left\{ R^2v_1''-Rv_1'\right\}=
R^{p-1} \left( c_1+c_2\cos\ln R^2+c_3\sin\ln R^2\right)\,,
\ee
for $v_0=R^p$, and of the same form for $v_0=R^2$ but with $p=2$ on the right hand side.
The $c_i$ are functions of $\lambda$ and the parameters $A,B,C$.  Comparing \eqref{v0eq}, we see that the general solution just reproduces the three $v_0$ solutions and is therefore rejected. It is as before the special solution we are after. For the two powers \eqref{pdef} it is simplest to express the result in terms of $p$, inverting \eqref{pdef} to express $\lambda$ as:
\be
\label{lambda-p}
 \lambda = 2-p-5/p\,.
 \ee
Combined with the leading order and setting $\eps=1$ we thus obtain:
\begin{multline}
\label{Rp}
v(R) = R^p + R^{p-1}\Big\{ \frac{3\left(384\pi^2[p^2-2p-25]A+p[p+30]\right)}{(p-3)(p^2-p-5)}\\
+\frac{768\pi^2(p^2-2p-5)\left[\gamma(-C,B)\cos\ln R^2+\gamma(B,C)\sin\ln R^2\right]}{(p^2-6p+13)(p^4-2p^3-5p^2+10p+25)}\Big\}+O(R^{p-2})\,,
\end{multline}
where we have written
\be
\gamma(x,y):=2(2p^2-4p-5)x+(p^3-4p^2-6p+15)y\,.
\ee
And for the remaining solution we obtain:
\be
\label{R2}
v(R)=R^2 + R\left\{ \frac{6(960\pi^2\lambda A-6\lambda+1)}{\lambda+4}+\frac{768\pi^2}{5}\left([B-2C]\cos\ln R^2+[C+2B]\sin\ln R^2\right)\right\}+O(1)\,.
\ee
Again this is as far as we can take the asymptotic expansion in closed form. The next and higher orders can be expressed as an integral of ratios of terms, similar to the asymptotic expansion for $f(R)$. Since \eqref{eigen} is linear in $v$, the denominator terms are only those generated by $f$. Therefore we obtain no further asymptotic restriction on parameters than  \eqref{safedisc}. 

So far we have assumed that $\lambda$ is some generic value. There are a host of special cases. In \eqref{v0eq}, if $\lambda=-5/2$ or $\lambda=2\pm\sqrt{5}$ then the powers are degenerate, since in the first case $p=2,\frac{5}{2}$, and in the second two cases the square-root vanishes in \eqref{pdef} and the two values of $p$ coincide as $p=\pm\sqrt{5}$ respectively. Thus in these cases there is a further solution $R^p \ln R^2$, where $p=2$ or $p=\pm\sqrt{5}$ respectively. Higher order terms for these new solutions can be developed as above.

If $\lambda=-4$ in \eqref{R2}, or such that it corresponds to $p$ being a root of one of the denominators in \eqref{Rp}, then these expansions are no longer applicable. Instead the special solutions need to be recomputed and have a new piece proportional to $R \ln R^2$. After this the higher order terms again may be computed as above.

We will ignore these special cases in what follows. The main property we will require is that there are three unique asymptotic solutions, \ie apart from an overall scale factor, fixed here so that the first coefficient in \eqref{Rp} and \eqref{R2} is set to one, there are no further parameters in them beyond $\lambda$ and the $A,B,C$ parameters inherited from $f(R)$. This is true for the special cases also. Using these as asymptotic boundary conditions to \eqref{eigen}, we thus find three unique solutions, which however can be expected to develop singularities at the fixed singular points. 

Since polynomial truncations result in complex eigenvalues \cite{Reuter1,Niedermaier:2006wt,Percacci:2007sz,Litim:2008tt,Reuter:2012id}, and there is no obvious reason for  \eqref{eigen} to exclude complex eigenvalues, we will not assume that $\lambda$ has to be real. In \eqref{eigen}, the coefficients of $v^{(n)}$ and $\lambda v^{(n)}$ are real. Therefore, if a complex $\lambda$ and thus complex $v$, satisfies the equation, then so does ${\bar\lambda}$ and ${\bar v}$. In this case we obtain the required real perturbation by adding the two $t$-dependent complex conjugate eigen-perturbations \eqref{eigenpert} together with complex conjugate coefficients $\alpha$.

If $\lambda$ is complex, then from \eqref{lambda-p}, evidently $p$ is also complex. Since \eqref{lambda-p} is invariant under $p\mapsto5/p$, if $p$ is one power corresponding to $\lambda$, then the other power is $5/p$.  From here on, we use this freedom to define $p$ such that $|p|\ge\sqrt{5}$. If $|p|>\sqrt{5}$ then this picks out uniquely one of the roots. If $|p|=\sqrt{5}$, then the other root is ${\bar p}$. In this case, writing $p=\sqrt{5}\exp i\theta$, we see that 
\be
\label{in-between}
\lambda=2-2\sqrt{5}\cos\theta
\ee
is real. In this case the two asymptotic solutions \eqref{Rp} should be added together with complex conjugate coefficients, yielding a real result where the complex powers can be re-expressed in terms of $R^{\sqrt{5}\cos\theta} \cos \left(\sqrt{5}\sin\theta \ln R\right)$ and $R^{\sqrt{5}\cos\theta}\sin\left(\sqrt{5}\sin\theta\ln R\right)$. 

Define now $w_+$ to be the unique solution whose behaviour as $R\to\infty$ satisfies \eqref{Rp} for our chosen $p$. Similarly define $w_-$ to be the unique solution whose behaviour matches \eqref{Rp} for power $p\mapsto 5/p$. Finally define $w$ to be the unique solution that matches \eqref{R2} as $R\to\infty$. The general solution for $v(R)$ is therefore given by:
\be
\label{vgeneral}
v(R) = \alpha_+w_+(R)+\alpha_-w_-(R)+ w(R)\,
\ee
where the $\alpha_\pm$ are two coefficients, and for the sake of argument we have chosen to normalise $v(R)$ so that as $R\to\infty$, its $R^2$ term has unit coefficient, as follows from \eqref{R2}. (This is allowed providing this coefficient does not vanish. The following arguments can be generalised as appropriate if this happens.)

For later use we note that if $\Real(p)>2$, $w_+(R)$ will be the dominant contribution at large $R$, so that $v(R)\approx \alpha_+ R^p$ in this regime, and from \eqref{lambda-p},
\be
\Real\lambda=2-\Real(p)-5\Real(p)/|p|^2<-10/|p|^2\,,
\ee
hence these correspond to irrelevant operators. 

On the other hand if $\Real(p)<2$, then $w(R)$ is the dominant contribution at large $R$, implying that in this regime $v(R)\approx R^2$. In this case, $\Real\lambda>-10/|p|^2$, so the relevancy or irrelevancy of these operators is not yet determined.

\subsection{Counting parameters in the eigen-operator solutions}
\label{eigencounting}

If $\lambda$ is real and satisfies 
\be
|\lambda-2|>2\sqrt{5}\,,
\ee
then $p$ is real and the $\alpha_\pm$ give us two real parameters. If $\lambda$ is real but does not satisfy the above, then it is of form \eqref{in-between} and the $\alpha_\pm$ are a complex conjugate pair. Therefore again the $\alpha_\pm$ supply two real parameters. If $\lambda$ is complex, then the $\alpha_\pm$ supply two complex parameters. In summary, in the appropriate field over which $\lambda$ is defined (\viz real or complex), the $\alpha_\pm$ supply two parameters. Therefore so far we have no restriction on the number of parameters in the solution $v(R)$ (or their domain) beyond the fact that we can eliminate one of them with a normalisation condition. As we have already mentioned in the introduction, in the LPA for scalar field theory the asymptotic behaviour of the eigen-operators divides into two classes: those that grow as a power of the field $\ph$ and those that grow as the exponential of a power. The latter are excluded by several arguments \cite{Morris:1996xq,Morris:1998da},  as already reviewed at the end of the introduction.
 Here there seems to be no reason to exclude some of the asymptotic behaviour in \eqref{Rp} or \eqref{R2}. We will explore this in more detail in sec. \ref{further}.

As we found in the fixed point equation, there are no fixed singularities coming from the denominators in \eqref{eigen}; they only arise from the coefficient of $v'''$. Since this is the same as the coefficient of $f'''$ in \eqref{fp2}, we have again three fixed singularities with locations $R=R_c=0,R_\pm$. Since the differential equation is linear in $v$, requiring that $v$ is smooth means that these single poles provide linear constraints on the derivatives of $v$ at $R_c$; they take the form 
\be
\label{constraint}
h_2\, v''(R_c)+h_1\, v'(R_c)+h_0\, v(R_c) =0\,. 
\ee
As is obvious from \eqref{eigen}, the coefficients have a linear dependence on $\lambda$ as $h_i = h^1_i \lambda + h^0_i$, and the $h^j_i$ have complicated dependence on $f(R_c), f'(R_c)$ and $f''(R_c)$. Unfortunately even the constraint at $R=0$ is too long to make it worthwhile displaying. The main point is that as a result, a normalised smooth solution $v(R)$ satisfies a constraint \eqref{constraint} at each singular point.

If we take the view that the applicable domain is $0\le R<\infty$, the first of the situations discussed in sec. \ref{counting}, then only the singular points at $R=0$ and $R=R_+$ are operative. For the same reasons as discussed in sec. \ref{whycount}, we can be certain that the corresponding constraints act independently. Therefore for each $\lambda$, the two parameters $\alpha_\pm$ in \eqref{vgeneral} are fixed to a discrete set of values. Regarding $\lambda$ as just another parameter, we have three parameters but only two constraints, therefore we will find solutions for continuous range(s) of $\lambda$ (two-dimensional domain(s) if $\lambda$ is complex). Now, given that the fixed point solution $f(R)$ is one of a line of fixed points, we know that one eigen-value is $\lambda=0$, corresponding to the perturbation that takes us to a new fixed point $f(R)+\delta\!f(R)$. There is no reason to expect this eigen-value to  be always at the end of a range, therefore typically the neighbourhood of $\lambda=0$ also provides solutions, and thus includes $\Real(\lambda)>0$. Thus we see that almost always, each fixed point on the line supports an uncountably infinite number of relevant eigenvalues. Since these would correspond to an uncountably infinite number of renormalised couplings, this is of course not an acceptable starting point for a predictive theory of gravity. The only way to escape this conclusion is if the $\lambda=0$ eigen-value lies at the end of a range. In this case however we will either again find that $\Real(\lambda)>0$ provides an uncountably infinite number of relevant eigen-values, or $\Real(\lambda)>0$ is excluded -- in which case we have no relevant operators at all and thus in the limit the theory has no adjustable couplings and is scale-invariant. As already discussed in the introduction and sec. \ref{counting}, the problem is that the eigen-operator  equation is not yet over-constrained.

As we discussed in the introduction and sec. \ref{counting}, this situation is resolved if we smoothly match into spaces with $R<0$, as we argue we physically are required to do in any case. Taking \eqref{fp2} as working example, we now have all three singular points ($R=0,R_\pm$) and these supply three independent constraints. Fixed points themselves now form a discrete set. As discussed in sec. \ref{numerics}, unfortunately for the current equation \eqref{fp2} this set is almost certainly empty. If there were such a fixed point then, given $\lambda$, the two parameters $\alpha_\pm$ are over constrained. Regarding $\lambda$ as one more parameter, we see that we would now find a discrete set of solutions and thus quantised renormalisation group eigen-values, as expected.

\subsection{Asymptotic properties of finite perturbations}
\label{further}
As already mentioned, \cf below \eqref{pdef}, the powers involved in the asymptotic solutions \eqref{Rp} and \eqref{R2} are not the ones expected from dimensional reasoning.  This can be traced to the fact that the quantum corrections in the flow equation do not decouple for large $R$ (see also the comments in the conclusions). On the other hand, in the analysis of the LPA for scalar field theory at large field, the recovery of the powers expected by dimensional reasoning is an important first step in proving that the eigen-perturbations have sensible properties when used as small but finite seeds for the renormalisation group flow out of the fixed point, distinguishing these from the non-power-law perturbations which are superfluous and should be discarded,
 as already reviewed at the end of the introduction, also see \cite{Morris:1996xq,Morris:1998da}. The question then arises whether all of the solutions \eqref{vgeneral} should be kept when viewed in this way or whether some of these should also be discarded. We have already seen that if $R<0$ is consistently included, the eigen-perturbations form the expected discrete spectrum, and we will see that the asymptotic regime paints a self-consistent picture for all of these operators when treated as finite perturbations.
 However the analytic study of the large field regime of the present equation \eqref{perturbation} is not as powerful as it is in the case of scalar field theory, the $t$-evolution ultimately being controlled by the finite singular points.

The separable solution \eqref{eigenpert} is valid providing $\alpha$ is small enough to ensure 
\be
\label{dfvsf}
\df(R,t)\ll f(R)
\ee
so that the linearisation step in \eqref{perturbation} is valid. 

We first consider the status of (marginally) irrelevant operators with small but finite couplings. These thus have $\Real\lambda\le0$. At the end of sec. \ref{asymptoticeigen} we saw that their asymptotic behaviour depends on whether $\Real\lambda$ is greater or less than $-10/|p|^2$. 

If we assume that\footnote{If $\lambda$ is real one can show that this corresponds to $\lambda>-2$.} $\Real\lambda>-10/|p|^2$, then recalling \eqref{asymp}, both the perturbation and the fixed point solution grow like $R^2$ at large $R$. Therefore, since these solutions are smooth, if \eqref{dfvsf} is valid for some finite $R$ it will be valid for all $R$. Since the $t$ dependence is given by a shrinking $\exp-2\lambda t$ (or a more general shrinking function if the operator is marginally irrelevant) if \eqref{dfvsf} is valid at the overall cutoff scale $t_{cut}=\ln k_{cut}/\mu$ then it is valid for all $t$ as it is lowered below the cutoff scale. We conclude that these operators correspond to genuine irrelevant perturbations that fall back into the fixed point following the already determined $t$-dependence, as the cutoff $k$ is lowered. 

Now consider those cases where $\Real\lambda<-10/|p|^2$. The perturbation $v(R)$ now grows faster, as $R^p$ where $\Real(p)>2$. Therefore for any finite choice of $\alpha$ and $t$, there is some $R$ beyond which the separable solution \eqref{eigenpert} no longer satisfies \eqref{dfvsf}. If we substitute $f(R,t) = f(R)+\df(R,t)$ directly into \eqref{flow2} and use the fact that $\df(R,t)\sim R^p$ dominates at large $R$, we see that actually in this case the quantum corrections on the right hand side can be neglected (growing no faster than $R^2$) and thus $\df(R,t)$ has to satisfy the left hand side of \eqref{flow2} on its own. Therefore in this regime, $\df(R,t)$ again takes the separable form \eqref{eigenpert} but with $\lambda$ now taking the value expected on dimensional grounds, namely $\lambda=2-p$. Comparing with \eqref{lambda-p},
we see that as we lower $t$, the large $R$ regime is controlled by an effective renormalisation group eigen-value $\lambda=2-p$ which is still irrelevant, but less so than the one derived at small $R$ and for truly infinitesimal perturbations. In summary, although some irrelevant perturbations, when considered small but finite, behave in an unusual way, all of them still fall into the fixed point as we integrate out modes by lowering $t$. 

We can therefore ignore the irrelevant perturbations now and consider what happens when  we build the renormalised trajectory by perturbing by some small but finite linear combination of the (marginally) relevant operators.  In this case each $v(R)\approx R^2$ for large $R$, therefore together we find that in this domain,  $\df(R,t)=g(t) R^2$ for some 
finite
function $g(t)$. If \eqref{dfvsf} is valid at finite $R$ it will thus also be valid as $R\to\infty$. As we noted below \eqref{pdef}, at the linearised level $g(t)$ is actually not determined by constraints in this regime. Similarly to the fixed point, \cf the discussion around \eqref{theps}, even for the full equations \eqref{flow2} this limit is singular and makes sense only when developed to higher order. At the linearised level $g(t)$ is determined by the matching at the singular points $R=0,R_\pm$ which couples together the other asymptotic solutions in \eqref{vgeneral}, as explained in sec. \ref{eigencounting}. Thus we learn that actually $g(t)$ must be expressed as 
\be
\label{relevant-t}
g(t)=\sum_n \alpha_n \exp -2\lambda_n t\,, 
\ee
where the sum is over the (marginally) relevant operators. As modes are integrated out, $t$ falls from $t_{cut}$, and $g(t)$ grows. For some $t\ll t_{cut}$, \eqref{dfvsf} will no longer be satisfied, however this will happen at the singular points and elsewhere at more or less the same $t$ as it happens in the asymptotic regime. From here on  the $t$ dependence changes from \eqref{relevant-t} as determined by the full non-linear equation \eqref{flow2} and in particular by a full non-linear matching at the finite singular points.

\section{Numerical solutions}
\label{numerics}

\subsection{General approach}\label{approach}
The numerical analysis of the fixed point equation \eqref{fp2} is dominated by the presence of the three fixed singularities
$R_+, 0, R_-$. It is therefore sensible to divide the real axis into the following intervals
\be
I_{-\infty} = (-\infty,R_-], \quad I_- = [R_-,0], \quad I_+ = [0,R_+], \quad I_\infty = [R_+,\infty).
\ee
A careful numerical search for solutions has to be carried out separately on each interval and,
as mentioned before, any possible solutions for $R \leq0$ have to be regarded as tentative solutions
since \eqref{fp2} was derived under the assumption $R >0$.
In order to bridge
the fixed singularities we Taylor expand any potential solution around them. In the case of $R_+$ we have
\be \label{tayexpRp}
f(R) =b_0 + b_1(R-R_+)+\sum_{n=2}^5 \frac{b_n(b_0,b_1)}{n!}(R-R_+)^n.
\ee
Similarly,
\be\label{tayexp0}
f(R) =a_0 + a_1 R + \sum_{n=2}^5 \frac{a_n(a_0,a_1)}{n!}R^n
\ee
is the Taylor expansion around $0$ and finally
\be\label{tayexpRm}
f(R) =\beta_0 + \beta_1(R-R_-)+\sum_{n=2}^5 \frac{\beta_n(\beta_0,\beta_1)}{n!}(R-R_-)^n.
\ee
bridges across $R_-$.

In each case, the requirement that the solutions $f(R)$ be regular at the singularity (\cf sec. \ref{parameters})
translates into the fact that 
only the first two coefficients are independent. All higher coefficients, starting with the second, can then be expressed
as functions of the first two. It should be noted that these expressions are unique for all coefficients except
$b_2$ and $\beta_2$ for which we find quadratic constraints. Their solutions can be uniquely written as
\be\label{b2equ}
b_2^\pm(b_0,b_1)=\frac{1}{720 R_+}\left(\, p(b_0,b_1) \pm \sqrt{q(b_0,b_1)}\,\right)
\ee
and
\be\label{beta2equ}
\beta_2^\pm(\beta_0,\beta_1)=-\frac{1}{720 R_+}\left(\, \tilde p(\beta_0,\beta_1) \pm \sqrt{\tilde q(\beta_0,\beta_1)}\,\right),
\ee
where $p, \, q$ and $\tilde p, \, \tilde q$ are polynomials in $b_0, \, b_1$ and $\beta_0,\, \beta_1$, respectively.
As we will see, both possibilities for $b_2$ and $\beta_2$ lead to fixed point solutions. 
As displayed above, the Taylor expansions we have used for our computations are of order $5$ with the higher
coefficients quickly becoming complicated expressions of the first
two.\footnote{In fact we extend \eqref{tayexp0} to sixth order for part of the analysis for reasons discussed at the end of sec. \ref{sec:aplane}.} Especially for the expansions around
$R_\pm$ it would be a challenging task to handle them for even higher coefficients.

Our search for numerical solutions was performed within the Maple(TM) package \cite{Maple} employing
the shooting method. We picked the numerical integration method ``dverk78'' which is a seventh-eighth
order Runge-Kutta integrator and is able to operate at arbitrarily high precision.

We now illustrate the individual steps that led to the discovery of solutions starting with the region $R \ge 0$.
If we started shooting down from some large $R_\infty$ to $R_+$ we would face the problem of having to
search through the $3$-dimensional parameter space of the asymptotic expansion \eqref{asymp}.
It is therefore more reasonable to shoot out from $R_+$ or $0$ where the parameter space is only two-dimensional.
Moreover, we start looking for solutions on $I_+$ as opposed to shooting from $R_+$ towards large $R$
as we expect the additional constraint at $0$ to reduce our parameter space by one dimension
which is a more severe condition than \eqref{safedisc} for the asymptotic series.
Most solutions we were able to find can be obtained by both shooting out from zero towards $R_+$ 
and from $R_+$ towards zero, in particular those which are valid for all $R\geq0$. 
The exceptions were solutions
valid only on $I_+$ with a moveable singularity appearing shortly after $R_+$, \eg at $R_c \approx 8$; in this case
integrating from $R_+$ to zero turned out to be difficult.\footnote{Similarly, integrating from $R_-$ to zero
was ill behaved for partial solutions with a moveable singularity at $R_c \approx -8$.}
For the purposes of illustration however, let us focus on shooting out from zero towards $R_+$ in the following.

\begin{figure}[h]
\begin{center}
 \includegraphics[scale=0.4]{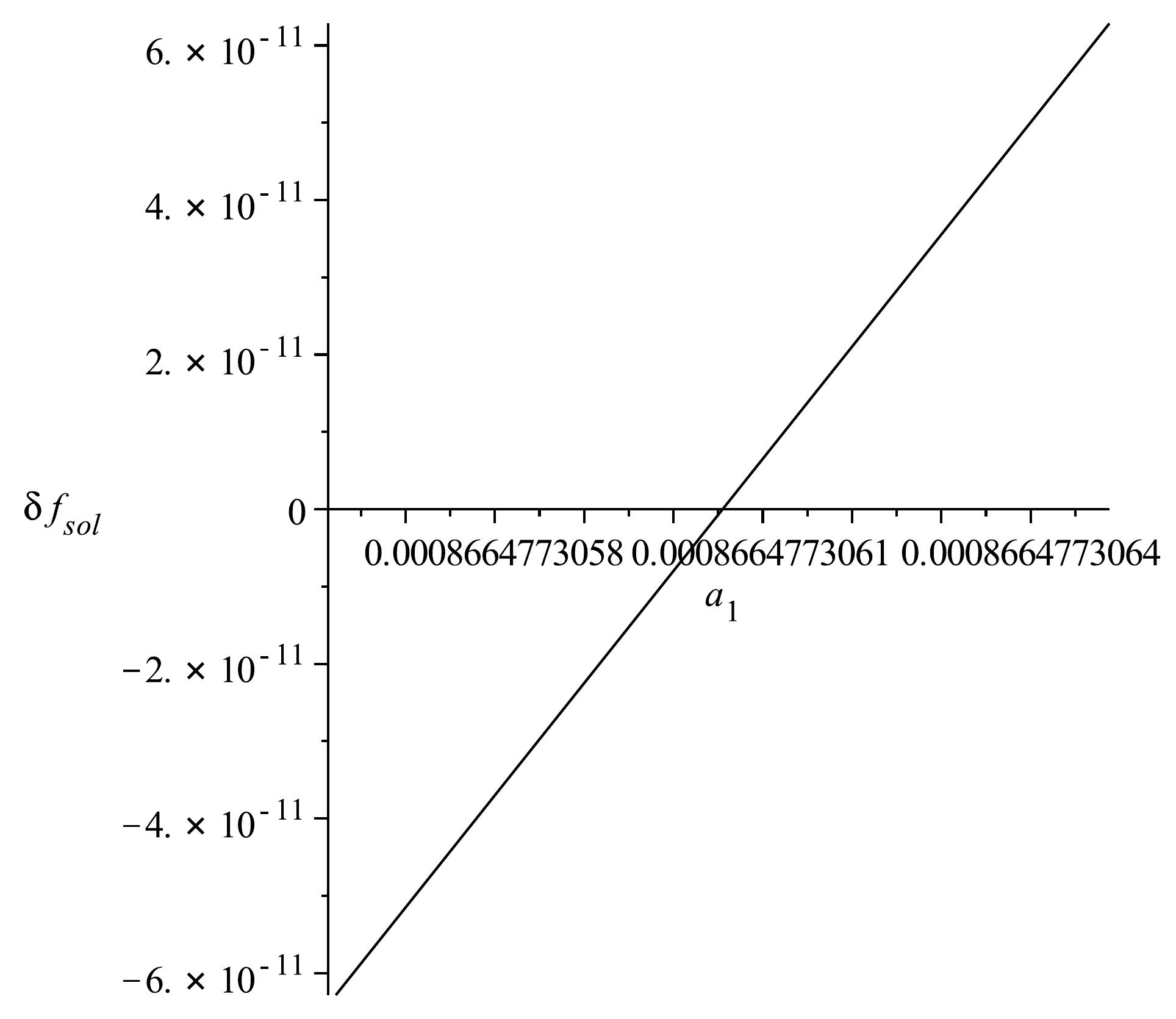}
 \caption{The quantity $\delta f_\text{sol}$ varies linearly across zero as a function of $a_1$ keeping $a_0$ fixed.
If this is the case we recognise the pair $(a_0,a_1)$ as a solution point for $I_+$. }
 \end{center}
 \label{fig:solcrit}
\end{figure}

Given any initial pair $(a_0,a_1)$ we use \eqref{tayexp0} to compute the three initial values 
\be \label{initconds0}
f(\eps_0), \quad f'(\eps_0), \quad f''(\eps_0)
\ee
at distance $\eps_0$ to the right of zero. Shooting out from zero towards $R_+$ we will find that most pairs $(a_0,a_1)$
end at a moveable singularity. For some however, the numerical integrator supplies us with the solution $f_n$
valid close enough to $R_+$ so that by using \eqref{tayexpRp} 
we can solve the system 
\be \label{bsystem}
f(R_+-\eps_1)=f_n(R_+-\eps_1), \quad f'(R_+-\eps_1)=f'_n(R_+-\eps_1)
\ee
at $\eps_1$ away from $R_+$ for $b_0$ and $b_1$. Using \eqref{tayexpRp} again, we are then in a position to compute
$f''(R_+-\eps_1)$ and compare it to the actual value $f''_n(R_+-\eps_1)$ via the quantity
\be \label{solcrit}
\delta f_\text{sol} = \frac{f_n''(R_+-\eps_1) - f''(R_+-\eps_1)}{f''(R_+-\eps_1)}.
\ee
This represents the additional matching condition at $R_+$ and implements the constraint coming from the singularity
at $R=R_+$. A solution is found if the two second derivatives agree to sufficient accuracy.
In fact, our criterion for a valid solution on $I_+$ will be that $\delta f_\text{sol}$
varies smoothly across zero upon variation of $a_1$ while keeping $a_0$ fixed.
This condition on $\delta f_\text{sol}$ is illustrated in fig. \ref{fig:solcrit}.
Small variations of $a_1$ lead to the expected linear behaviour
of $\delta f_\text{sol}$. Varying $a_1$ by a larger amount brings the non-linear form of $\delta f_\text{sol} (a_1)$ to light
whereas if we go to very small variations of $a_1$ we find numerical fluctuations superimposed on the linear behaviour.
There are good reasons for this solution criterion and why we should not just require $|\delta f_\text{sol}|$ to be small,
as we will explain in detail in sec. \ref{sec:aplane}.

A priori there is no special region in the $a$-plane of points $(a_0,a_1)$
where solutions can be found.
However, there is an important
structure to the $a$-plane which is connected
to the Taylor expansion \eqref{tayexp0}. The coefficients $a_2(a_0,a_1),\, a_3(a_0,a_1), \dots$ become singular
along the dashed lines $\gamma_i$ in fig. \ref{fig:as} which implies that the expansion
\eqref{tayexp0} is not valid anymore.
For reasons that will be explained in sec. \ref{sec:aplane},
along these lines there are no solutions except at finitely many points.
It is therefore reasonable to search for the first solution 
between these singular lines. We can do this by using similar graphs as in fig. \ref{fig:solcrit} where we
plot $\delta f_\text{sol} (a_1)$ in the range between two singular lines while keeping $a_0$ fixed or, if moveable 
singularities prevent the solver from reaching $R_+$, instead of $\delta f_\text{sol} (a_1)$
we can plot the maximum $R$ reached in a similar manner.
In this way it is possible to find the first solution point in the $a$-plane.
Once we have a solution $(a_0,a_1)$, we create two new $a$-vectors
$(a_0+\delta a_0,a_1+\delta a_1)$ and $(a_0+\widetilde{\delta a_0},
a_1+\widetilde{\delta a_1})$, where the two variations $(\delta a_0,\delta a_1)$ and
$(\widetilde{\delta a_0},\widetilde{\delta a_1})$ are orthogonal to each other and small enough so that
we are still able to integrate from zero to $R_+$. Generically $\delta f_\text{sol}$
will be large at these new points and so we repeatedly apply the secant method to $\delta f_\text{sol}(a_0,a_1)$
until we find a new pair $(a_0,a_1)$ with the behaviour shown in fig. \ref{fig:solcrit}.
We have automated this process by implementing it in a procedure in Maple and it is this method of obtaining
a new solution point in the $a$-plane starting from an existing one that lies at the heart of how we found our
solution lines. Given at least two solution points $(a_0,a_1)$ and $(\tilde a_0,\tilde a_1)$ we can optimise
the procedure we have just described by using their difference as a guide 
 for the direction in which to look
for the next solution point, i.e. we define
\be
(\delta a_0,\delta a_1) \propto  (\tilde a_0 -a_0,\tilde a_1-a_1).
\ee
These two methods constitute the way by which we extended into lines of solutions starting with only one solution point.

Clearly, the numerical approach to the search for solutions on $I_-$ proceeds 
in a similar way.
We start off with a pair $(a_0,a_1)$  and  
use \eqref{tayexp0} at $\eps_0$ to the left of zero, 
 to determine
the initial conditions which are needed for shooting towards $R_-$. If the (partial) solution makes it to $R_-+\eps_1$,
we solve the corresponding version of \eqref{bsystem} here 
and adapt \eqref{solcrit} to judge whether we are dealing with a solution.
We can employ the same methods as for $I_+$ to find the very first solution on $I_-$ and to extend into
solution lines.

When it comes to finding solutions on $I_\infty$ the degree of difficulty of the problem at hand has already been lowered
by the fact that we can restrict ourselves to check if any of the solutions valid on $I_+$ will make it out to infinity,
\ie we are left with a one dimensional
parameter space. The strategy therefore is to select a possible
candidate pair $(b_0,b_1)$ and to exploit \eqref{tayexpRp} at $R_+ + \eps_1$ to find the initial values. 
We then integrate up to some large $R_\infty$ where we finally match to the asymptotic expansion \eqref{asymp}. 
The accuracy of the three asymptotic parameters $A,B,C$ will depend on the value of $R_\infty$
which is why it should be taken reasonably large.\footnote{We defer a discussion of this important point to
sec. \ref{errorana}.} If the constraint \eqref{safedisc} is satisfied
the existence of the solution is guaranteed for all $R>R_\infty$. On the other hand, if the asymptotic parameters $A,B,C$
are such that \eqref{safedisc} is not fulfilled, we are dealing with a partial solution only since
a moveable singularity is bound to appear at some $R>R_\infty$, \cf sec. \ref{asymptotics}.
In principle this same method can be used to find solutions on $I_{-\infty}$ but as we will see, none of the solutions
we found on $I_-$ extend to $I_- \cup I_{-\infty}$.

\subsection{The solutions} \label{sec:sols}
After extensive numerical searches using the approach outlined in sec. \ref{approach} we were able to obtain 
a total of five 
solution lines valid on the range $R\geq0$, four additional solution lines representing solutions on $I_-$, and we found evidence for more solution lines. 
For clarity of presentation we devote a separate section to each set of solution lines.

\subsubsection{Solutions for $R\geq0$}
\begin{figure}[h]
\begin{center}
\includegraphics[width=0.6\textwidth,height=250pt]{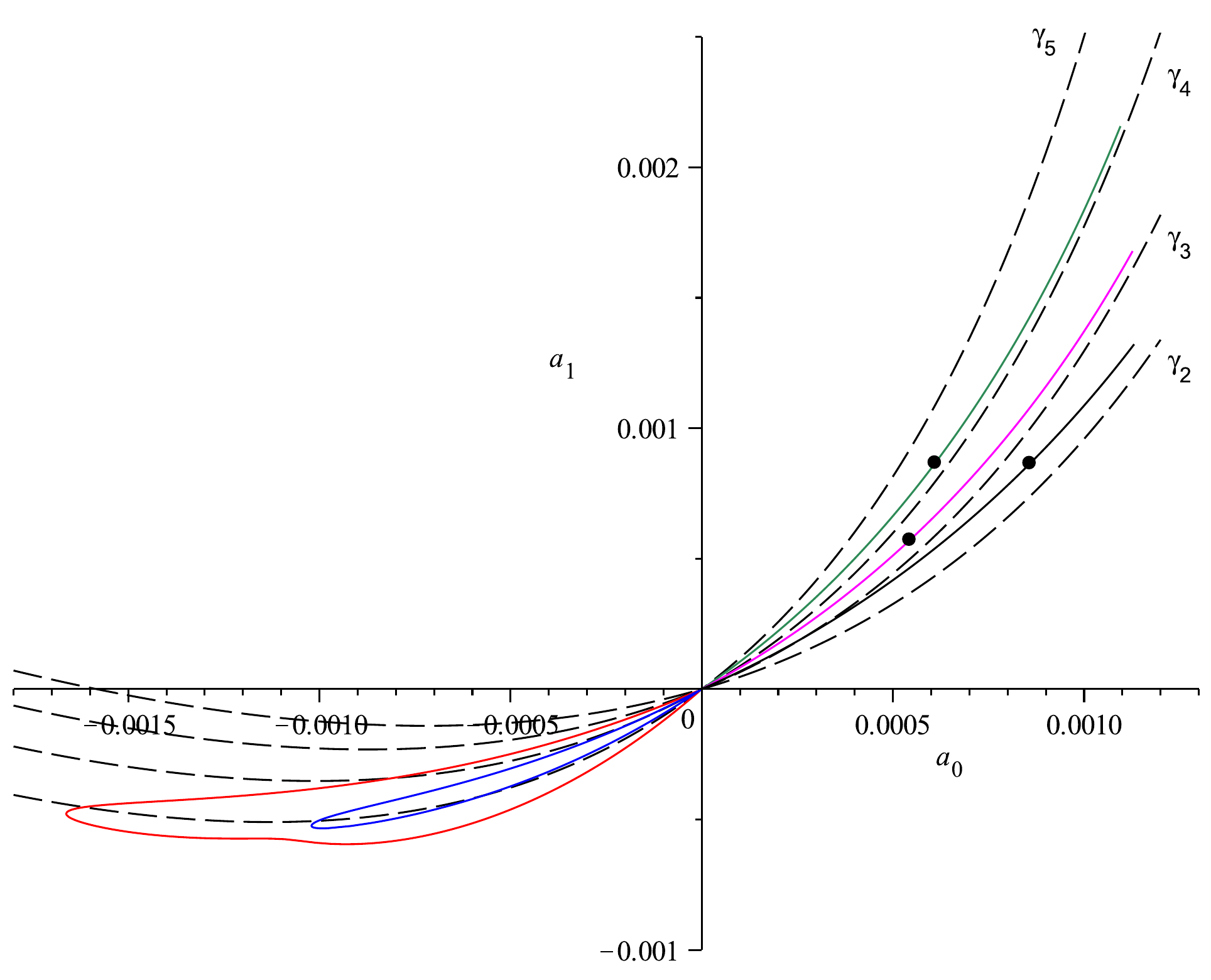}
\end{center}
\caption{Five lines of solutions in the $a$-plane corresponding to the $b$-lines in fig. \ref{fig:bs}.
Different solution lines are represented consistently in
 different colours (these are red and blue in the range $a_0<0$ and, from bottom to top, 
 black, magenta and green in the
range $a_0>0$).
The black points represent example solutions which are plotted in fig. \ref{fig:expls}.
Along each of the black dashed lines a coefficient in \eqref{tayexp0} becomes singular (\cf sec. \ref{sec:aplane}).
}
\label{fig:as}
\end{figure}

\begin{figure}[h]$
\begin{array}{cc}
\includegraphics[width=0.4\textwidth,height=150pt]{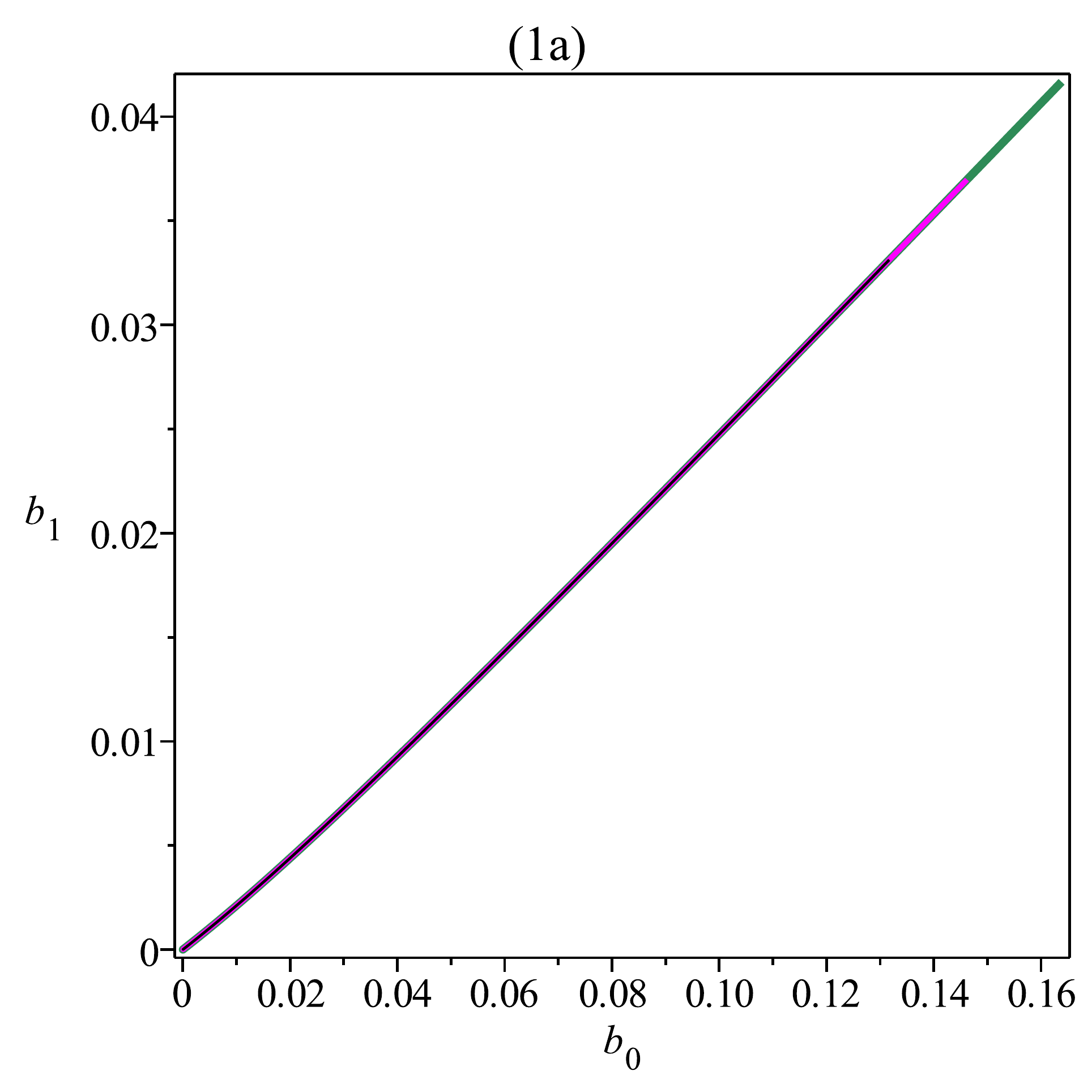} &
\includegraphics[width=0.55\textwidth,height=150pt]{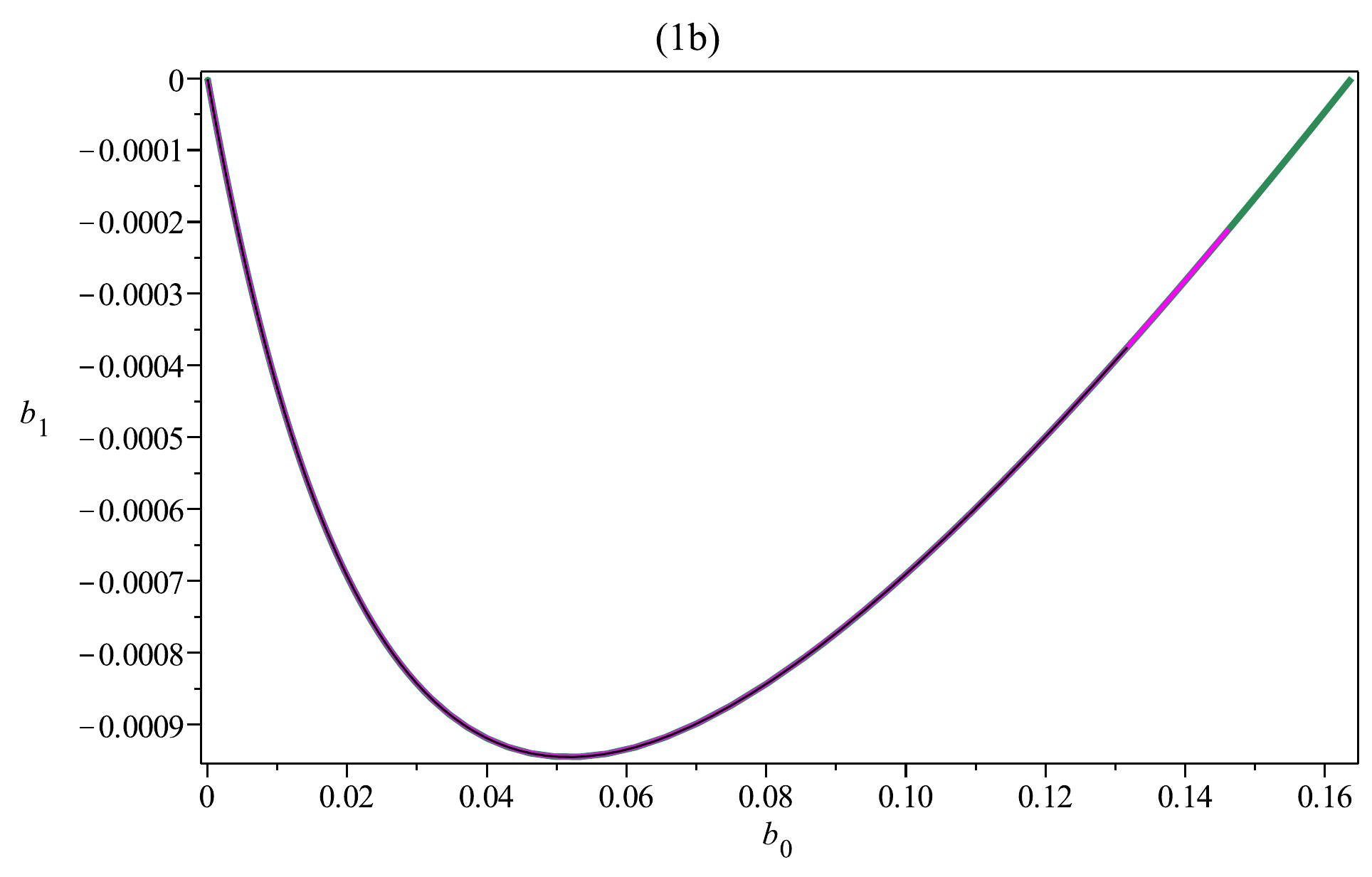} \\
\includegraphics[width=0.4\textwidth,height=150pt]{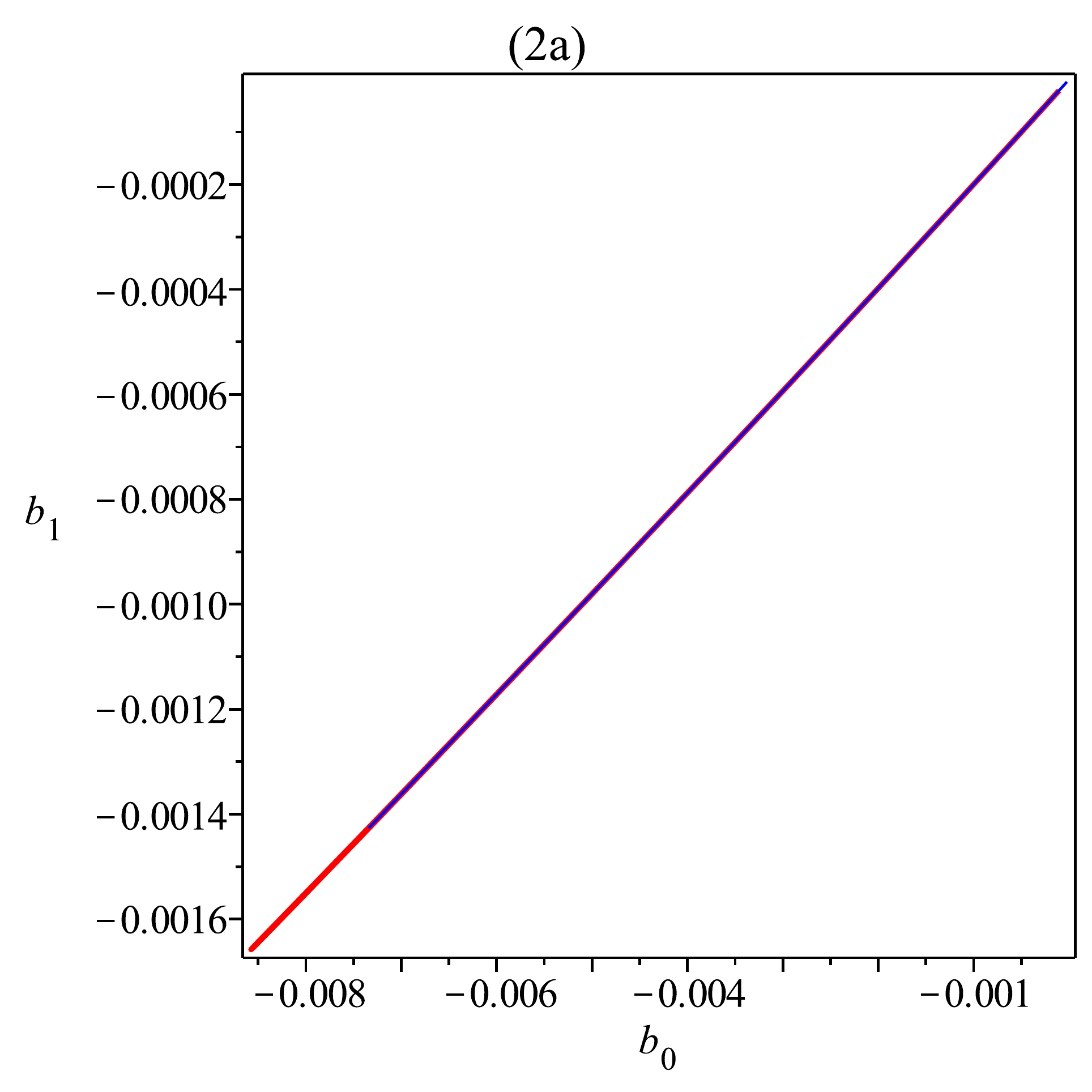} &
\includegraphics[width=0.55\textwidth,height=150pt]{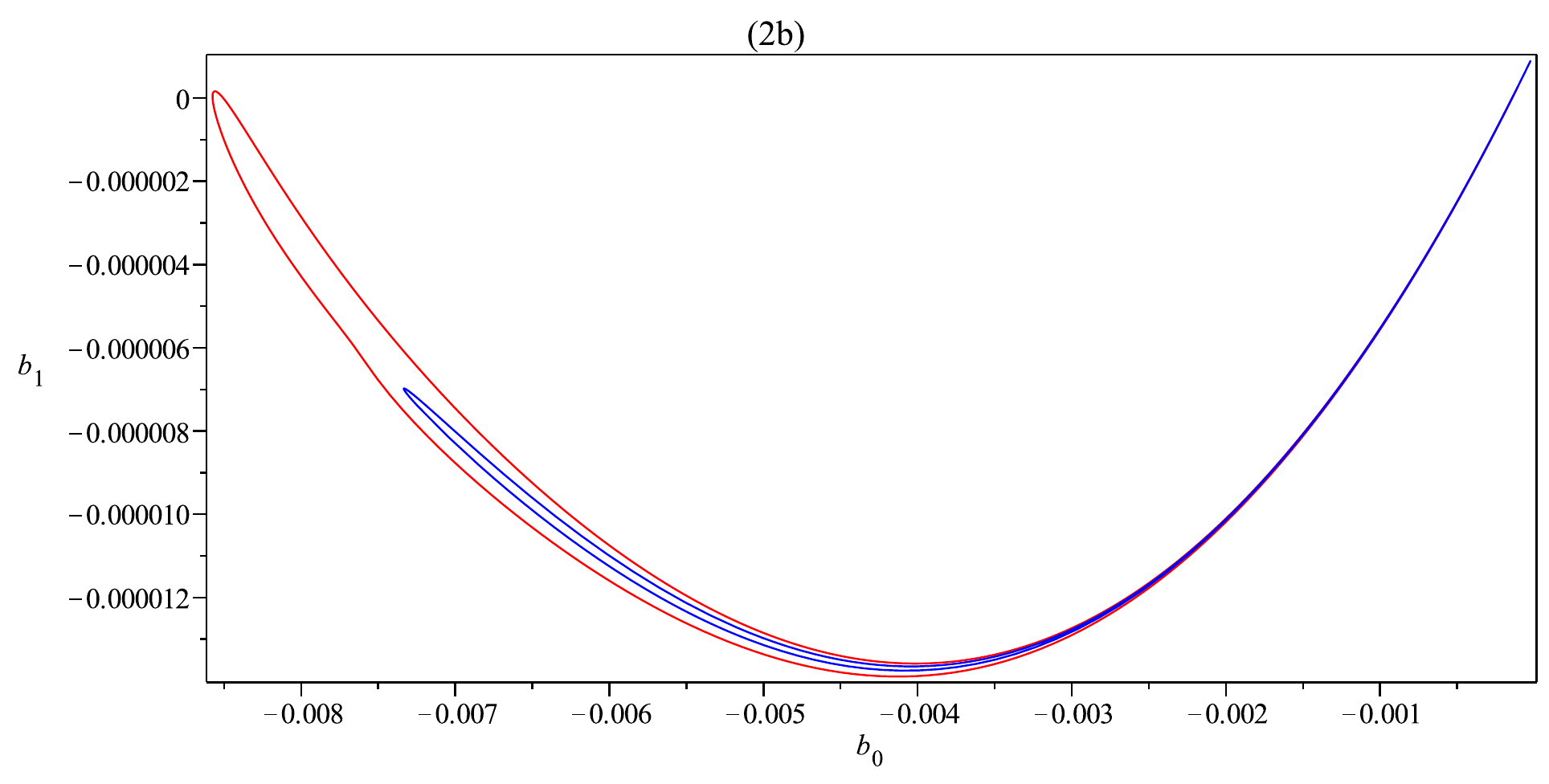} 
\end{array}$
\caption{Three lines of solutions on $I_+ \cup I_\infty$ in the range $b_0>0$ in (1a) and two additional solution lines
in the range $b_0<0$ in (2a). In both cases we have subtracted the straight line given by 
$b_1=0.254352\cdot b_0 -2.5\cdot 10^{-7}$ and plotted the result on the right.}
\label{fig:bs}
\end{figure}

The lines of solutions defined on the original domain of validity $I_+ \cup I_\infty$ of the fixed point equation
\eqref{fp2} are shown in fig. \ref{fig:as}. We have three lines of fixed point solutions
in the range $a_0>0$, all starting at the origin and moving out between two singular lines of the expansion 
\eqref{tayexp0}.\footnote{Further such solution lines are expected between singular lines $\gamma_n$ with $n\ge5$, as we discuss in sec. \ref{sec:aplane}.} 
In fact, the bottom two solution lines are both situated between the two singular lines
$\gamma_3$ and $\gamma_4$ in the region close to the origin,
\cf fig. \ref{fig:als}, but the black solution line crosses the singular
line $\gamma_3$ in its early stages and moves out between $\gamma_2$ and $\gamma_3$.
In order to find these fixed point solutions in the region of positive $a_0$
it was crucial to use the $b^+_2(b_0,b_1)$ solution in \eqref{b2equ}. Furthermore, all these solutions are valid for all $R\geq0$, \ie satisfy the asymptotic constraint \eqref{safedisc}, and their coefficient $A$ of the dominant
asymptotic component is positive.
As explained in the Discussion and Conclusions, $A>0$ is a necessary (but not sufficient) condition for the solutions to make physical sense. 
The corresponding solution lines in the $b$-plane are shown in fig. \ref{fig:bs} (1a), (1b).
All three start at the origin and approximately 
follow a straight line towards larger values of $b_0$.
Even after subtraction of this linear component, the result being shown in fig. \ref{fig:bs} (1b), it is 
impossible to distinguish them by eye. The distance between these solution lines varies 
but can be as small as $\Delta b_1/b_1= 10^{-10}$.

In the range $a_0<0$ there are two more solution lines but they behave differently, see fig. \ref{fig:as}.
Leaving the origin they
run towards the left but eventually both turn around to approach the origin again. In contrast to the solutions
with $a_0>0$, these solutions 
all have $A<0$, \ie they assume negative values for large enough $R$.
In order to satisfy the constraint arising from the fixed singularity at $R_+$ it was necessary
to employ the $b_2^-(b_0,b_1)$ solution in \eqref{b2equ} in this region of the $a$-plane.
If we look at the corresponding solution lines in the $b$-plane shown in fig. \ref{fig:bs} (2a), (2b)
we again observe a linear behaviour. Unlike in the case for $b_0 >0$ however, we can distinguish
the two solution lines clearly after subtracting their straight line approximation.
It then also becomes clear that, just like the solution lines in the $a$-plane, after moving
away from the origin they turn around and approach it again. In both the $a$- and the $b$-plane
the blue solution line is contained inside the region bounded by the red solution line.

The fact that the solution lines in the $b$-plane can be approximately described by a straight line
can be attributed to the following observation. 
If we rescale according to
\be
f(R)=b_1 \bar f(R) =b_1 \left(\frac{b_0}{b_1} + (R-R_+) + \frac{b_2}{b_1}(R-R_+)^2 + \dots \right)
\ee
and consider the limit $b_1 \to 0$, the fixed-point equation \eqref{fp2} becomes  
\be
\left.{\tilde \cT}_2 + \cT_1 +\cT^{\text{np}}_0+{\tilde \cT}^{{\bar h}}_0 \right|_{f= \bar f} =0 \,.
\ee
This follows from the fact that the right hand side in \eqref{fp2} is invariant under rescaling of $f$
but the left hand side is linear. Hence we would expect that, close to the origin of the $b$-plane,
any solution line should show predominantly linear behaviour.
The same conclusions can be drawn for solution lines in the $a$-plane. However, computing the slope
of the solution lines in this plane as we approach the origin, we find the slopes vary linearly, implying quadratic behaviour of the solution lines close to the origin.
Given the huge magnification factor implied above, which occurs
as we integrate from $R_+$ to zero,
we would expect to recover linear behaviour of the solution lines at distances much closer to the origin
of the $a$-plane than we have probed here.

The non-linear components of the solution lines in the $b$-plane in fig. \ref{fig:bs} where obtained
by subtracting the same straight line from both the solution lines with positive and negative $b_0$.
This is a first indication that it could be possible to match these solution lines across the origin.
If one computes the second coefficient $b_2$ along the solution lines the two possibilites \eqref{b2equ}
also seem to match across the origin. However, since the $b$-lines are so close to each other it is
difficult to give a definite answer to this question in the $b$-plane.
The solution lines in the $a$-plane can be separated clearly and computing their slopes
close to the origin it seems to be possible to match the lines with positive $a_0$ uniquely to 
a solution line with negative $a_0$, \eg the black solution line matches to the upper part of the red
solution line and the solution
line in magenta can be matched to the upper part of the blue solution line.
The lower part of the blue solution line can be continued by the green solution line whereas the lower part of the 
red solution line seems to belong to an as yet undeveloped line on the other side of the origin.
Although these pairings are confirmed by matching the other coefficients of the $a$-series \eqref{tayexp0}
across the origin, none of these methods would be able to distinguish solution lines that become
tangential as they approach the origin.

The solution lines on the right hand side in fig. \ref{fig:as} can be continued towards larger values of $a_0$
but the numerical integration becomes ever more time consuming.
We have also found a first solution point on four additional solution lines between $\gamma_4$ and $\gamma_5$, 
on both sides of the origin. Judging by these first solutions they consist of partial solutions only.
They could also complicate the matching of solution lines across the origin since three of them are
very close to each other compared to the solution lines we know of and could thus be tangential at the origin.

\begin{figure}[ht]
\begin{center}$
\begin{array}{ccc}
\includegraphics[width=0.4\textwidth,height=115pt]{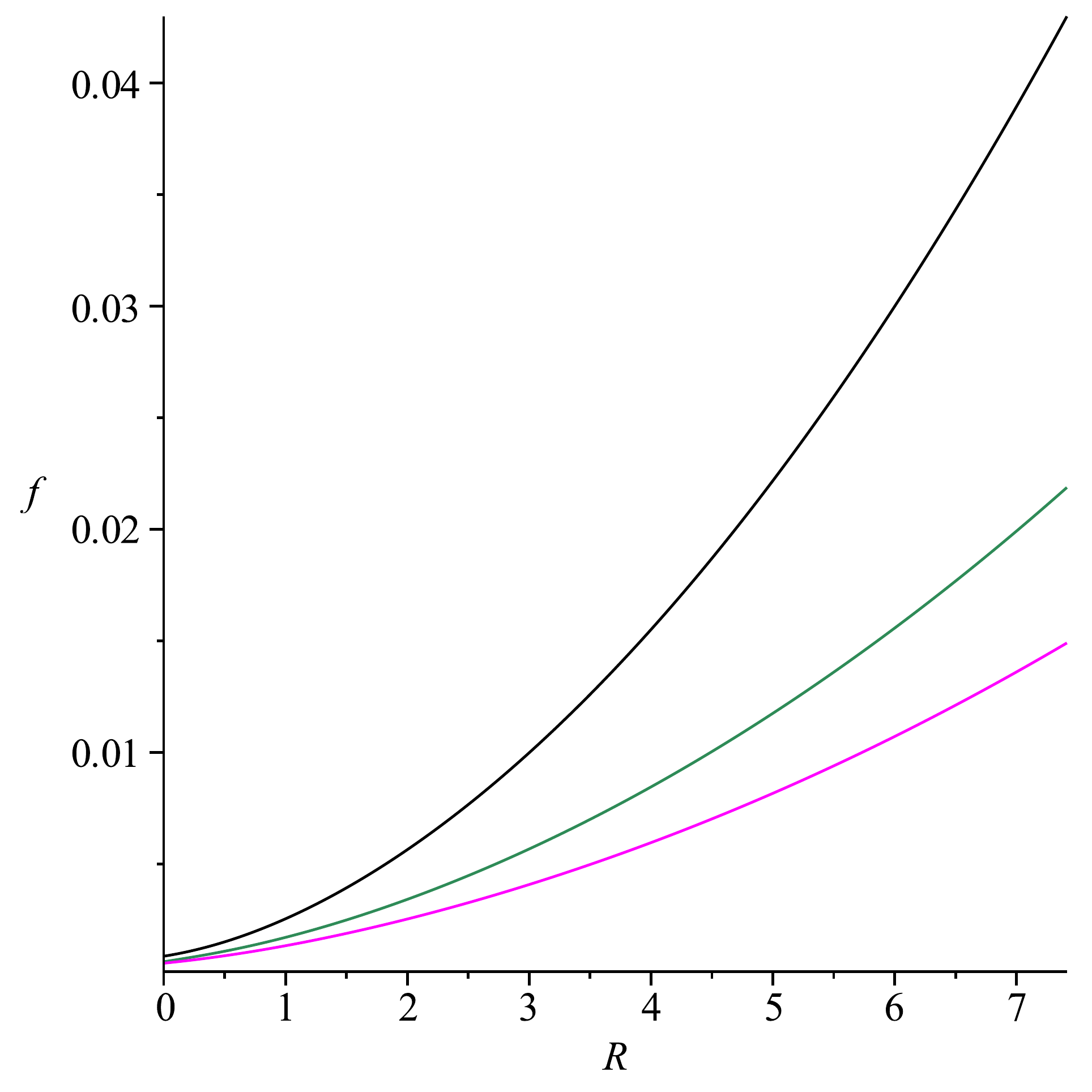} & & \includegraphics[width=0.4\textwidth,height=115pt]{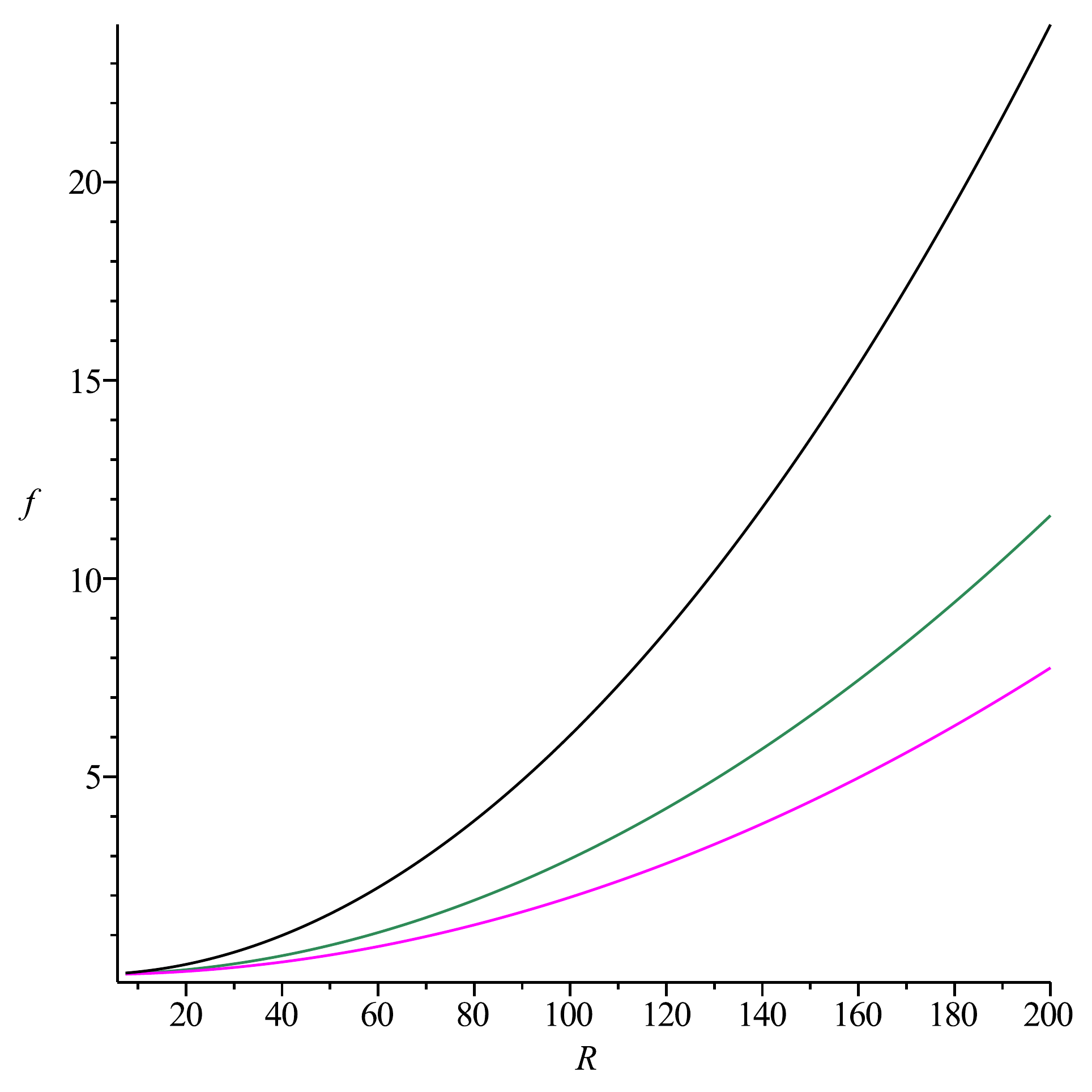} \\
 \includegraphics[width=0.4\textwidth,height=115pt]{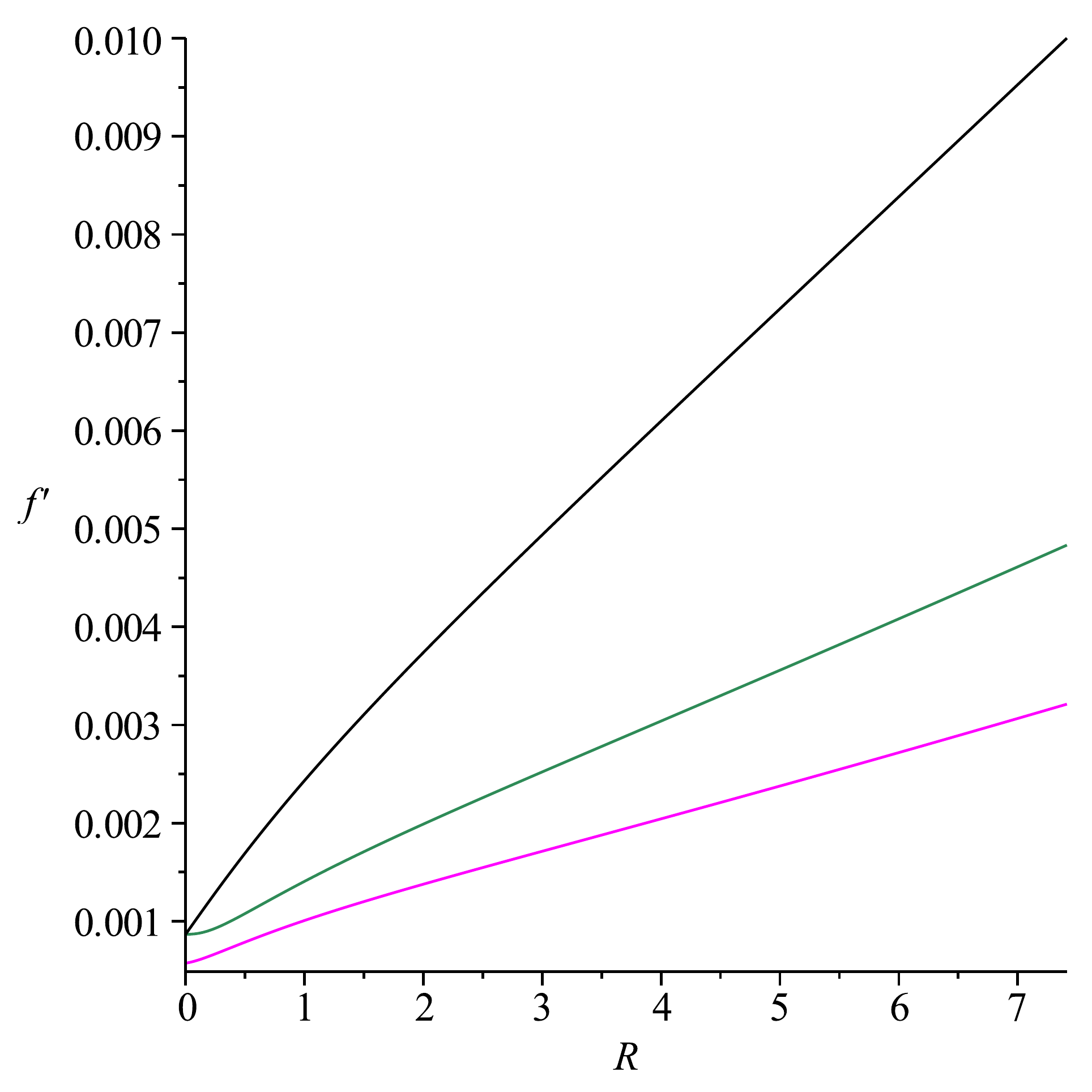} & & \includegraphics[width=0.4\textwidth,height=115pt]{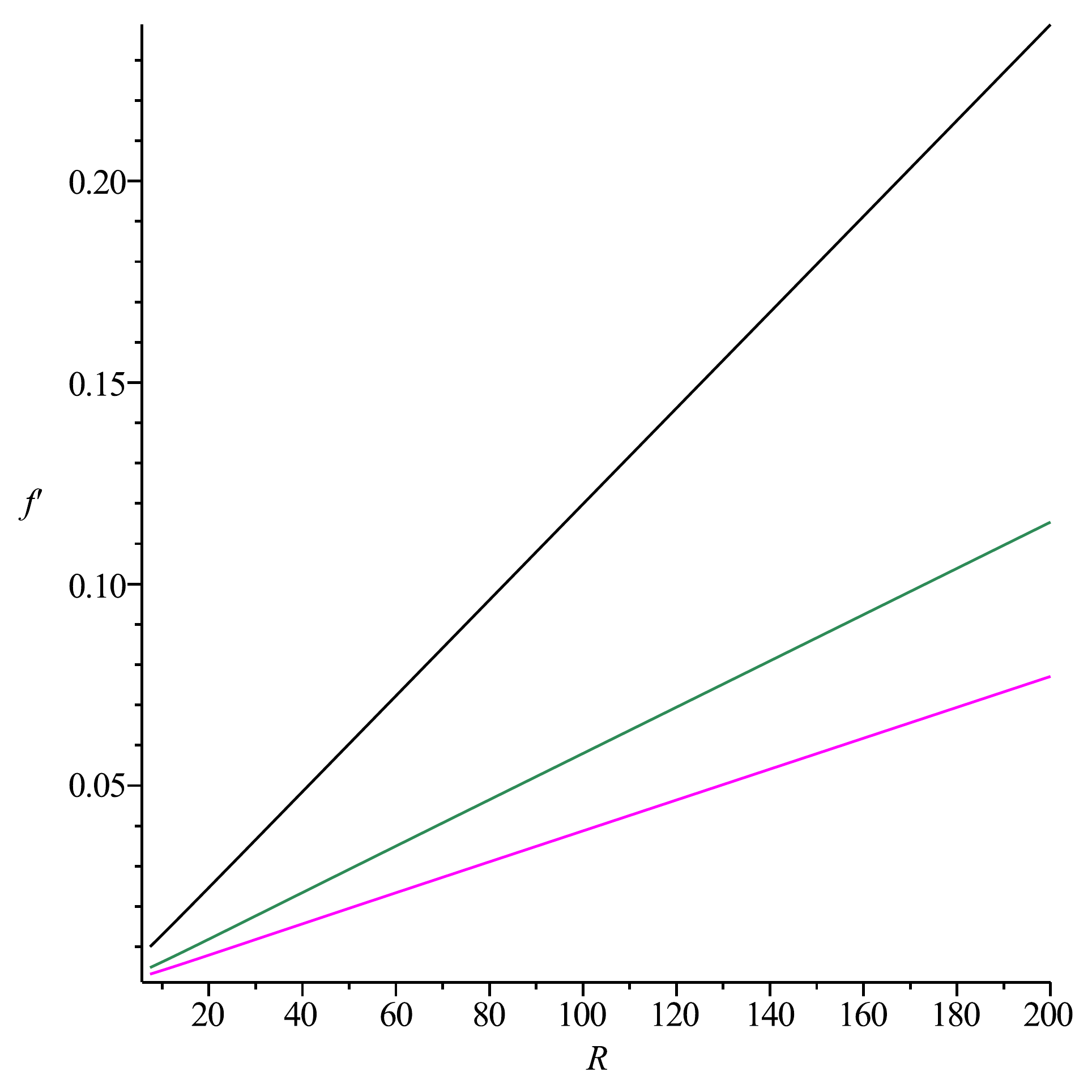} \\
\includegraphics[width=0.4\textwidth,height=115pt]{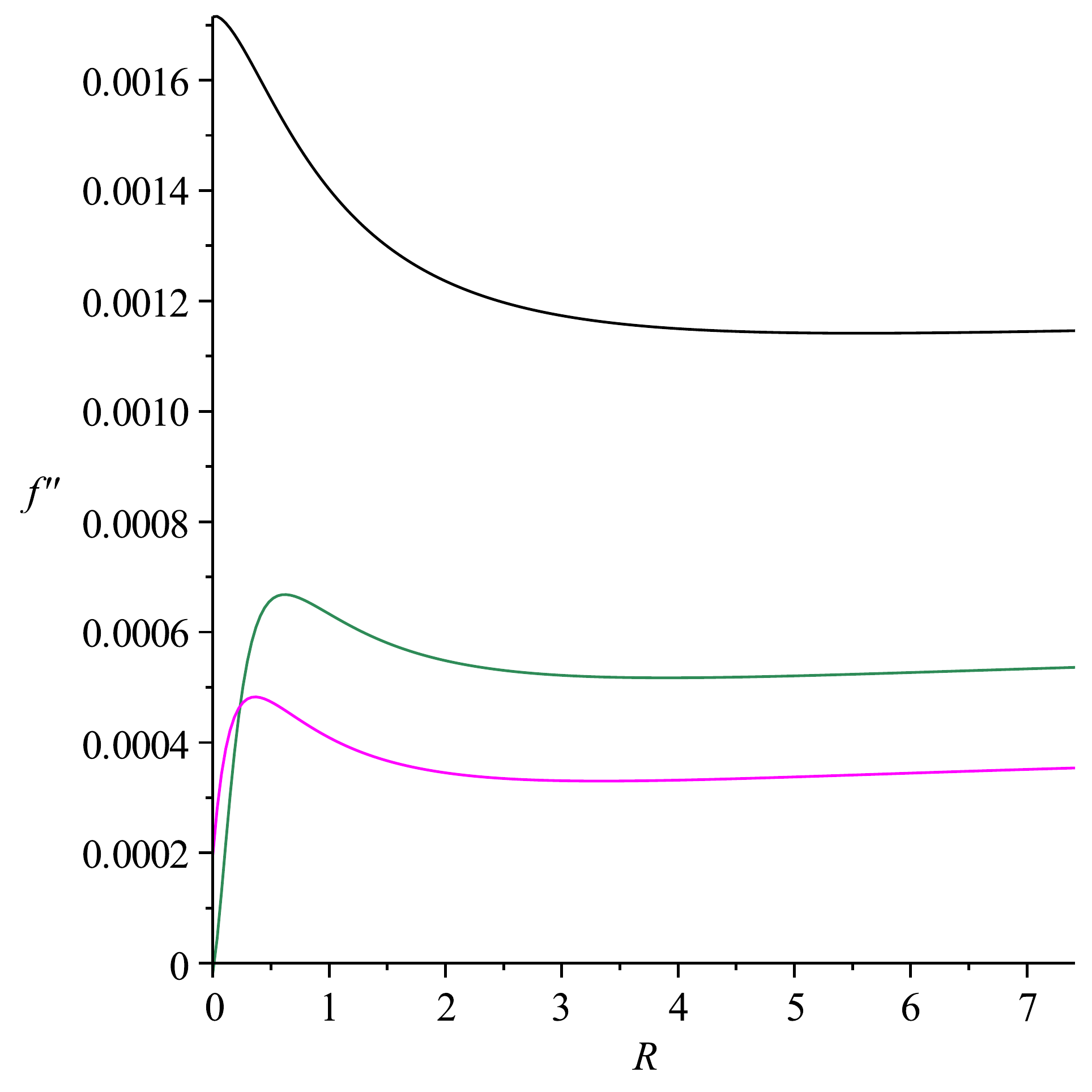} & &
\includegraphics[width=0.4\textwidth,height=115pt]{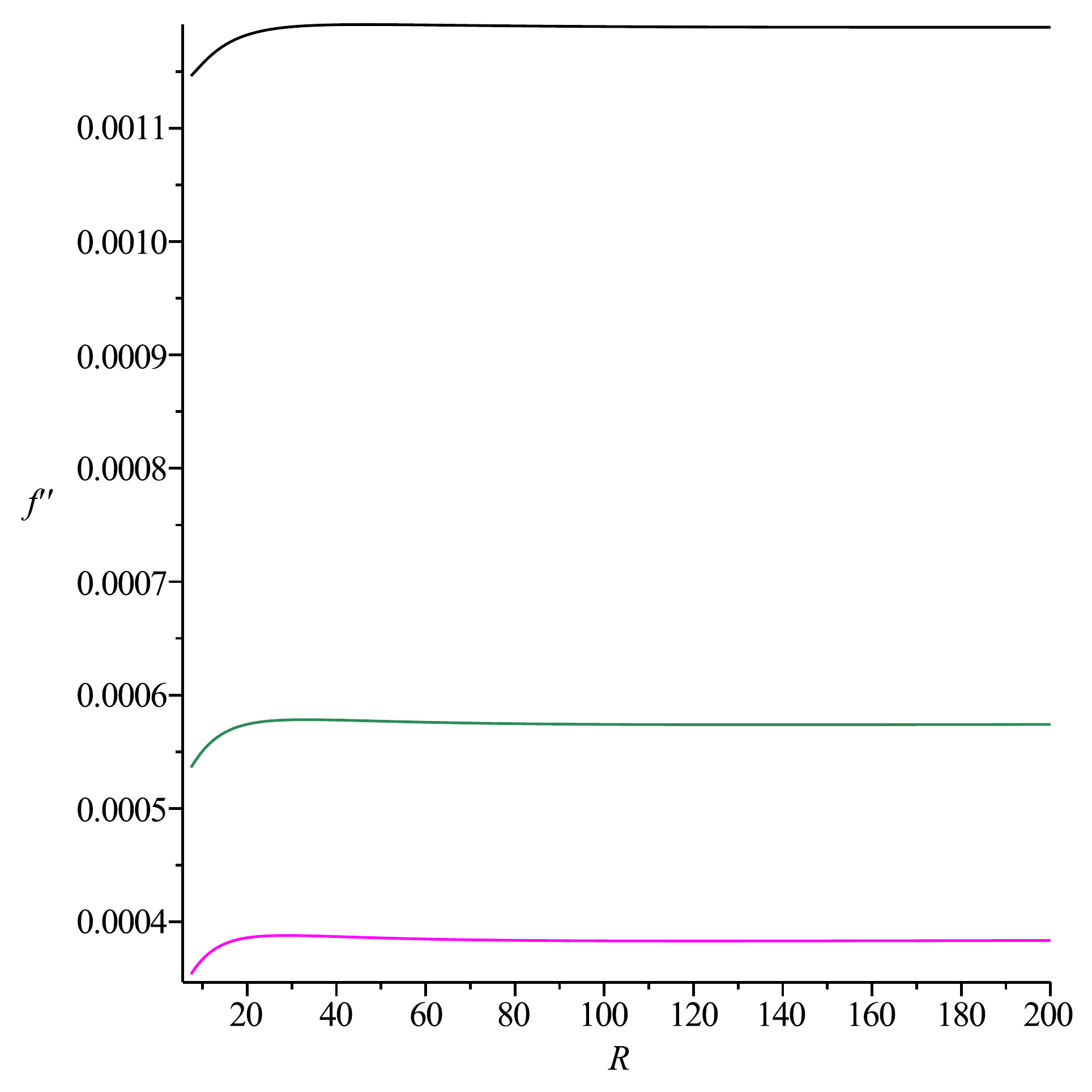}
\end{array}$
\end{center}
\caption{Fixed point functions and their first and second derivatives represented by the points shown in fig. \ref{fig:as} in corresponding colours (from bottom to top these are magenta, green, black)
plotted over the interval $I_+$ in the left column and over the range 
$R_+ \leq R \leq R_\infty$ on the right.}
\label{fig:expls}
\end{figure}

As indicated by the points in fig. \ref{fig:as}, we have selected one example solution
on each of the solution lines for $R \geq 0$ with $A>0$
and plotted the fixed point function $f(R)$ and its first and second derivatives in fig. \ref{fig:expls}.
All three fixed point functions assume positive values only. 
The first and second derivatives already show that we approach quadratic
behaviour close to $R_+$ as the plots on the right hand side in fig. \ref{fig:expls} confirm.
Nevertheless, in agreement with what is already built into the asymptotic expansion \eqref{asymp},
we can also discern the oscillatory pattern from the plots of the second derivative.
The parameter values for these solutions are listed in table \ref{tab:par}. In the Discussion and Conclusions,
we comment on the physical interpretation of these solutions and also the solutions with $a_0<0$. 

\begin{table}[h]
\begin{center}
\begin{tabular}{|c|c|c|c|c|c|c|c|}
 \hline
  FP & $a_0 \, [10^{-4}]$ & $a_1$ $[10^{-4}]$ & $b_0$ $[10^{-3}]$ & $b_1$ $[10^{-3}]$ & $A$ $ [10^{-5}]$ & $B$ $ [10^{-5}]$ & $C$ $ [10^{-5}]$ \\
 \hline
  green & 6.1053 & 8.6648 & 21.869 & 4.8337 & 28.741 & -3.2321 & -1.3074 \\
 \hline
  magenta & 5.4421 & 5.7113 & 14.899 & 3.2125 & 19.207 & -3.3397 & -0.47817 \\
 \hline
 black & 8.5841 & 8.6461 & 42.986 & 10.002 & 59.479 & -1.5161 & -2.1547 \\
 \hline
\end{tabular}
\end{center}
\caption{Parameter values for the fixed point functions plotted in fig. \ref{fig:expls}.
Note that many of these parameters are known to higher accuracy than given here, \cf sec. \ref{errorana} and table \ref{tab:errs}.}
\label{tab:par}
\end{table}

\subsubsection{Solutions for $R_-\leq R \leq 0$}
The fact that the fixed point equation \eqref{fp2} admits the uncountably infinite number of solutions
presented in the previous section can be expected from the parameter counting method. Although the derivation
of \eqref{fp2} was carried out with spaces of positive scalar curvature $R$ we have nevertheless extended
our analysis to the range $R\leq0$ in the hope this would constrain the solution space to at most a finite number 
of physically sensible fixed point functions valid for $-\infty < R < \infty$.
Since in this regard we are interested in fixed point solutions with positive quadratic asymptotic component, \ie $A>0$,
we have concentrated our efforts on finding solutions valid on $I_-$ which are situated in the $a_0>0$ half-plane.

\begin{figure}[h]
\begin{center}
\includegraphics[width=0.49\textwidth,height=200pt]{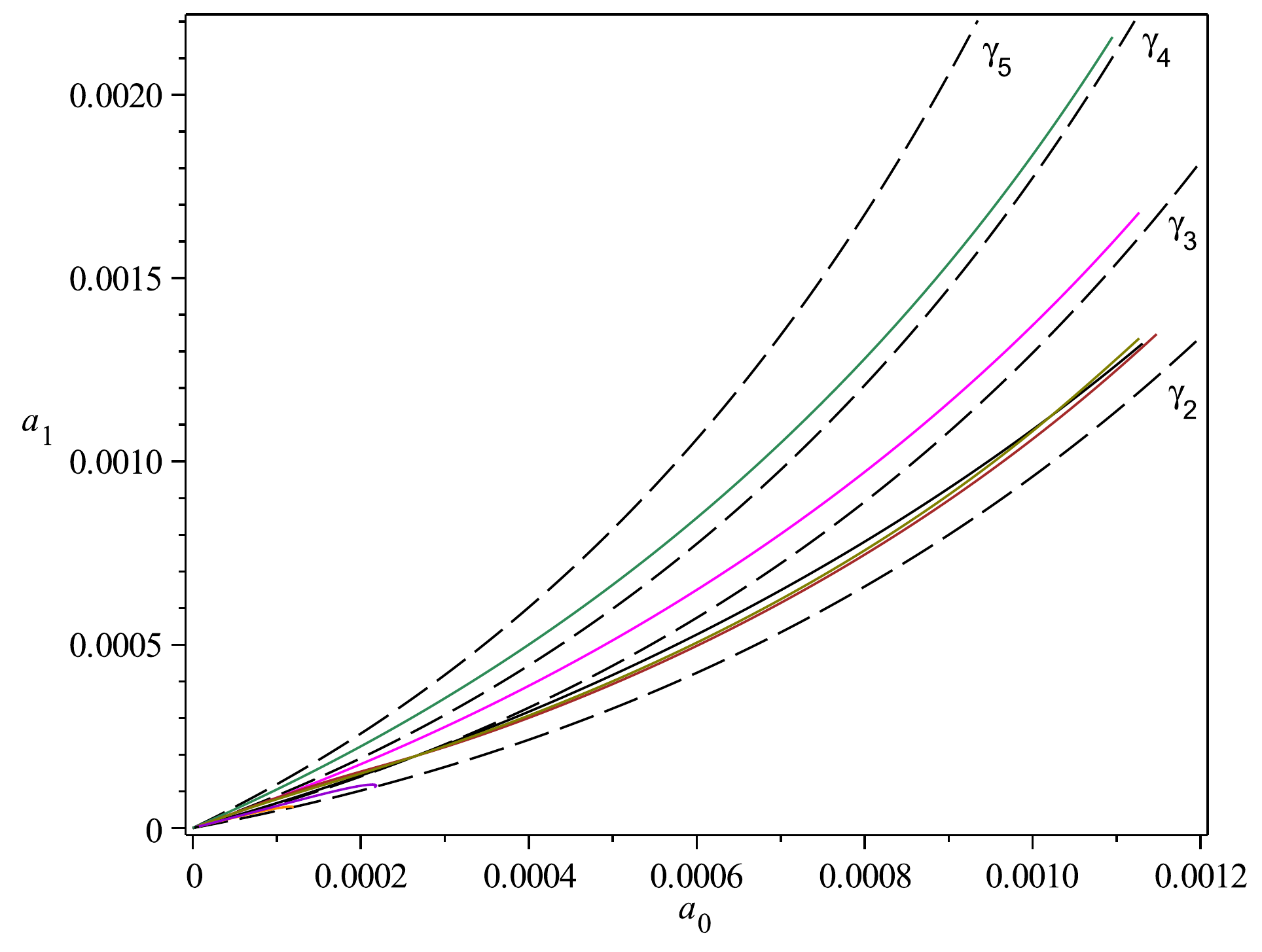}
\includegraphics[width=0.5\textwidth,height=200pt]{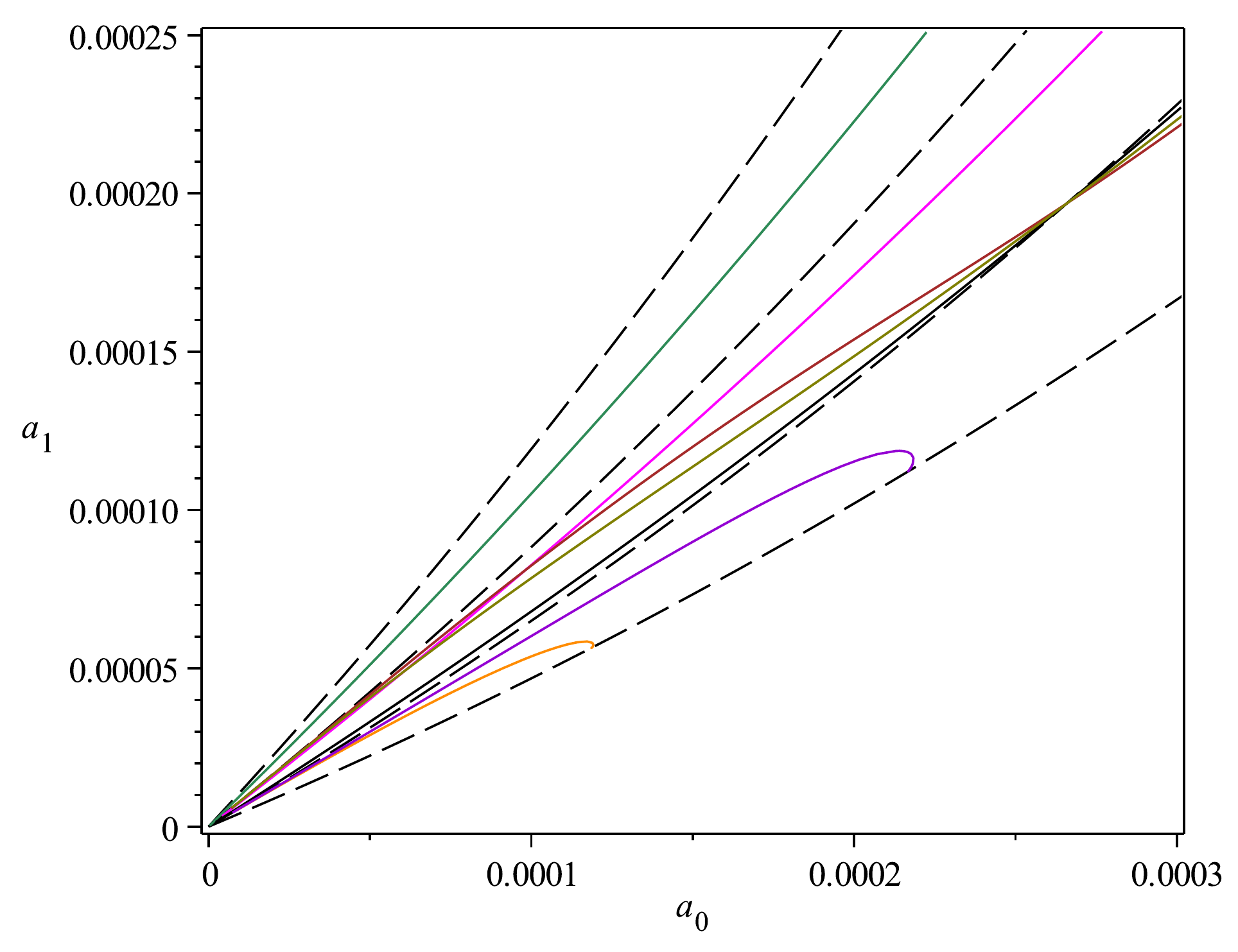}
\end{center}
\caption{Four solution lines for $I_-$ in the $a$-plane together with three of the solution lines
displayed in fig. \ref{fig:as} on the left and a magnification of the region close to the origin on the right.
In the magnification the colouring from bottom to top for the solution lines on $I_-$
is orange, violet, olive and brown.
The black dashed lines are the same singular lines $\gamma_i$ as in fig. \ref{fig:as}.}
\label{fig:als}
\end{figure}

We have been able to find four solution lines in this range whose form in the $a$-plane is displayed in fig. \ref{fig:als}.
Two of these solution lines
extend as far as the previous solution lines for $R\geq0$ whereas the other two turn around 
while still close to the origin running alongside the bottom singular line $\gamma_2$. Because the latter two lines are
so close to $\gamma_2$ we have refrained from developing them all the way back to the origin but we have
checked that they are still present at $a_0=10^{-4}$ where they are approximately $2\cdot 10^{-9}$ away
from $\gamma_2$. 

The situation is different for the brown and olive solution line in fig. \ref{fig:als}. They start out
between the same singular lines $\gamma_3$ and $\gamma_4$ as the two solution lines for positive $R$ shown in black and magenta 
and both cross $\gamma_3$ where the black solution line does but approach the black solution line 
for large values of $a_0$. The olive solution line then intersects the black solution line and thus gives rise
to a solution valid on the interval $R_- \leq R < \infty$. Although the brown solution line does not cross
the black solution line again 
in the range we have developed it in, it is to be expected that it will do so 
if the two lines are continued further. Close to the origin we find two more solutions valid on $R_- \leq R < \infty$:
the solution lines in brown and olive both intersect the solution line in magenta for $I_+$ exactly once (this is barely discernible in
fig. \ref{fig:als}). The former at
$(a_0,a_1) = (9.8\cdot 10^{-5},8.1\cdot 10^{-5})$ and the latter at $(a_0,a_1)= (6.07 \cdot 10^{-5},4.9\cdot 10^{-5})$. 

\begin{figure}[h]
\begin{center}
\includegraphics[width=0.49\textwidth,height=200pt]{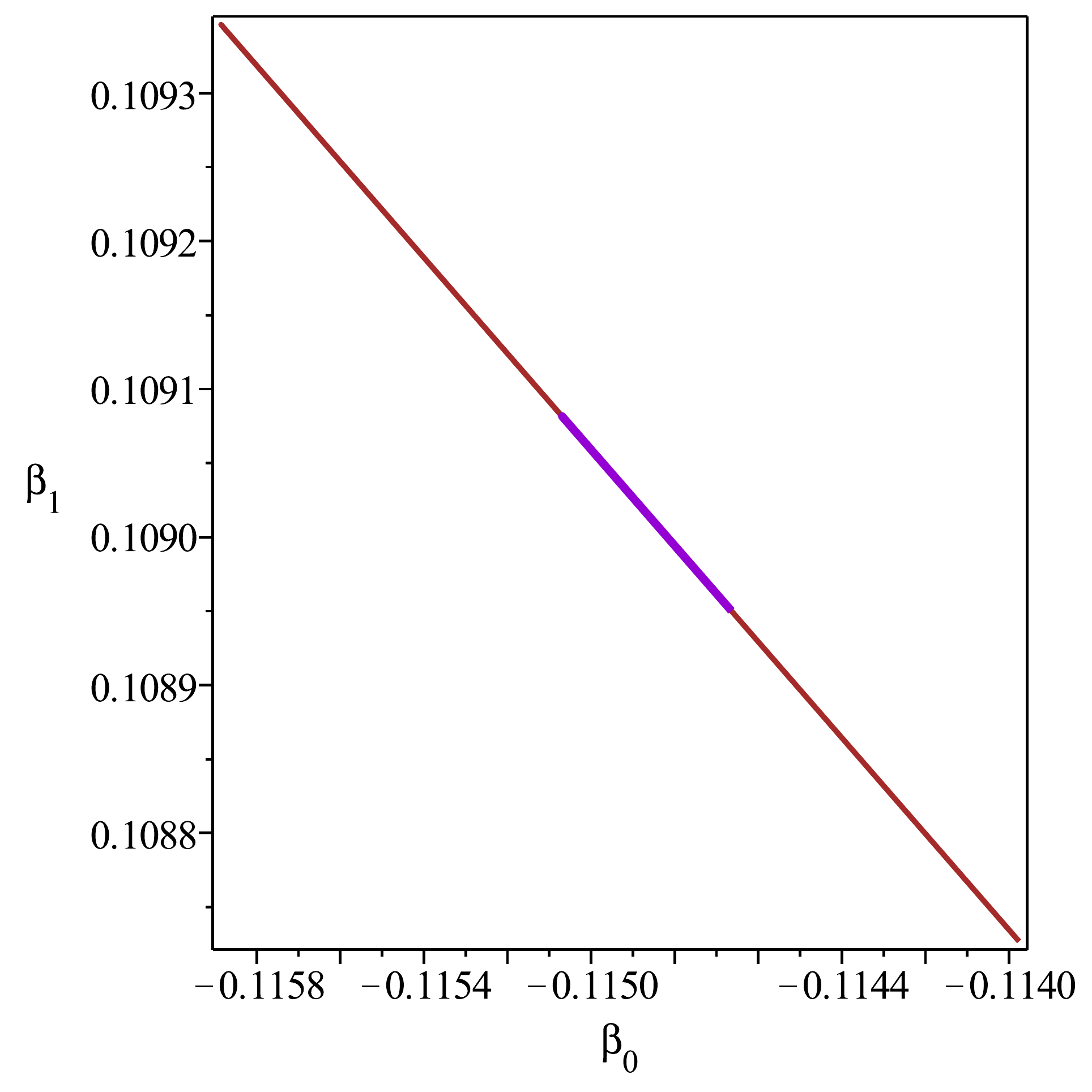}
\includegraphics[width=0.5\textwidth,height=200pt]{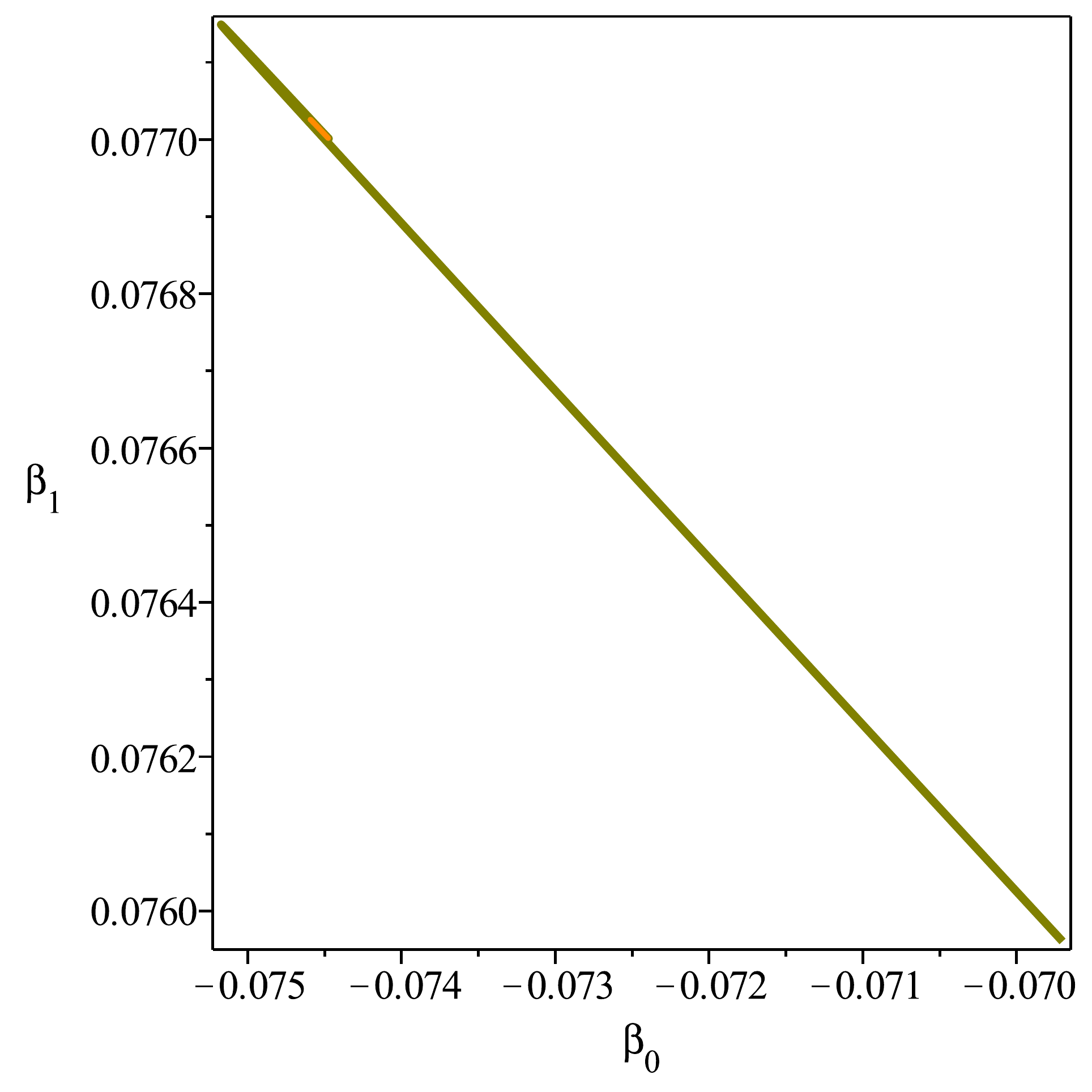}
\end{center}
\caption{The four solution lines for $I_-$ in the $\beta$-plane in the same colours as
the corresponding lines shown in fig. \ref{fig:als}.}
\label{fig:bts}
\end{figure}

The corresponding solution lines in the $\beta$-plane are shown in fig. \ref{fig:bts}. Like the solution lines
in the $b$-plane they are governed by an underlying linear behaviour with superimposed non-linear variations
which in the case of the left plot are of order $10^{-8}$ and for the plot on the right they can be as large
as $3\cdot10^{-5}$. Note the two different ways in which these four lines are paired: in the $a$-plane
the two long and the two short solution lines are paired together whereas in the $\beta$-plane it is always
one $\beta$-line corresponding to a long solution line in the $a$-plane that is paired with a $\beta$-line
corresponding to a short solution line in the $a$-plane. As we would expect from continuity, varying the initial pair $(a_0,a_1)$ along a smooth curve from one of the short solution lines in the $a$-plane to the other leads to a smooth curve of  points $(\beta_0,\beta_1)$ in the $\beta$-plane connecting the two corresponding solution lines in the two different regions of the $\beta$-plane. However, this connecting curve does not consist of solutions since
the constraint at $R_-$ is not satisfied.
Curiously, despite the fact that the solution lines in the $a$-plane approach the origin, the corresponding lines 
in the $\beta$-plane do not. Another difference to the situation for $R\geq0$ is that we have not found solution lines
between the upper two singular lines $\gamma_4$ and $\gamma_5$.

We have also found evidence for a line of solutions valid on $I_-$ between the middle two singular lines $\gamma_3$
and $\gamma_4$ in the range of negative $a_0$. As we would have to match any such solution at $R=0$ to a solution with 
negative asymptotic parameter $A$, these solutions are physically less relevant.

It has to be emphasised that none of the solutions for $I_-$ presented here extend to $I_{-\infty} \cup I_-$.
Instead they all end at a moveable singularity $R_c \approx 8$.

\subsection{Error analysis}
\label{errorana}

The actual numerical computations have been carried out with $25$ significant digits in Maple
and we used the values $\eps_0=5 \cdot 10^{-5}$ and $\eps_1=10^{-3}$ in \eqref{initconds0}
and \eqref{bsystem} and also in the matching for $R\le0$.
We will now go through a detailed error analysis for the example solutions plotted in fig. \ref{fig:expls}.

\begin{figure}[h]
 \begin{center}
  \includegraphics[width=0.4\textwidth,height=170pt]{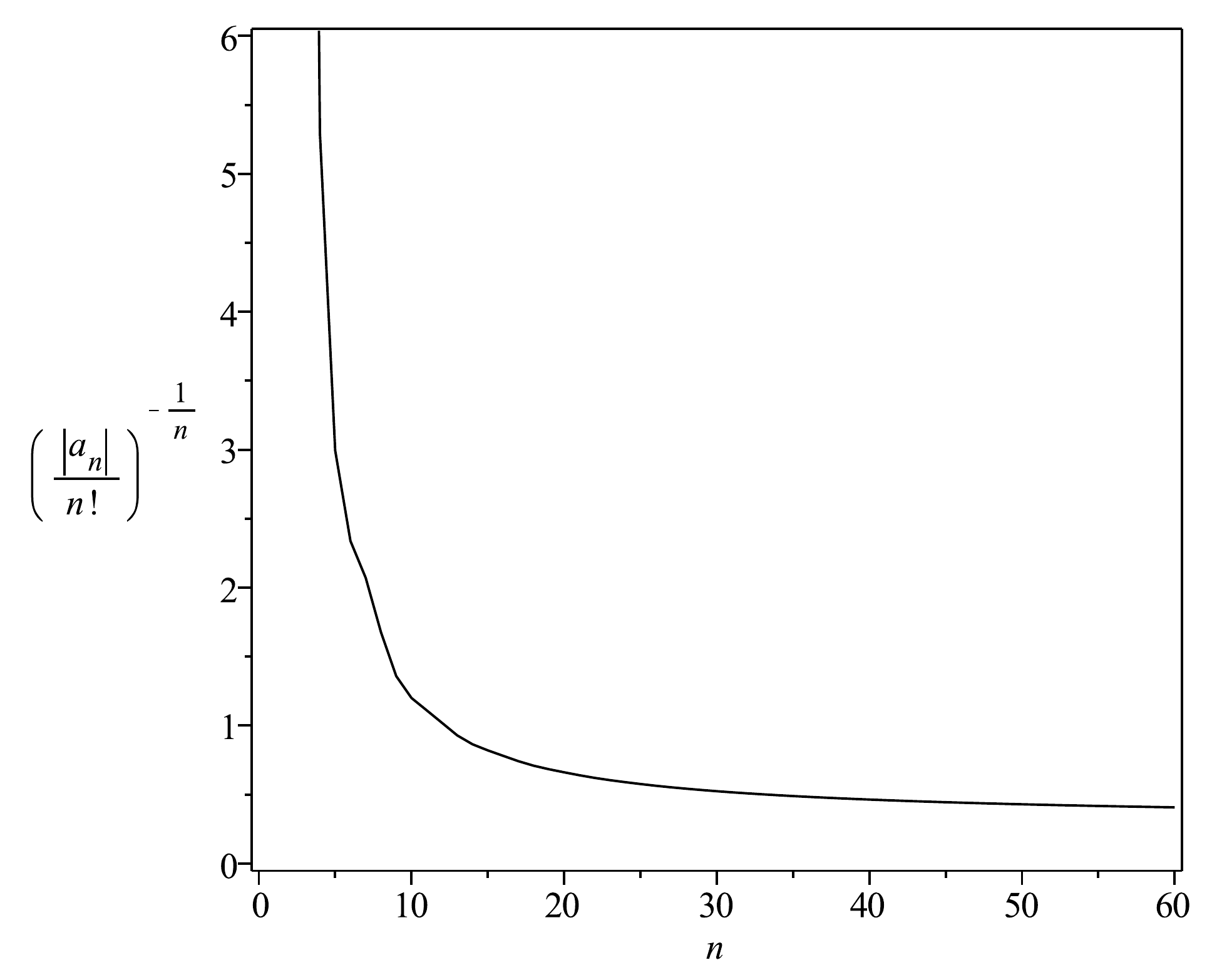}
\hspace{1cm}
  \includegraphics[width=0.425\textwidth,height=170pt]{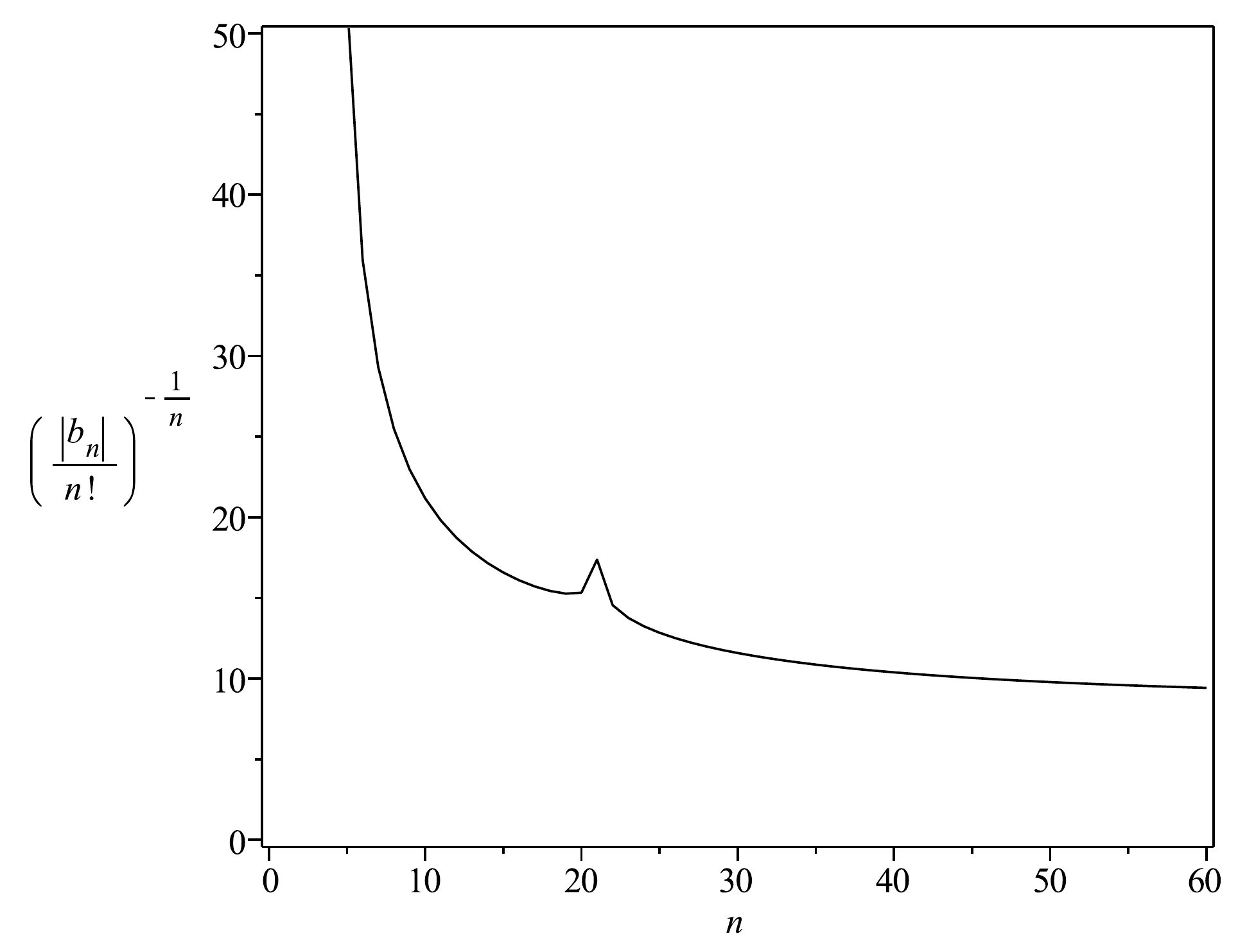}
 \end{center}
\caption{The result of a root test applied to the first $60$ coefficients of \eqref{tayexp0} on the left
and of \eqref{tayexpRp} on the right for the green fixed point solution in fig. \ref{fig:expls}.}
\label{fig:roottest}
\end{figure}

The first source of error in our solution strategy comes from the truncated Taylor
expansions \eqref{tayexpRp}-\eqref{tayexpRm}. We investigated the radius of convergence of \eqref{tayexp0}
and \eqref{tayexpRp} by performing a root test on the first $60$ coefficients for each of the example solutions.
The result for the green example solution is shown in fig. \ref{fig:roottest}.
The root test on the left hand side points to a convergence radius of $r_0 \approx 0.4$ for the Taylor
series around zero. The plot on the right pertains to a root test for the Taylor series
around $R_+$ and seems to indicate that the convergence radius of that series is $r_+ \approx 9$.
The two convergence radii have to satisfy the relation $r_+ \leq R_+ + r_0 $. Since $R_+ \approx 7.4$,
this is roughly satisfied for the above values for $r_0$ and $r_+$.

\begin{figure}[h]
 \begin{center}
  \includegraphics[width=0.5\textwidth,height=170pt]{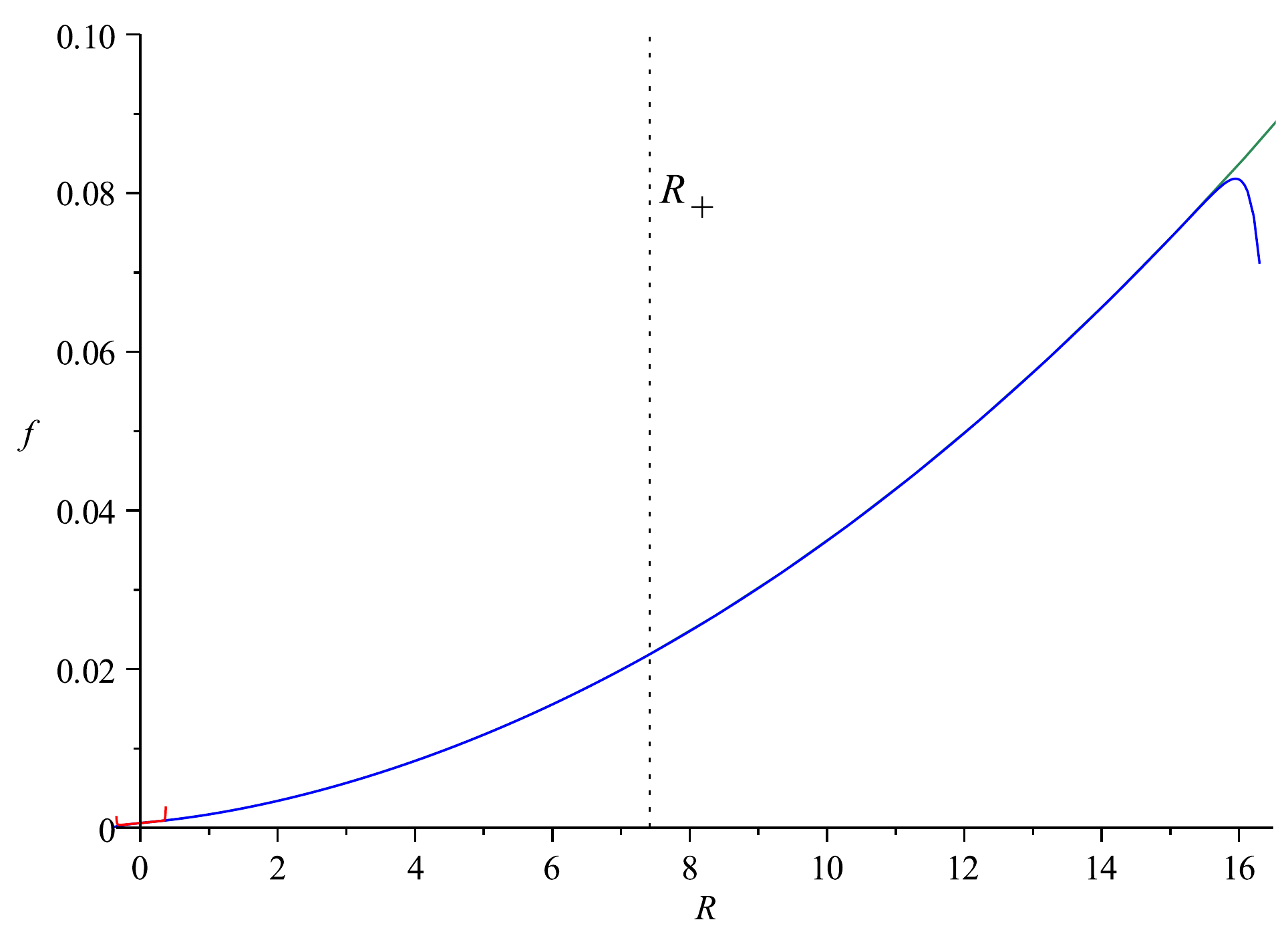}
  \includegraphics[width=0.4\textwidth,height=170pt]{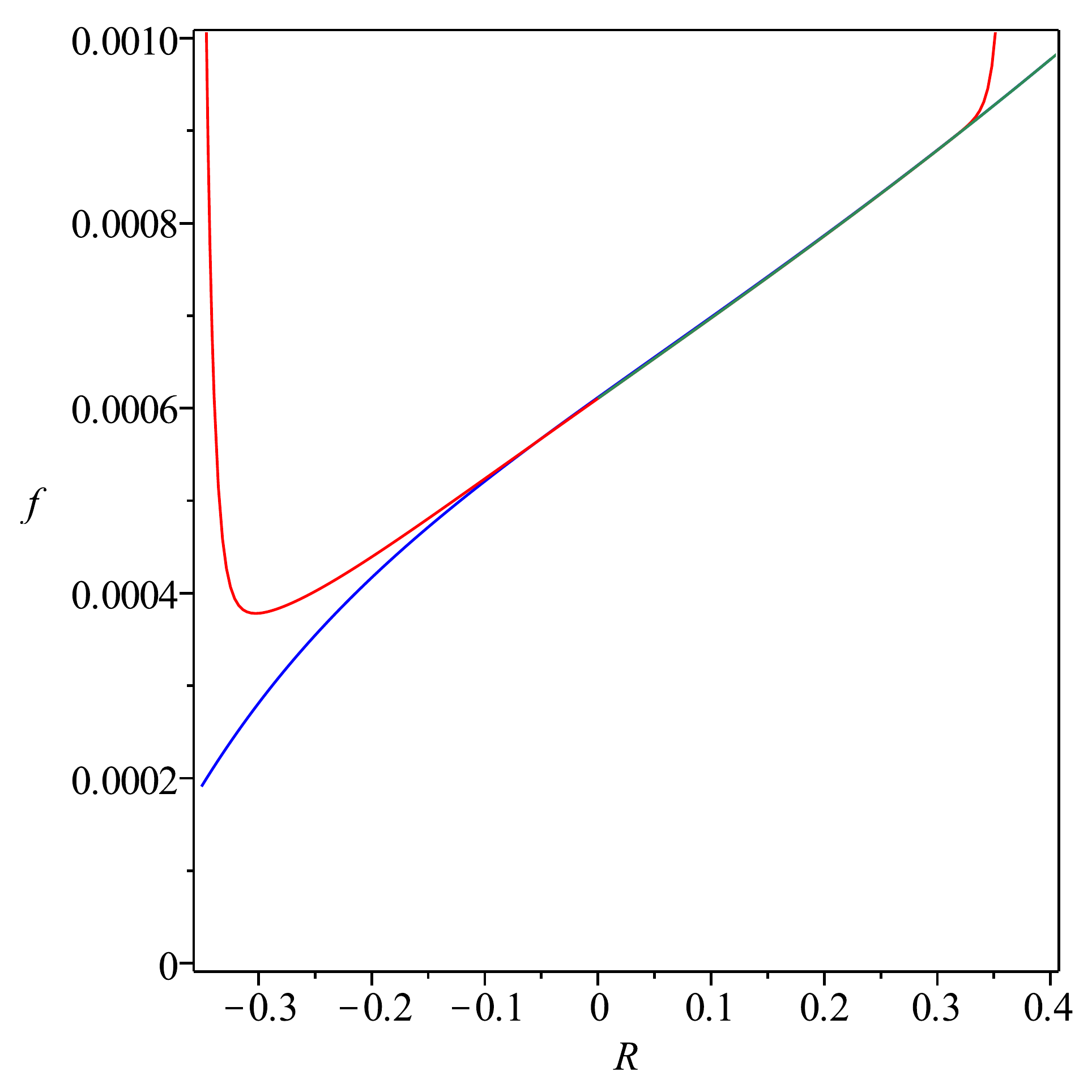}
 \end{center}
\caption{The green example solution together with its Taylor approximations of order $60$ around $R_+$ 
and zero in blue and red, respectively. The magnification on the right shows their behaviour round $R=0$.}
\label{fig:TaylorAndSol}
\end{figure}

If we use these Taylor approximations and plot them together with the numerically computed solution
we get fig. \ref{fig:TaylorAndSol}. We notice the rapid divergence of the Taylor expansions
from the numerical solution towards the end of the respective convergence interval but in between the numerical
solution is indistinguishable from the Taylor approximation.

Zooming in on the Taylor approximations in the region around $R=0$, \cf fig. \ref{fig:TaylorAndSol} (right),
we can see that for negative values of $R$ the $b$-series approximation in blue starts diverging
earlier than the red $a$-series approximation. This plot is consistent with a radius of convergence
$r_+ \approx R_+ + 0.4$ at $R=R_+$ which we would need to satisfy the aforementioned 
relation between $r_+$ and $r_0$. 

The analogous plots to fig. \ref{fig:TaylorAndSol} for the first and second
derivatives of $f(R)$ for the green solution example show equally good agreement with the only difference that
the Taylor expansions start diverging slightly earlier as the number of derivatives increases.
Altogether these results present strong evidence that we can trust the Taylor expansions used to bridge the singularities.

With this in mind, the error analysis proceeds along the following lines.
We estimate the relative error $\tilde \Delta_{a,s}$ in truncating the Taylor series \eqref{tayexp0},
representing the relative error on the initial conditions \eqref{initconds0},
by using the modulus of the last term in the Taylor expansion. If we now suppose the true initial conditions \eqref{initconds0} 
are within this error of the ones we actually have, 
we can determine the amount by which we have to vary
the two parameters $(a_0,a_1)$ in order to obtain this variation $\tilde \Delta_{a,s}$ of the initial conditions
and then take the result $\Delta_{a,s}$ as the corresponding error on the pair $(a_0,a_1)$.

We also have to consider the intrinsic error of the solver, \ie even if we started with
exact initial conditions at $\eps_0$, 
the integrator will induce a small error in each step
which leads to a numerical solution which is different from the true solution.
The size of this error $\Delta_{a,n}$ can be determined by using plots like
fig. \ref{fig:solcrit} and varying $a_1$ by ever smaller amounts. At some point we will be able to see numerical
fluctuations appearing and we have reached the maximum accuracy. These fluctuations limit the accuracy to which
we can determine the zero of $\delta f_\text{sol}$ we are looking for. The value of $\Delta_{a,n}$
can then be read off  from the $a_1$-axis of this plot.
Since these two errors are uncorrelated we obtain the total error on $(a_0,a_1)$ according to
\be \label{toterra}
\Delta_a=\sqrt{\Delta_{a,s}^2 + \Delta_{a,n}^2}.
\ee
To obtain the resulting error on the corresponding solution point $(b_0,b_1)$ 
we now evolve this total error $\Delta_a$ with the numerical integrator
by starting from the modified pair $\left(1+\Delta_a\right)\cdot(a_0,a_1)$ 
and compute the error $\Delta_{b,n}$ on the pair $(b_0,b_1)$.
We then also have to include the truncation error of the Taylor series at $R_+$, \viz \eqref{tayexpRp}, 
in a similar manner as for the truncation error at zero, 
and finally compute the total error
$\Delta_b$ on the pair $(b_0,b_1)$ with the corresponding form of \eqref{toterra}.

By similar methods we can assess the error on the asymptotic parameters $A,B,C$. At $R_\infty$ we have
two sources of errors. Firstly, we have to take into account that
the error on the point in the $b$-plane will have evolved to a new error $\Delta_{\infty,n}$
on the initial conditions at $R_\infty$ and hence on the asymptotic parameters $A,B,C$.
Secondly, we also have a truncation error $\Delta_{\infty,s}$ produced by the asymptotic series \eqref{asymp}. 
Again, we combine these two errors into the total error $\Delta_\infty$ on any of the parameters $A,B,C$ by using
the analogous form of \eqref{toterra}.
Compared to the errors on solution points in the $a$- and $b$-plane the error on the asymptotic
parameters is relatively large. This is to be expected from the fact that we have to integrate over
a comparatively long range, namely from $R_+$ to $R_\infty=200$, in order to determine these parameters and thus
the error on the solution point in the $b$-plane can evolve to larger values. Additionally,
we truncate the asymptotic series at order $R^0$, \ie we keep only the first three orders, which also contributes
to the size of $\Delta_\infty$.
Despite the magnitude of $\Delta_{\infty}$, the essential message delivered by the asymptotic parameters, that 
the constraint \eqref{safedisc} is fulfilled, remains unaltered for all example solutions.

The various values for the errors on points in the $a$- and $b$-plane as well as on the asymptotic parameters
are summarised in table \ref{tab:errs}.

\begin{table}[h]
\begin{center}
\begin{tabular}{|l|c|c|c|}
 \hline
  FP & green & magenta & black \\ 
  \hline
 \hline $\Delta_{a,s}$ & $4\cdot 10^{-11}$ & $3\cdot 10^{-12}$ &$3\cdot10^{-14} $ \\
 \hline $\Delta_{a,n}$ & $1\cdot 10^{-14}$ & $1\cdot 10^{-14}$ & $1\cdot 10^{-12}$  \\
 \hline $\Delta_a$ & $4\cdot 10^{-11}$ & $3\cdot 10^{-12}$ &  $1 \cdot 10^{-12}$  \\
 \hline \hline $\Delta_{b,s}$ & $2.8\cdot 10^{-7}$ & $1.4\cdot 10^{-7}$ & $1\cdot 10^{-9}$ \\
 \hline $\Delta_{b,n}$ & $3\cdot 10^{-9}$ & $2\cdot 10^{-10}$ & $3\cdot 10^{-10}$ \\
 \hline $\Delta_b$ & $2.8\cdot 10^{-7}$ & $1.4\cdot 10^{-7}$ & $1\cdot 10^{-9}$ \\
 \hline \hline $\Delta_{\infty,s}$ & $3.5\cdot 10^{-5} $ &  $5\cdot 10^{-5}$ & $2\cdot 10^{-5}$  \\
 \hline $\Delta_{\infty,n}$ & $1.8\cdot 10^{-5}$ &  $1.3\cdot 10^{-5}$ &  $3\cdot 10^{-7}$ \\
 \hline $\Delta_{\infty}$ & $4\cdot 10^{-5}$  & $5.2\cdot 10^{-5}$ & $2\cdot 10^{-5}$\\
\hline
\end{tabular}
\end{center}
\caption{Error values for the parameters given in table \ref{tab:par} of the fixed point solutions shown in fig. \ref{fig:expls}. 
$\Delta_{a,s}$, $\Delta_{a,n}$ and $\Delta_a$ are the $a$-series error, the numerical error and the total error on the solution point
in the $a$-plane. $\Delta_{b,s}$, $\Delta_{b,n}$ and $\Delta_b$ have the analogous meaning for the solution point in the $b$-plane
and $\Delta_{\infty,s}$, $\Delta_{\infty,n}$ and $\Delta_{\infty}$ for the asymptotic
parameters $A,B,C$. All errors are relative errors.}
\label{tab:errs}
\end{table}

We have verified that the position of these example solutions in the $a$- and $b$-plane remains unaltered if we decrease
the distances $\eps_0$ and $\eps_1$ used to compute the initial 
conditions close to zero and $R_+$.
This can be used to shrink the series errors $\Delta_{a,s}$ and $\Delta_{b,s}$ in table \ref{tab:errs} to smaller values.
In fact, the value for $\Delta_{b,s}$ in the case of the black example solution has been computed with $\eps_1=10^{-4}$
instead of the usual value $\eps_1=10^{-3}$ which has been used in all other cases. In this way we can make sure that
the different solution lines in the $b$-plane are not an artefact of truncating the $b$-series and would still be present
if we used an arbitrarily long Taylor series at $R_+$ despite the fact that they can get very close to each other.

\subsection{Structure of the $a$-plane} \label{sec:aplane}
From the configuration of lines in the $a$-plane on the right hand side in fig. \ref{fig:als} it might seem that the point where
the brown and the olive line from $I_-$ intersect the black solution line from $I_+$ on the singular line $\gamma_3$ represents
a fixed point function $f(R)$ defined on $[R_-,\infty)$. As we will see now this is not the case.
In performing the Taylor expansion \eqref{tayexp0} around zero we obtain expressions $a_n(a_0,a_1)$, $n=2,3,...$
such as \eqref{expra2}, relating the higher coefficients to the first two. As is already the case for $a_2$ these expressions have denominators that vanish along lines in the $a$-plane which are
shown as dashed black lines in fig. \ref{fig:as} together with the solution lines.
We have plotted the first four singular lines $\gamma_i$ corresponding to singular denominators of
the coefficients $a_i$ for $i=2,3,4,5$.
For example, the singular curve $\gamma_2$ is given by
\be \label{asing2} 
a_1=\gamma_2( a_0 ) = -\frac{3 a_0\left(64\pi^2 a_0 +1\right)}{192\pi^2 a_0-7}. 
\ee
Each coefficient $a_{n+1}$ receives an additional factor in its denominator compared to $a_n$ and so we get a series
of singular curves. The next one is $\gamma_3$,
\be \label{asing3} 
\gamma_3 ( a_0 )= -\frac{3a_0 \left(256 \pi^2 a_0+5\right)}{768\pi^2 a_0-25},
\ee
and similar expressions hold for all other singular curves.

It is indeed straightforward to derive a general expression for the $n$-th singular curve $\gamma_n$. 
Substituting the Taylor series \eqref{tayexp0} into the normal form of \eqref{fp2} we get $\gamma_n$ by setting
the coefficient of $a_n$ at order $R^{n-2}$ to zero, $n=2,3,\dots$.
The relevant 
terms can be picked out 
by hand from the normal form and one obtains the following condition:
\be
\frac{3}{(n-3)!}(a_1+a_0) + \frac{4}{(n-2)!}\left(192\pi^2 (a_1+a_0)a_0 -7a_1 +3a_0 \right)=0
\ee
This expression is also valid for $n=2$ if we adopt the convention to keep only the second term in that case.
If we solve this condition for $a_1$ we obtain a general expression for $\gamma_n$.

Let us now carefully look at $\gamma_3$.
The constraint equation we get from plugging \eqref{tayexp0} into \eqref{fp2} at first order in $R$ is given by
\be \label{a3cons}
 9\left( \left( -768 \pi^2 a_0 + 25 \right)a_1 -768 \pi^2 a_0^2 -15 a_0 \right) a_3
 + \alpha_{22}a_2^2 + \alpha_{21}a_2 + \alpha_{20}=0,
\ee
where $a_2$ is known from zeroth order as in \eqref{expra2} and the $\alpha_{ij}$ depend on $a_0$ and $a_1$ only:
\begin{align}
 \alpha_{22} &= 54 \left(-128\pi^2 a_0 +3\right)\\
 \alpha_{21} &= 6 \left(-576 \pi^2 a_1^2 + 3\left(-640\pi^2 a_0 +3\right)a_1 -640\pi^2a_0^2 -7a_0\right)  \\
 \alpha_{20} &= 8 \left( -144\pi^2 a_1^3 - \left( 480\pi^2 a_0+1\right) a_1^2 -6\left(48\pi^2 a_0+1\right)a_0a_1\right)
\end{align}
As explained before, the singular curve $\gamma_3$ is then simply given by setting the coefficient of $a_3$ in \eqref{a3cons} to zero.
For any solution $f(R)$ represented by a point on $\gamma_3$ we must also have
\be
\alpha_{22}a_2^2 + \alpha_{21}a_2 + \alpha_{20}=0.
\ee
Using \eqref{expra2}, we can determine a discrete set of solutions for $a_0$ and $a_1$  
from the last equation together with the condition that we are on $\gamma_3$. Solving for them and calculating $a_2$ gives
\be \label{crosspt}
 p_3 = (a_0,a_1)=\left(2.6589\cdot 10^{-4},1.9684\cdot 10^{-4}\right), \qquad a_2=1.9913\cdot 10^{-4}. 
\ee
There are actually five possible solutions for the present case but we have singled out the unique non-zero solution
corresponding to the point $p_3$ where the solution lines cross $\gamma_3$ in fig. \ref{fig:als} (right).
All other solutions have $a_0>7\cdot10^{-3}$ and are thus far beyond the range we have developed the solution lines in. 
This entails that any solution line crossing $\gamma_3$ has to do so at discrete points such as 
$p_3$. And indeed, this is exactly
what we can see in fig. \ref{fig:als}: all three solution lines, the two from $I_-$ and the one
from $I_+$ intersect $\gamma_3$
at $p_3$, the only point on $\gamma_3$ in the plot 
which can represent a solution. At this stage, the third coefficient $a_3$
is left undetermined at $p_3$, 
and all higher coefficients $a_4,a_5 \dots$ will
be unique functions of $a_3$. The parameter counting
arguments now tell us that we expect a discrete set of solutions on $I_+$ and another discrete set of solutions on $I_-$.
This is a heuristic argument why we should not expect to find a solution at $p_3$ which is valid on $I_-\cup I_+$
as these two discrete sets will in general not overlap. 

There is a way to determine the unique Taylor expansion at zero of any fixed point solution located on a solution
line running through $p_3$ which proceeds without determining the free parameter $a_3$ by numerical integration.
We should expect that the initial conditions at zero depend continuously
on the initial conditions at $R_+$. We can then exploit \eqref{a3cons} once more to determine $a_3$. The condition
of continuity implies that if we solve \eqref{a3cons}
for $a_3$ and take the limit as $a_0,a_1$ approach $p_3$ we will get a finite value.\footnote{whereas in the neighbourhood of $p_3$ the same process
for any point $p \in \gamma_3$, will produce a divergent expression.}
We can then proceed to obtain
a unique Taylor expansion at $p_3$ by exploiting the higher order constraints by which the coefficients $a_4,a_5,...$
are fixed. Since we are taking limits of an expression depending on two variables in which both numerator 
and denominator become zero, it is not surprising to find the limit value depends on the direction
along which we approach $p_3$. If $m$ denotes the slope of the straight line along which we take the limit, the function
\be
m \mapsto \lim_{(a_0,a_1) \rightarrow p_3} a_3(a_0,a_1,m)
\ee
depends smoothly on $m$. Thus there is no reason to expect that the limits along the solution lines from $I_+$ 
and $I_-$ have to coincide. In fact, computing numerically the values of the slopes in each case and taking the limit,
one finds the three different values
\be \label{a3vals2}
a_{3+}=-1.297\cdot 10^{-4}, \quad a_{3-}^b=9.518\cdot 10^{-4} \quad \text{and} \quad a_{3-}^o=7.239\cdot 10^{-4}
\ee
for the solution valid on $I_+$, the brown solution and the olive solution on $I_-$.

What we have illustrated here in the case of the singular curve $\gamma_3$ holds analogously for all the other singular
lines as well. This implies that the singular lines furnish a certain structure on the $a$-plane
since solution lines are not free to cross them at any point. On each singular line there are only a few
points that can be exactly determined where a solution line can go through. In general, the precise
Taylor expansion at zero of the solution located at one of these points will depend on the slope with which
the solution line crosses the singular line.

\begin{figure}[h]
\begin{center}
 \includegraphics[scale=0.6]{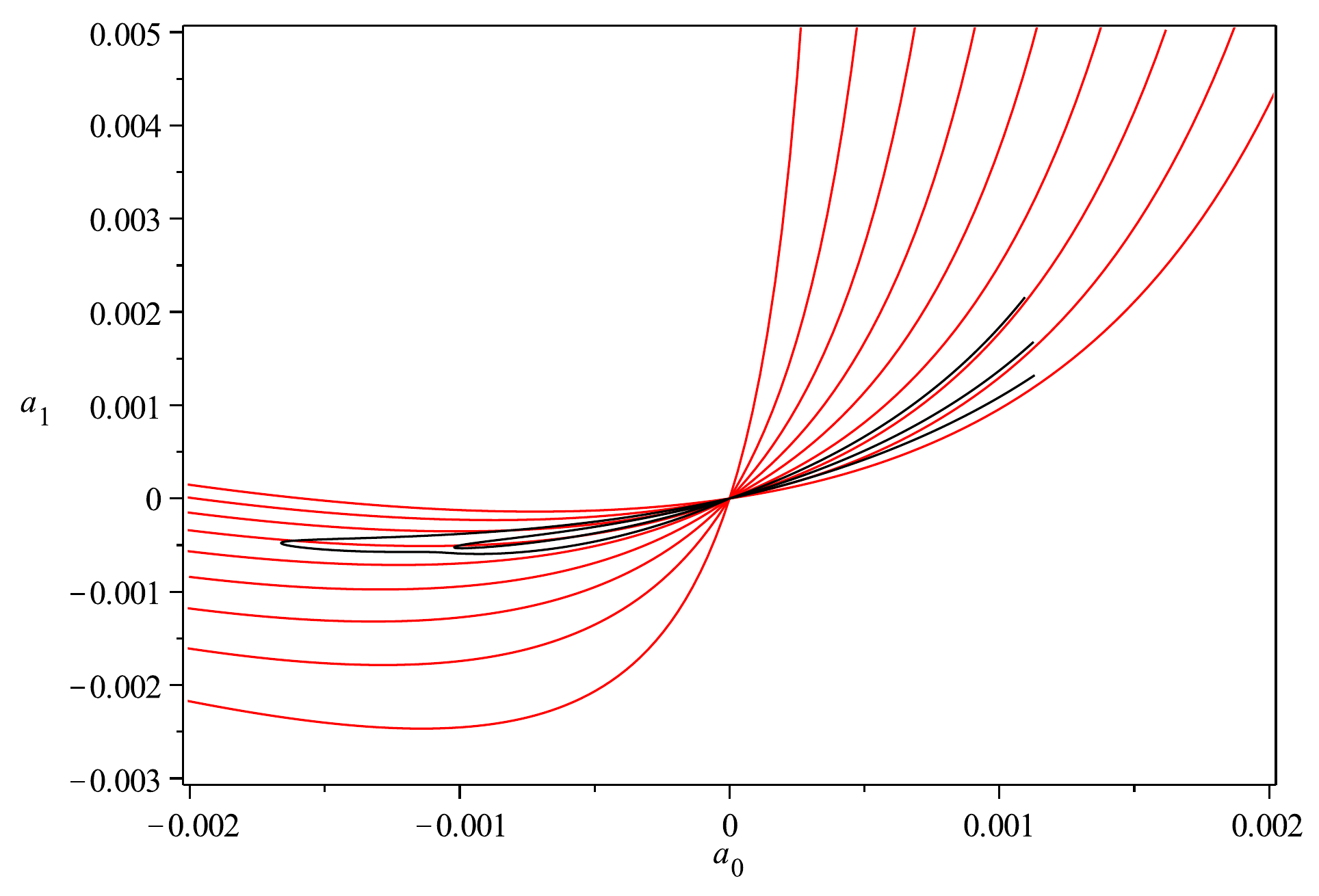}
\end{center}
 \caption{The solution lines (black) of fig. \ref{fig:as} in the $a$-plane together with the singular lines (red).
From bottom to top in the range of negative $a_0$ the singular lines are $\gamma_2, \gamma_3,\dots,\gamma_{10}$
stemming from the coefficients $a_2,a_3,\dots,a_{10}$ of the $a$-series \eqref{tayexp0}.}
\label{fig:singlinesbig}
\end{figure}

Let us finally comment on the reason why it is imperative to use \eqref{solcrit} with the behaviour shown in
fig. \ref{fig:solcrit} as a criterion to decide if we count a point as a solution point.
Consider searching for fixed point solutions on $I_+$ by integrating from $R_+$
along the lines described in sec. \ref{approach}, exploiting the corresponding form of \eqref{solcrit}
at zero instead of $R_+$. Furthermore, let us first focus on the range $a_0>0$ of the $a$-plane
in fig. \ref{fig:singlinesbig}, where we have plotted the first nine singular lines and the solution lines
for $I_+$.

According to its definition the quantity $\delta f_\text{sol}$ can
become arbitrarily small as we approach a solution line and shows singular behaviour as we cross
a singular line of the $a$-series. More precisely, since the $n$-th singular line is only relevant
for coefficients $a_k(a_0,a_1)$ with $k\geq n$ and we are working with an $a$-series of order $5$, 
$\delta f_\text{sol}$ will only show singular behaviour close to the lines $\gamma_2,\dots,\gamma_5$.
Away from these two special cases $\delta f_\text{sol}$
has a typical size that varies across the $a$-plane (as long as we stay above $\gamma_2$
in fig. \ref{fig:singlinesbig} numerical integration from $R_+$ to zero is generically possible despite
the threat posed by moveable singularities). Speaking in orders of magnitude, $\delta f_\text{sol}$
starts at $10^{-3}$ between $\gamma_2$ and $\gamma_3$ and drops to $10^{-8}$ in the area
between $\gamma_4$ and $\gamma_5$. If we now take into account the next order in the Taylor
expansion around zero, \ie we include the term with coefficient $a_6(a_0,a_1)$, we find that its size
is always several orders of magnitude smaller than $\delta f_\text{sol}$ in these regions
and thus will have no effect on existing solution lines.
This changes once we consider the region above $\gamma_5$ towards $\gamma_6$ and beyond.
Here, the critical quantity $\delta f_\text{sol}$ becomes comparable in size to the $6$th order correction
of the $a$-series, both evaluating to around $10^{-14}$ as we approach $\gamma_6$.
When searching for solutions in a region above the singular curve $\gamma_n$ 
it is therefore essential to use a Taylor series at zero of order at least $n+1$.
For example, using a Taylor series of order $6$ at zero we found a first solution point
of a solution line situated between $\gamma_5$ and $\gamma_6$ in the region of positive $a_0$
and we would expect that more solution lines can be found continuing this pattern.

Coming back to why we should not just require $|\delta f_\text{sol}|$ to be smaller than some upper bound
for a point to be regarded a solution point, it is clear from this discussion that there is no sensible
way to find such a bound.
Moreover, the value of $|\delta f_\text{sol}|$ depends sensitively on $\eps_0$. For example,
with a $5$th order Taylor series at zero and $\eps_0=5\cdot10^{-5}$ a scan of the type in fig. \ref{fig:solcrit}
between $\gamma_5$ and $\gamma_6$ at $a_0 = 7\cdot 10^{-4}$ produces values of $\delta f_\text{sol}$
between $10^{-12}$ and $10^{-14}$ but $\delta f_\text{sol}$ never changes sign. Decreasing $\eps_0$
in several steps to $\eps_0=10^{-6}$, the same scan shows correspondingly decreasing values of $|\delta f_\text{sol}|$
reaching $10^{-18}$ to $10^{-21}$ in the last step. During none of these steps there is a sign change
of $\delta f_\text{sol}$ and thus no evidence for a solution line. This again shows that just imposing
an upper bound on $|\delta f_\text{sol}|$ cannot be sufficient. It also confirms the necessity
of the $6$th order term in the $a$-series if one intends to find solutions between these two singular lines.

What we have described here in the region $a_0>0$ of the $a$-plane in fig. \ref{fig:singlinesbig} holds 
analogously in the region of negative $a_0$. Indeed, the bottom part of the large solution line on this side
has been developed with a Taylor series of order $6$ at zero from the beginning.

One might also wonder if there are similar singular lines in the $b$-plane arising from
the expansion \eqref{tayexpRp}. An investigation into the expressions for its coefficients indeed shows
that such singular lines are present. However, the solution lines we have found in the $b$-plane
are all a safe distance away from them and we therefore do not expect a similar picture.

\section{Discussion and Conclusions}
\label{conclusions}
In this section we highlight the main technical developments, then the main discoveries and their physical interpretation, and finally discuss the extent to which we should trust these discoveries given the methods used to derive the $f(R)$ approximations in the first place.

Our main overall conclusion is a technical one: the parameter-counting arguments, introduced in scalar field theory at the LPA and higher levels of derivative expansion \cite{Morris:1994ki,Morris:1994ie,Morris:1994jc,Morris:1996xq,Morris:1998da} and adapted here to this context, remain a very powerful way to determine what to expect from the equations in terms of fixed points and perturbations. The counting argument is the statement that any restrictions to the dimension of the parameter space arising from fixed singular points and the asymptotic behaviour, act independently, and therefore, if there are any global (or partial) solutions, the dimensionality of their parameter space  can be determined by counting  up the number of such restrictions.  Although the application in this paper is to the asymptotic safety programme in quantum gravity, this simple counting argument should be applicable far more generally, providing crucial insight whenever one is searching for smooth solutions to highly non-linear renormalisation-group-type equations with little or no inherent symmetry (\cf sec. \ref{whycount}).

That this insight is crucial is underlined by our numerical investigation in sec. \ref{numerics}. The basic method of attack is as follows:
series expansions are developed analytically around the fixed singular points and around infinite field, and used to `patch' onto numerical
solutions in between. Apparent violations of the counting rules arise quite often during numerical investigations of the resulting parameter
space of partial solutions. If one trusts the rules, this motivates further careful numerical and analytical studies at these points.
As explained in sec. \ref{sec:aplane}, we found a whole region in which $\delta f_\text{sol}$, as given by
the corresponding version of \eqref{solcrit} at zero, fell to less than $10^{-14}$. By decreasing $\eps_0$ this fell to $10^{-21}$, but never changed sign. We insist that a sign change is required in order to identify a solution. 
This phenomenon 
 led to the recognition of the importance of higher terms in the $a$-series and the corresponding curves in the $a$-plane where these coefficients diverge
  (except at special points).  As a result, we uncovered a beautiful structure which is effectively responsible for the lines of solutions and guides them through the $a$-plane, \cf fig. \ref{fig:singlinesbig}.

Important for the application of the counting argument is a thorough understanding of the analytic asymptotic expansion of the fixed point solutions and their eigen-perturbations. We showed how to check that the full parameter space of asymptotic behaviour has been uncovered, and how to develop systematically the higher orders, even in cases such as the current one where the differential equations and the asymptotic series are highly involved. Again, these technical developments are valid much more generally, thus we expect them to be important in applications other than to asymptotically safe quantum gravity.

Now we turn to the discoveries we made by applying these methods to the $f(R)$ approximations for asymptotically safe quantum gravity as introduced in refs. \cite{Machado,Codello,Benedetti}. Using the counting argument we showed that the $f(R)$ approximations derived in refs. \cite{Machado,Codello} have no global smooth solutions, and thus are unsatisfactory. 

The reason can be traced to fixed singular points introduced via their choice of cutoff \cite{Benedetti}. In ref. \cite{Benedetti}, the cutoff was chosen so that these were avoided, however three fixed singular points remain: $R=0$ and $R=R_\pm$. We showed that the asymptotic series for large $R$ developed in ref. \cite{Benedetti} was only part of the full parameter space of asymptotic solutions. The full asymptotic series solution provides no restriction on the dimension of the parameter space but only restricts the three parameters, $A,B,C$, to lie inside a cone \eqref{safedisc}. 

The $f(R)$ approximations are derived on four-spheres, and thus we have $0<R<\infty$. Actually the point $R=0$ must also be included to satisfy the quasi-locality requirement of the Wilsonian renormalisation group as discussed in sec. \ref{fpeqns}. Since only the $R=0$ and $R=R_+$ singular points are operative, the counting arguments show that continuous lines of fixed points should exist. We confirmed this in sec. \ref{numerics}, by uncovering five
 separate lines of fixed points, and evidence for a sixth
 line. As we saw, there are very possibly more lines, but uncovering these will require further increases in the number of significant digits kept during the numerical calculations.

A fixed point solution on one of these lines is determined for example by providing the initial conditions $f(0)=a_0$ and $f'(0)=a_1$. We showed how the existence and position of these lines in $(a_0,a_1)$ space can be understood from the singularity structure of the Taylor series coefficients at $R=0$.
We have uncovered lines with $a_1<0$ as well as lines with $a_1>0$, corresponding to both possible signs of
the coefficient for the $\sqrt{g} R$ term in Minkowski signature. Similarly, the cosmological constant term, corresponding to $a_0$, is negative on some of the lines of fixed points we analysed and positive on others.
However, the sign of $-a_1$ and $a_0$ have no direct connection to the sign of the physical Newton's gravitational constant $G$ and cosmological constant $\Lambda$ that we would predict from this theory. To compute these latter quantities would require adding all the (marginally) relevant perturbations about the fixed point and integrating out along the corresponding renormalised trajectory, taking the limit as $k\to0$ in such a way as to obtain an effective ${\tilde f}({\tilde R})$ parametrised by the finite (marginally) relevant couplings. The predicted Newton's constant and cosmological constant then follow from its Taylor expansion (see for example the corresponding situation for scalar field theory in ref. \cite{Morris:1996xq}).

Of more importance is the fact that we have found lines of solutions with either sign for $A$. Only the solutions with $A>0$ have a chance to be physically sensible since these
 correspond to fixed point solutions $f(R)$ that are bounded below (in both Euclidean and Minkowski signature) and which in unscaled physical variables having completed the functional integration (\ie in the $k\to0$ limit) result in a positive ${\tilde f}({\tilde R})=A {\tilde R}^2$. (Note that this does not mean that the fixed points correspond to an $R^2$ type theory: infinitely many equally important interactions have been sent to zero in this limit.)

We developed the full asymptotic expansion for infinitesimal perturbations around these fixed points in sec. \ref{asymptoticeigen} and furthermore analysed the asymptotic behaviour of finite perturbations in sec. \ref{further}. Unlike in the LPA for scalar field theory, we saw no reason to exclude some of this asymptotic behaviour. This is not surprising, given that here the asymptotic expansion of the fixed point itself does not restrict the dimension of the parameter space. In this case the counting arguments show that each fixed point supports a continuous infinity of operators, and operator dimensions are not quantised. Therefore we are forced to conclude that all these fixed points are unsatisfactory.

We saw that the situation can be resolved if we also include consistently spaces with $-\infty<R<0$, as we are required to do, since  quantum gravity should also make physical sense for such spaces. We therefore searched for smooth global solutions over the full range $-\infty<R<\infty$. The counting arguments now tell us that any such fixed points are discrete in number and support a quantised eigen-operator spectrum. However, after developing a detailed understanding of the structure of the parameter space, and performing an extensive numerical search, we failed to uncover any solutions in this case. The closest we came to finding such solutions numerically, were  partial solutions which do not extend to $R=-\infty$ but which are defined on the interval $[R_-,\infty)$.

To what extent can we trust all these discoveries as uncovering the real physics of the situation rather than artefacts of the approximations and methods used? Although, as we pointed out in the introduction, truncation to general functions $f(R)$ is still severe, breaking in an uncontrolled way both scheme invariance and the underlying diffeomorphism invariance of the quantum field (expressed through modified BRS Ward identities \cite{Reuter1}), and further followed by smoothing approximations, we will see that some of the effects clearly have their origin in some generic qualitative features and whose presence can be justified on physical grounds.

As we have sought to demonstrate, our methods for analysing the resulting differential equations are powerful and exhaustive and we feel confident therefore that the parameter space has been fully explored, except in the following respects. Firstly, as already mentioned, there are almost surely further lines of fixed points valid in $I_+\cup I_\infty$ whose development requires additional terms in the $a$-series.
Secondly, at the expense of time-consuming numerical integration, it would be possible to
extend some of the known solution lines further into the region of larger values of $a_0$
where their exact behaviour is still to be determined.

The fact that we need to consider consistently also $R<0$ surely reflects the real physics of the situation. However, we are now using the equations outside the four-sphere region in which they were derived. As we noted in sec. \ref{introduction}, really we should use an $f(R)$ approximation that consistently takes into account throughout its derivation, spaces in which $R$ can be negative. This task is beyond the scope of the present paper.

Even if we restrict ourselves to $R\ge0$, we uncovered many novel features and these are consequences of the fixed singular points $R=0,R_+$ and the novel asymptotic behaviour. 

As  discussed at the end of sec. \ref{review}, the $R=0$ singular point arises as a consequence of choosing cutoffs that amount to modifying the Laplacian, \cf \eqref{typeI}, thus forcing the cutoff to depend on $f''$. This also turns the fixed point equation from a second order ordinary differential equation, as would have been expected from the structure of \eqref{erg}, into a third order one. The parameter counting is thus unaltered, suggesting that the $R=0$ singular point is not an unwanted artefact but is in fact necessary, reflecting the true physics of the situation  \cite{Benedetti}. 

In the LPA we find that the leading asymptotic behaviour is determined by dimensions, and this is because quantum fluctuations can be neglected at large field $\ph$ \cite{Morris:1994ki, Morris:1998da}. Here we find as in ref. \cite{Benedetti}, that the large field behaviour of the fixed point solution is again as expected by dimensions. But the reason is more subtle, as discussed below \eqref{theps}. This is reflected in the fact that for the perturbations, the asymptotic behaviour is not determined by dimensions -- see below  \eqref{v0eq}. 

Recalling that we multiplied through by $R^2$ in \eqref{flow2}, we see that the quantum fluctuations for the fixed point solution are $O(1)$ in this regime, whilst for perturbations they vanish as $O(1/R)$. In scalar field theory, the quantum fluctuations in the sharp cutoff case are also $O(1)$ (up to logarithmic corrections) \cite{Morris:1994ki} whilst for smooth cutoffs they vanish as $\ph\to\infty$, see \eg \cite{Morris:1994ie}. The crucial difference is that here the left hand side of \eqref{flow2} is multiplied by the volume which is vanishing in the large field limit, as $1/R^2$, making the classical contribution subdominant in the large field regime. This points to a fundamental physical reason for the novel asymptotic behaviour: at the same time as $R\to\infty$, where one might na\"\i vely expect to be able to neglect quantum corrections in the Wilsonian renormalisation group, the volume of the space is vanishing and this forces quantum fluctuations to remain important. It would be desirable to strengthen this conclusion through a  more thorough investigation of the asymptotic dependence of quantum fluctuations on choice of cutoff. 

This is all the more important since the physical scalar quantum fluctuations determine even the leading asymptotic behaviour of the perturbations, as we saw from the discussion surrounding \eqref{v0eq}. As we noted below \eqref{theps}, the leading $R^2$ behaviour for the fixed point itself comes about by balancing the physical scalar quantum corrections against the non-physical scalar quantum corrections.  It is tempting to speculate that these effects are somehow a reflection of the instability present in the conformal mode, which at first sight leaves no imprint on the exact renormalisation group \cite{Reuter1}.

The final key feature was the presence of the fixed singular point at $R=R_+$. As discussed in ref. \cite{Benedetti} and reviewed at the end of sec. \ref{review}, this arises from the failure to integrate out the constant physical scalar mode. More study seems called for in order to establish whether this is really a generic feature of cutoff choices that are otherwise acceptable \cite{Benedetti}, and to understand the consequences, both from the renormalisation group point view and more generally, of leaving behind such an unintegrated mode. 

Note that intuitively it is clear that the underlying diffeomorphism invariance (as expressed through modified Ward identities) does not allow complete freedom in choosing separate cutoffs for the component fields in \eqref{TT} and the ghosts and auxiliary fields, otherwise we would fail completely to recover the fact that the components must all effectively combine back into the quantum field $h_{\mu\nu}$. However the extent to which the cutoff choices are constrained, and the consequences of not obeying such constraints, has yet to be explored in the literature.

Returning to the earlier forms of the $f(R)$ approximation \cite{Codello,Machado}, we have already seen that their choice of cutoff resulted in fixed singular points which in turn ruled out the existence of any global smooth fixed point solutions.  Similar, but $f(R)$-dependent, single poles are responsible for the unphysical termination of certain renormalised trajectories in polynomial truncations  (\eg termination of type IIIa trajectories as described in ref. \cite{Reuter:2001ag,Machado}). It has been conjectured that these terminations are an artefact of the truncations used \cite{Machado,Codello} and indeed it has been shown that they can be avoided in a suitably chosen larger space of operators, but this is at the expense of introducing non-polynomial functions of $R$ which violate quasi-locality \cite{Machado}. In ref. \cite{Codello}, it was noted that the termination of these trajectories can be avoided if so-called type III cutoffs are used. This mechanism bears similarities to the avoidance of fixed singularities in \eqref{flow2} by use of a cutoff of type II (see the discussion below \eqref{Sigma2} and ref. \cite{Benedetti}). 

Clearly this all points to the fact that care needs to be exercised over the choice of cutoff.  Apparently sensible choices of cutoff can result in a break-down of the Wilsonian renormalisation group, perhaps only in certain regions of parameter space, and this is evidenced by unphysical behaviour. In fact such a situation is well established in simpler systems and in some cases rigorously proved \cite{Sokal:1994un}.

We see from these investigations that unsurprisingly the form of the cutoff plays a crucial r\^ole in the nature of the solutions. 
In any case, in view of the novelty and number of unexpected features that we have uncovered, more investigations using different cutoffs and different types of approximation at this LPA-like level of truncation, or beyond, are evidently called for.

\section*{Acknowledgments}
We thank Dario Benedetti for useful discussion and alerting us to an error in the published paper
and first arXiv version of ref. \cite{Benedetti}.

\section*{Appendix. An early version of the fixed point equation of ref. \cite{Benedetti}}
\renewcommand{\thefigure}{A.\arabic{figure}}
\renewcommand{\theequation}{A.\arabic{figure}}
\addcontentsline{toc}{section}{Appendix. An early version of the fixed point equation of ref. \cite{Benedetti}}
At an earlier stage of this work, before we became
aware of an error in the first version of
ref. \cite{Benedetti}, we had carried out an analysis of the uncorrected version of
the fixed point equation \eqref{fp2}.
This first form of the equation is obtained by deleting the term $36(R^2+2)(f'+6f'')$
in the contribution of the physical scalar modes in \eqref{Tbh} and the corresponding term
 in \eqref{th}. 
This change does not affect the technical advances in
 this work such as the validity
of the parameter counting method, the structure of the asymptotic expansion 
and eigen-perturbations, or the conclusions drawn
from these. The main features of the set of solutions of the fixed point equation
we had uncovered also resemble the results presented in sec. \ref{numerics}.
We describe these solutions here
to illustrate the general applicability of the methods used in this work for analysing such equations as \eqref{fp2}.

In total we found four solution lines in the range $I_+ \cup I_\infty$ and one solution line 
on the interval $I_-$.
\begin{figure}[h]
\begin{center}
\includegraphics[width=0.7\textwidth,height=210pt]{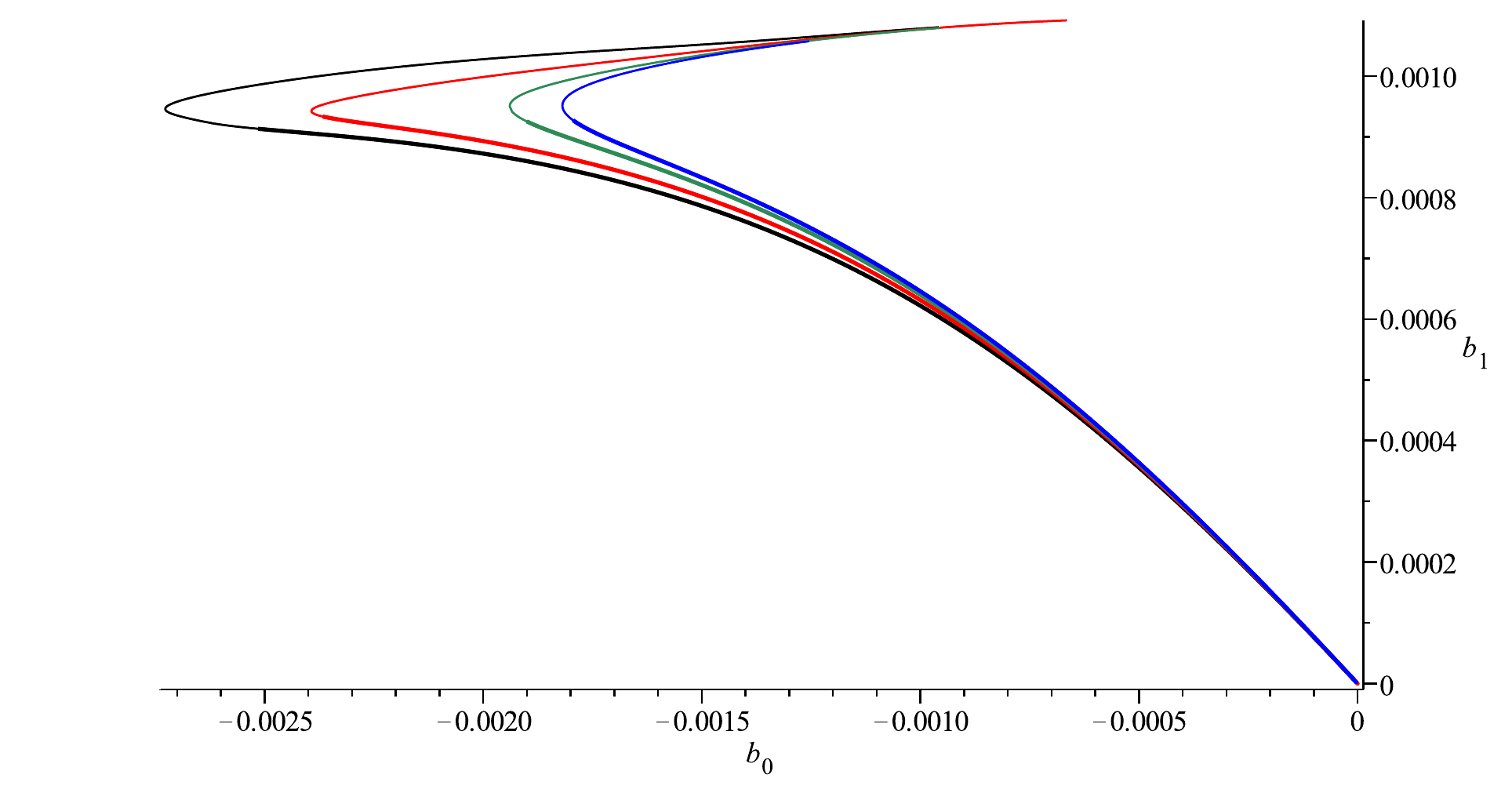} 
\end{center}
\caption{The four lines of solutions in the $b$-plane. Different solution lines
are represented consistently in different colours (from left to right these are black, red, green and blue).
The bold part of each line distinguishes solutions valid for all $R\geq0$ from
those valid only on $I_+$.}
\label{fig:bsold}
\end{figure}

\begin{figure}[h]
\begin{center}
\includegraphics[width=0.7\textwidth,height=210pt]{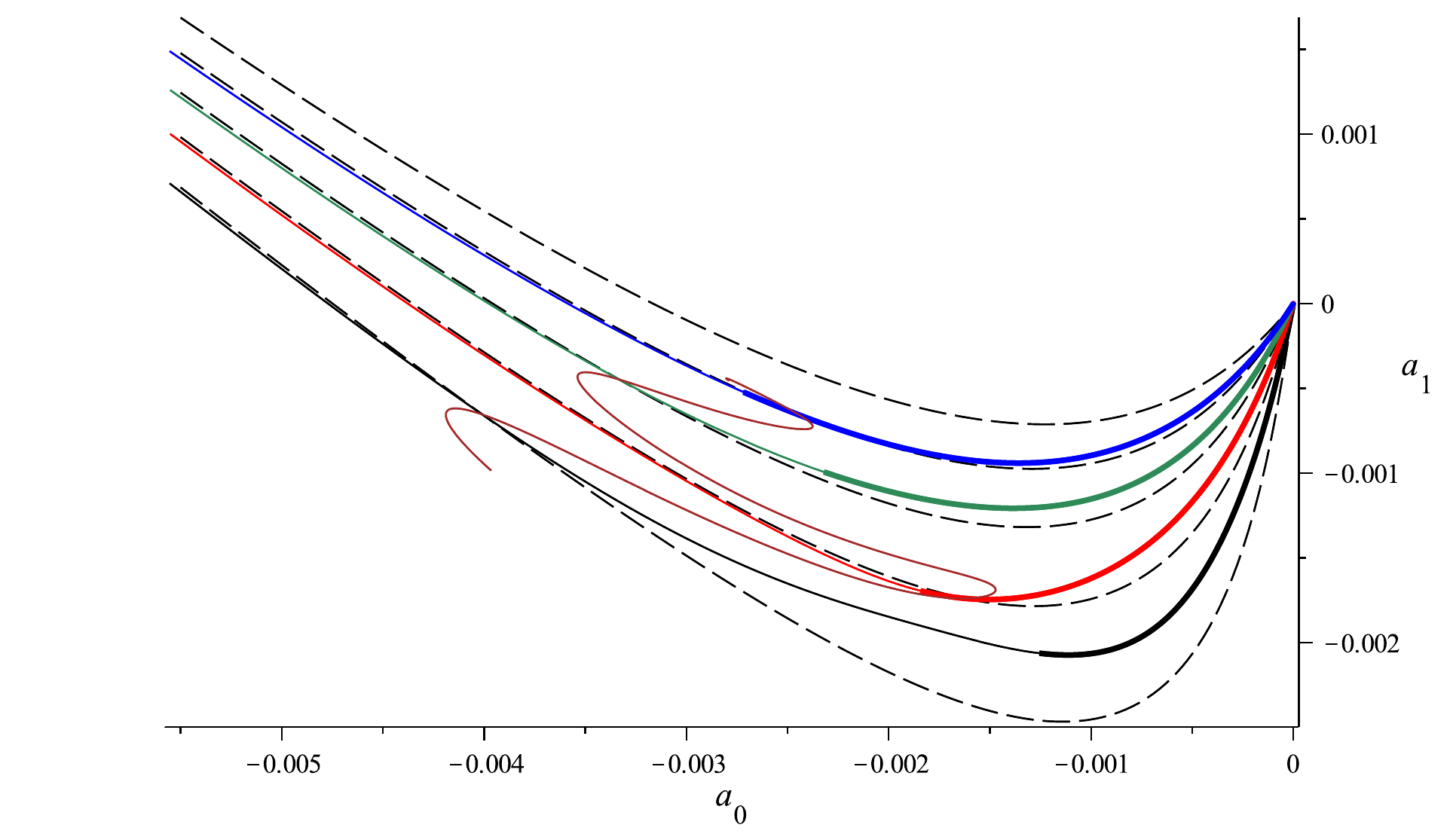}
\end{center}
\caption{The four lines of solutions in the $a$-plane corresponding to the $b$-lines in fig. \ref{fig:bsold}
(from bottom to top in black, red, green and blue) together with the only solution line for $I_-$ in brown.
The solutions valid for all $R\geq 0$ are again given by the bold parts of the lines.
Along each of the black dashed lines a particular coefficient in
the Taylor series around zero becomes singular.}
\label{fig:asold}
\end{figure}

\begin{figure}[h]
\begin{center}
 \includegraphics[width=0.4\textwidth,height=150pt]{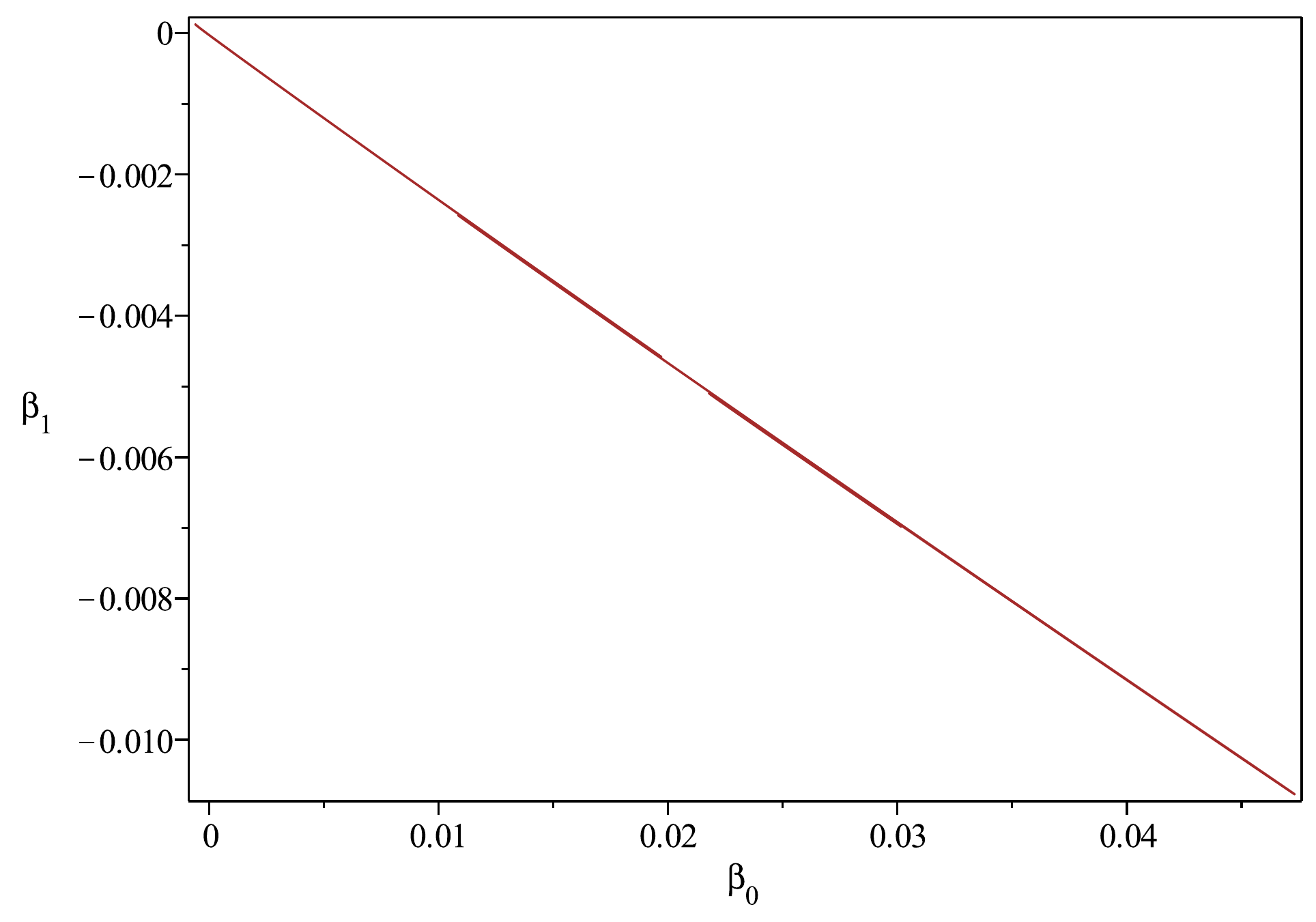}
 \includegraphics[width=0.5\textwidth,height=150pt]{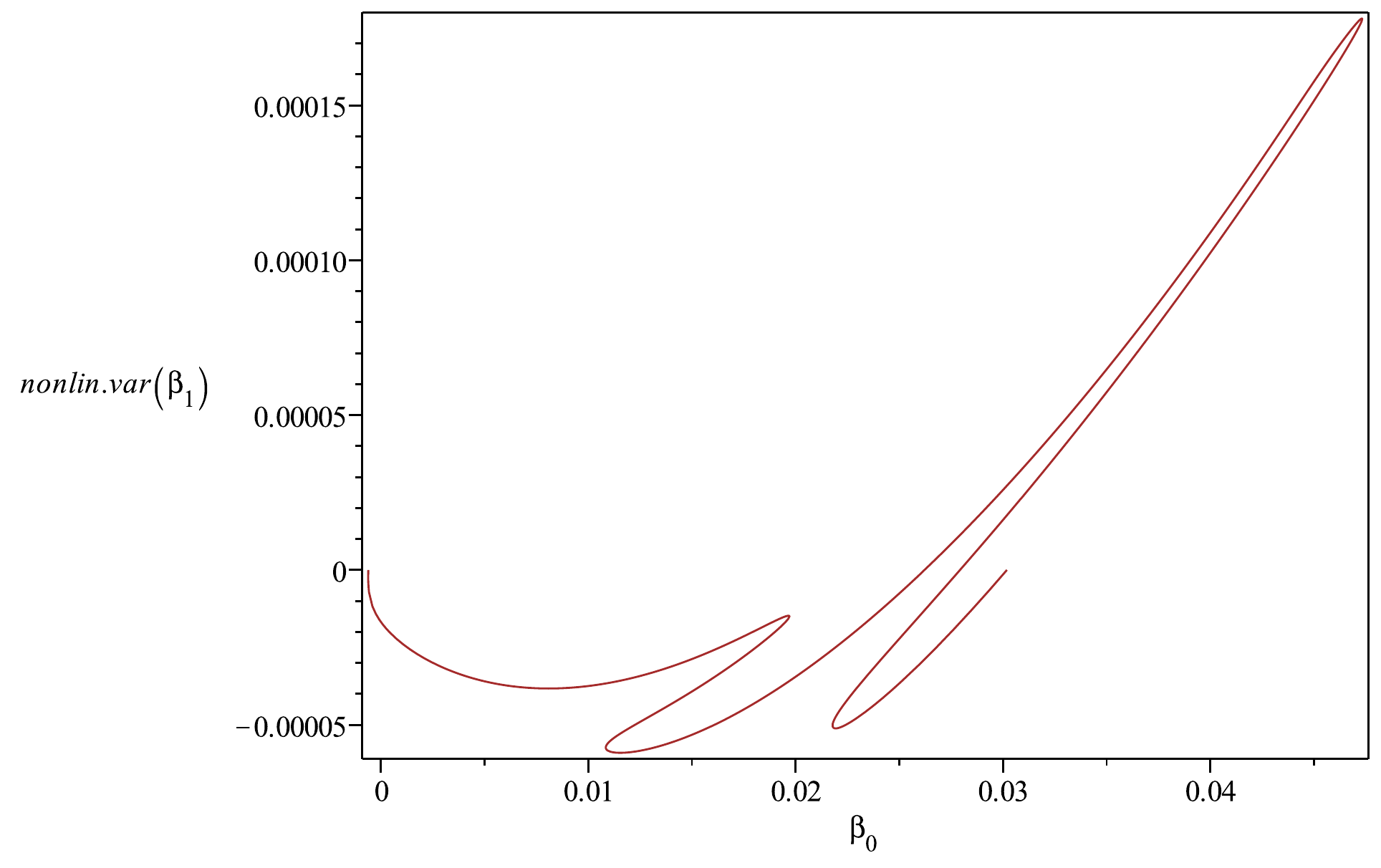}
\end{center}
 \caption{The (only) solution line in the $\beta$-plane is shown on the left together with its nonlinear variations
 on the right. The straight line $m \cdot \beta_0+c$ that has been subtracted from the line on the left has slope
 $m=-0.2312$ and $c=1.14\cdot 10^{-5}$.}
 \label{fig:btsold}
\end{figure}

They are shown in fig. \ref{fig:bsold} and \ref{fig:asold}. 
One main difference to the results in sec. \ref{numerics} is the existence of a compact subset for each solution
line for $I_+$ containing only those points that define fixed point solutions valid on $I_+ \cup I_\infty$
with asymptotic parameters $A,B,C$ satisfying a
 modified version of \eqref{safedisc},
\be
\label{safediscold}
\frac{49}{20}A^2 > B^2 + C^2 \,.
\ee
With regard to solutions for negative values of $R$ the only solution line we were able to find
is also shown in fig. \ref{fig:asold}. The corresponding solution line in the $\beta$-plane is displayed
in fig. \ref{fig:btsold}. It does not contain a bold part which means that no solutions valid
for all $R\leq0$ were found. Like the solution lines in the $\beta$- and $b$-plane of sec. \ref{numerics}, 
the solution line in the $\beta$-plane is mostly linear with only small non-linear variations superimposed.
In this analysis we did not use both roots  
of the quadratic constraint for the second 
coefficient of the $b$- and $\beta$-series which may be why 
the solution lines end at the origin
of their respective plane.
 
Just as in sec. \ref{numerics}, we also have singular lines associated with the Taylor expansion around zero
shown as black dashed lines in fig. \ref{fig:asold}. Furthermore, all the conclusions drawn in connection
with singular lines in sec. \ref{sec:aplane} hold in the present situation. In particular,
solution lines are allowed to cross singular lines at only a finite number of points on them as
is beautifully illustrated by the brown solution line for $I_-$ in fig. \ref{fig:asold} cutting through
the singular lines precisely where the solution lines for $I_+$ do.
Similarly, we do not obtain fixed point solutions valid on $I_- \cup I_+$ at these points since
the value of the coefficient in the $a$-series associated with the singular line depends 
on the slope with which the solution lines cross the singular line.


\end{document}